\documentclass[preprint,12pt]{elsarticle}

\usepackage{amssymb}

\usepackage[margin=0.75in]{geometry}
\usepackage[hyphens]{url}
\usepackage{adjustbox}
\usepackage{lipsum}
\usepackage{graphicx}
\usepackage[utf8]{inputenc}
\usepackage[T1]{fontenc}
\usepackage{newtxtext}
\usepackage{newtxmath}
\usepackage{bm}
\usepackage{latexsym} 
\usepackage{color}
\usepackage{amsmath}
\usepackage{amsfonts}
\usepackage{amssymb}
\usepackage{supertabular}

\usepackage{multirow}
\usepackage{setspace}
\usepackage[percent]{overpic}
\usepackage{textcomp}
\usepackage{multicol}
\usepackage{booktabs}
\usepackage{tabularx}
\usepackage{units}
\usepackage{tablefootnote}
\usepackage[USenglish]{babel}
\usepackage{times}
\usepackage{etoolbox}
\usepackage{blindtext}
\usepackage{pifont}
\usepackage{diagbox}
\usepackage{appendix}
\usepackage{threeparttable}
\usepackage[normalem]{ulem}
\usepackage{lscape}
\usepackage{subcaption}
\usepackage{caption}
\usepackage{array}
\usepackage{hyperref}
\usepackage{lineno}
\usepackage{floatrow}
\usepackage{picture}

\biboptions{round,authoryear}

\journal{Journal of Fluids and Structures}

\begin{document}

\begin{frontmatter}


\title{Modeling and Simulation of a Fluttering Cantilever in Channel Flow}



\author{Lu{\'i}s Phillipe Tosi\corref{cor1}}
\ead{ltos@caltech.edu}
\cortext[cor1]{Graduate Research Assistant, Mechanical Engineering}
\author{Tim Colonius\corref{cor2}}
\cortext[cor2]{Professor, Mechanical Engineering}
\address{Division of Engineering and Applied Science \\California Institute of Technology, Pasadena, CA 91125}

\begin{abstract}
Characterizing the dynamics of a cantilever in channel flow is relevant to applications ranging from snoring to energy harvesting. Aeroelastic flutter induces large oscillating amplitudes and sharp changes with frequency that impact the operation of these systems.  The fluid-structure mechanisms that drive flutter can vary as the system parameters change, with the stability boundary becoming especially sensitive to the channel height and Reynolds number, especially when either or both are small.  In this paper, we develop a coupled fluid-structure model for viscous, two-dimensional channel flow of arbitrary shape. Its flutter boundary is then compared to results of two-dimensional direct numerical simulations to explore the model's validity.  Provided the non-dimensional channel height remains small, the analysis shows that the model is not only able to replicate DNS results within the parametric limits that ensure the underlying assumptions are met, but also over a wider range of Reynolds numbers and fluid-structure mass ratios.  Model predictions also converge toward an inviscid model for the same geometry as Reynolds number increases.  


\end{abstract}


\end{frontmatter}



\section{Introduction}

The stability of an elastic member within a channel, or as part of the channel, has been studied for many decades \citep{Johansson1959,Miller1960,Inada1988,Inada1990,Nagakura1991}.  Applications include wind instruments \citep{Sommerfeldt1988,Backus1963}, human snoring \citep{Balint2005,Tetlow2009}, vocalization \citep{Tian2014}, and enhanced heat transfer \citep{Shoele2015,Hidalgo2015}.  Recently, this geometry has also been used for flow-energy harvesting, with devices specifically targeting power generation for remote sensor networks \citep{Sherrit2014,Sherrit2015,Lee2015,Lee2016}.
For most of these applications, the flutter instability boundary is the essential result sought, as the functional requirements of engineering designs (i.e. flow-energy harvester, heat management systems) or the manifestation of sound in natural systems (i.e instruments, snoring, voice) depend on it. A particular challenge with characterizing the flutter onset is its dependence on the specific flow regimes, which vary between and within
applications noted.  In this paper, we target the prediction of the flutter instability over laminar and turbulent regimes, and show that with specific restrictions to fluid-to-structure mass and throat-to-length ratios, our model agrees well with both viscous numerical simulations and inviscid modeling results.  

With recent advancements, it has become possible to directly simulate the fluid-structure interaction (FSI) numerically by solving the Navier-Stokes equations coupled to a model of the structure. Two dimensional FSI algorithms have been used to study the stability of an elastic member within channel flow \citep{Tetlow2009,Shoele2015}, and more recently to assess the effect of Reynolds number on the ensuring flutter boundary \citep{Cisonni2017}.  Yet, one of the challenges with fluid-structure systems is the large number of parameters necessary to describe the subset of possible system regimes; 
fully coupled computational approaches require considerable computing time to capture the flutter instability, often only being able to span a fraction of this parameter space.  
A more tractable but less accurate (or versatile) approach involves modeling of the structure displacement and velocities, with fluid forces approximated via simplified equations of motion. The early works of \citeauthor{Miller1960} \citep{Miller1960} and \citeauthor{Johansson1959} \citep{Johansson1959} address the divergence instability in this light specifically targeting channels within nuclear reactor cooling systems. More recent work by \citeauthor{Guo2000} \citep{Guo2000} takes an inviscid approach to understanding the onset of flutter in a symmetric channel. \citeauthor{Alben2015} \citep{Alben2015} solves for the inviscid flutter boundary using a vortex sheet model, and \citeauthor{Shoele2015} \citep{Shoele2015} extend the plane wake vortex sheet method by \citeauthor{Alben2008} \citep{Alben2008,Alben2015} to asymmetric channel flow. 

As recently evident from \citep{Cisonni2017}, the inviscid treatment of channel flow for small throat-to-length ratios
is unable to predict stability boundaries that more generally depend on the Reynolds number.  Large deviations are observed between their results and the inviscid models, particularly when both Reynolds number and throat-to-length ratio are small. Moreover, \citeauthor{Cisonni2017} \citep{Cisonni2017} are unable to address the bridge between the viscous and inviscid stability regimes due to the Reynolds number restriction in their numerical simulations.  

Apart from fully-coupled numerical simulations that include viscous effects, it is possible to include viscosity in an approximate way in reduced-order models.   \citeauthor{Nagakura1991} \citep{Nagakura1991} and \citeauthor{Wu2005} \citep{Wu2005} employ models with a viscous formulation of fluid forces for elastic beams in channel flow.  This form for fluid forces was originally proposed by \citeauthor{Inada1988} \citep{Inada1988} for rigid plates in converging or diverging channels, with \citeauthor{Fujita1999} further expanding the analysis to flexible, cylindrical constant channel flows \citep{Fujita1999,Fujita2001,Fujita2007}.  Yet, the parametric bounds for validity of this model remain largely untested.  This includes not only viscosity effects, but also geometrical parameters (i.e. throat-to-length ratio), and fluid-to-structure mass and stiffness ratios. 


This paper addresses these parameter bounds first by deriving a fully coupled fluid-structure model that considers flow confinement in arbitrarily shaped channels through a set of assumptions grounded in parametric limits.  We then compare the model predicted flutter stability boundaries to those from fluid-structure direct numerical simulations (DNS), as well as to results of the inviscid model from \citeauthor{Shoele2015} \citep{Shoele2015}. Though derived differently, our model accounts for fluid forces in a similar fashion to \citep{Nagakura1991,Wu2005} when the constant channel geometry is considered, and to \citeauthor{Inada1988} \citep{Inada1988} when a rigid beam is considered. The difference comes as a factor 
in the advection term, and careful attention to the parametric validity established through the control volume analysis and subsequent approximations of the geometry and flow field.

Our DNS algorithm uses an immersed-boundary projection formulation developed by \citeauthor{Taira2007} in \citep{Taira2007,Colonius2008}.  The complementary FSI algorithm is a strongly-coupled formulation between the flow and the structure developed by \citeauthor{Goza2017} \citep{Goza2017}.  Both modeling and simulations only consider two-dimensional flow, with the modeling further simplifying the problem into a quasi-1 dimensional framework.  An extension to three-dimensional flow is possible \citep{Tosi2018thesis}, and will the subject of forthcoming publications.


The paper is organized as follows.  We first derive the quasi-1 dimensional fluid-structure model, which captures the linearized incompressible Navier-Stokes and solid equations of a cantilever beam in an arbitrarily shaped channel configuration.  
Next, the fluid-structure direct numerical simulation algorithm is described, along with the dynamic mode decomposition employed to extract the frequency, growth rate, and beam shapes of unstable modes.  Lastly, modeling and simulation results for a cantilever beam in constant channel flow are compared over a broad range of parameters.  Our model results are then compared to those from the inviscid model as Reynolds number is varied.

\section{Quasi-1 Dimensional Fluid-Structure Model} \label{sec:lubeclosure}

\begin{figure}[H]
    \centering
    \begin{subfigure}[t]{0.45\textwidth}
    \centering
   	\includegraphics[width=.95\textwidth]{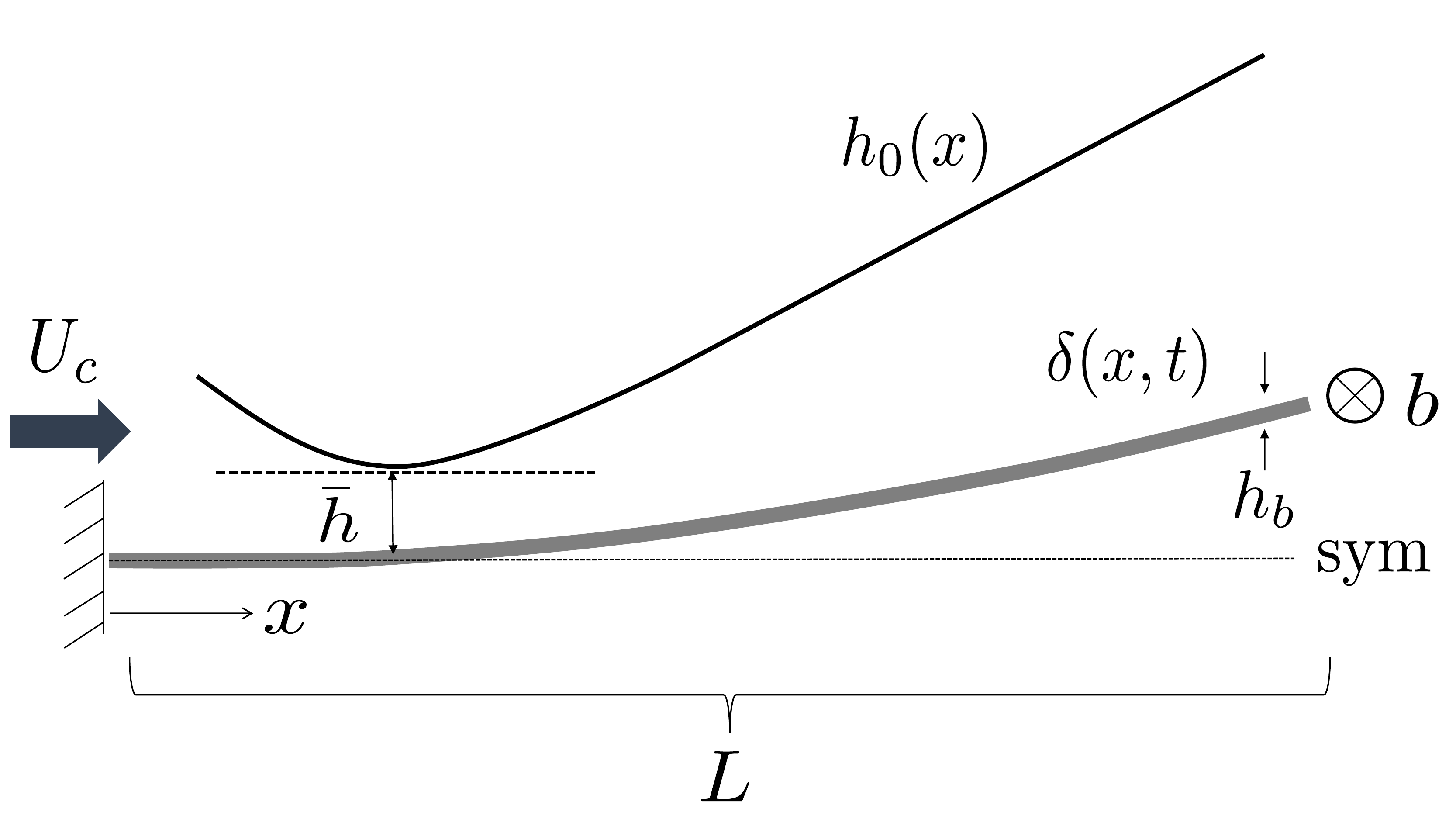}
    \caption{}  
    \label{fig:NozzleDiffuserGeometryl}  
    \end{subfigure}%
    ~ 
    \begin{subfigure}[t]{0.45\textwidth}
    \centering
   	\includegraphics[width=.65\textwidth]{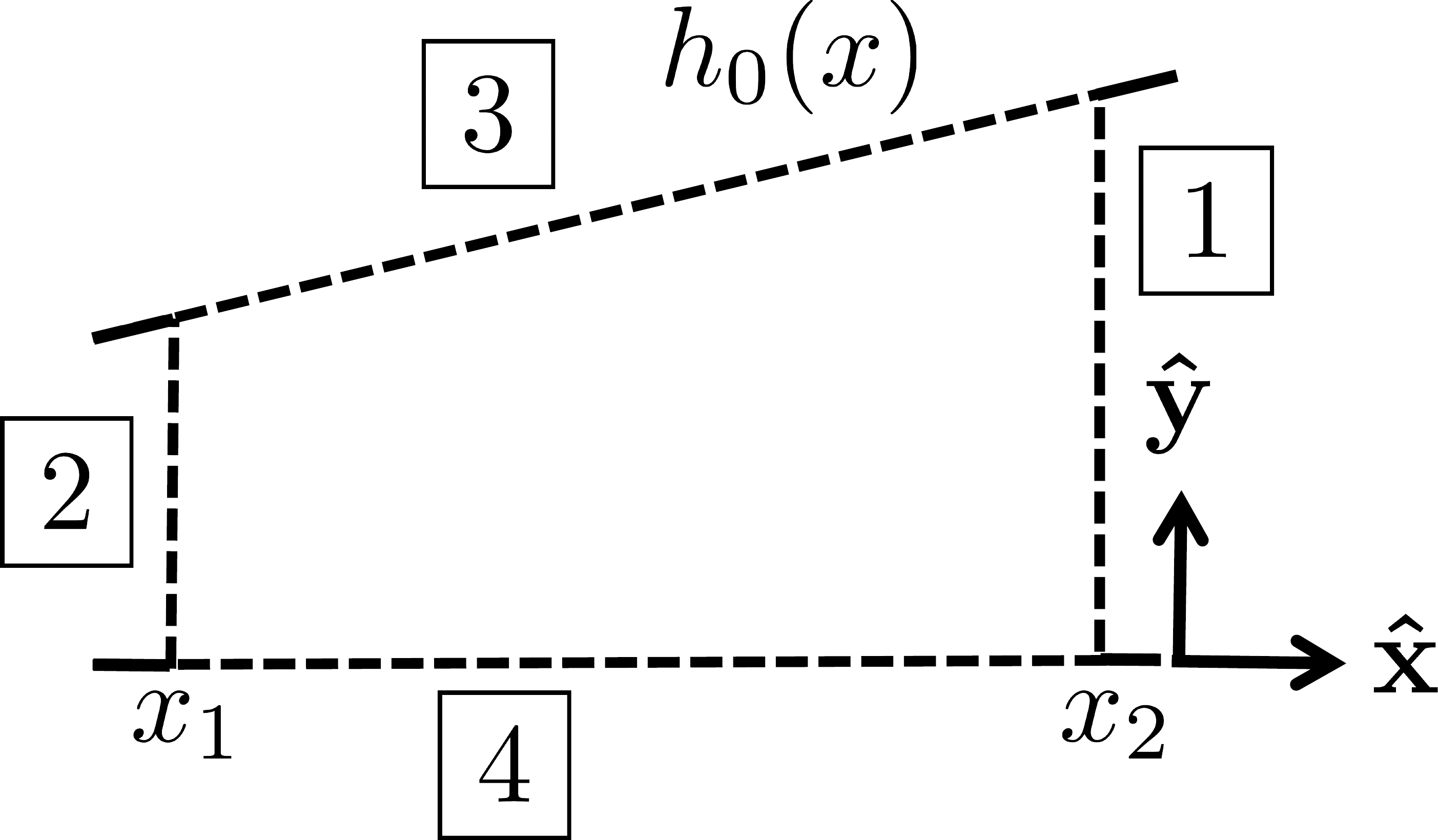}
    \caption{}  \label{fig:2DControlVolume}  
    \end{subfigure}
    \caption{Two-dimensional channel geometry (left) and control volume (right).} 
\end{figure}

The geometry illustrated in figure \ref{fig:NozzleDiffuserGeometryl} and dimensional parameters in table \ref{tab:dimensionpars} are inspired by the flow energy harvester configurations in \citep{Sherrit2014,Sherrit2015}.
We consider the beam displacement as the output of a system defined by a characteristic velocity (or flow rate), geometrical parameters, and material properties, for a total of 10 possible nondimensional groups that determine its dynamics.  This large number of parameters makes a thorough numerical or experimental investigation difficult.  The purpose of this section is to provide a simple model that allows us to analytically identify the most important properties affecting the stability of the system.

\begin{table}[hbt!] 
\centering
\caption{Table of fluid-structure dimensional parameters.} \label{tab:dimensionpars}
\begin{tabular}{| c  c  c  | }  \hline 
\textbf{Variable} & \textbf{Description} & \textbf{Dimension} \\ 
\hline
 $\delta$ & beam displacement & $l$ \\
 $x$ & beam length coordinate & $l$\\   
 $t$ & time & $t$\\  
 $U_c$ & characteristic velocity & $l*t^{-1}$ \\  	
 $L$ & beam length & $l$\\  				
 $h_b$ & beam thickness & $l$\\  			
 $b$ & beam width & $l$\\  					
 $\bar{h}$ & throat height & $l$\\  			
 $\rho_f$ & fluid density & $m*l^{3}$ \\  				
 $\mu_f$ & fluid viscosity & $m*l^{-1}*t^{-1}$ \\  		
 $\rho_s$ & beam density & $m*l^{-3}$\\  				
 $E$ & Young's modulus & $m*l^{-1}*t^{-2}$\\  			
 $\nu$ & Poisson's ratio & ND  \\ \hline					
 \end{tabular} 
\end{table}

\subsection{Structure equations of motion}  \label{sec:FSIOperator2D}

Considering the cantilever in figure \ref{fig:NozzleDiffuserGeometryl} in transverse vibration, we apply the undamped, two-dimensional Euler-Bernoulli beam equation \citep{Inman2008},
\begin{equation} \label{eq:EulerBernoulliBeam2D}
\rho_s h_b \frac{\partial^2}{\partial t^2} \delta(x,t) + 
\frac{\partial^2}{ \partial x^2} \left( E I_{\mathrm{2D}} \frac{\partial^2 }{\partial x^2} \delta(x,t)  \right) = 
P^{\mathrm{bot}}(x,t) -  P^{\mathrm{top}}(x,t),
\end{equation}
where the two-dimensional area moment of inertia is 
\begin{equation} \label{eq:MomofIntertiaI2D}
I_{\mathrm{2D}} = \frac{h_b^3}{12 (1-\nu^2)}.
\end{equation}
$P^{\mathrm{bot}}(x,t)$ and $P^{\mathrm{top}}(x,t)$ are the pressures acting on the bottom and top of the beam, respectively (i.e. per superscript).  The beam is clamped at its leading edge and free at its trailing edge, so the boundary conditions are
\begin{equation} \label{eq:StructureBC_ClampedFree}
\delta(0,t) = 0, \ \frac{\partial }{ \partial x} \delta(0,t) = 0, \ 
\frac{\partial^2 }{ \partial x^2} \delta(L,t)= 0, \ 
\frac{\partial^3 }{ \partial x^3} \delta(L,t)= 0.
\end{equation}
In considering the flow separately in the top and bottom channels, we write the geometrical constraint,
 \begin{equation} \label{eq:BeamShapeConstraint}
\delta(x,t) = \delta^{\mathrm{top}}(x,t) = -\delta^{\mathrm{bot}}(x,t).
\end{equation}

\subsection{Fluid Equations of Motion} 
\label{sec:EquationsofMotion}

To develop a relation between pressure, beam displacement and its derivatives, we consider the control volume defined in $\mathbf{\hat{x}}$ and $\mathbf{\hat{y}}$ illustrated in figure \ref{fig:2DControlVolume}, which corresponds to a small section of one of the channels in figure \ref{fig:NozzleDiffuserGeometryl}.  The channel is formed between the moving cantilever boundary at a variable height $\delta(x,t)$ and the upper surface at a specified location, $h_0(x)$.

We apply mass and momentum conservation to this control volume under the simplifying assumptions of constant fluid density and a gradually-varying channel in the streamwise direction, $h_0'^2 \ll 1$ and $\delta'^2 \ll 1$, such that $\sqrt{1 + h_0'^2} \approx 1$ and  $\sqrt{1 + \delta'^2} \approx 1$ for $x \in [0,L]$.  From mass conservation, we obtain
\begin{equation} \label{eq:MassCons2D_final}
-\frac{\partial \delta}{ \partial t} + \frac{\partial Q_x}{\partial x} = 0,
\end{equation}
where
\begin{equation} \label{eq:FlowRatexdef}
Q_x = \int_{\delta}^{h_0} u dy,
\end{equation}
is the volume flow rate.  
From momentum conservation, 
\begin{equation} \label{eq:MomConsInt_sub}
\resizebox{.9 \textwidth}{!}{$
\begin{aligned}
\frac{\partial}{\partial t} \left( \int_{\delta}^{h_0} \begin{bmatrix}
u \\ v \end{bmatrix} dy \right) +
\frac{\partial}{ \partial x} \left( \int_{\delta}^{h_0} \begin{bmatrix}
u^2 \\ uv \end{bmatrix} dy \right)  = 
-\frac{1}{\rho_f} \left\{ 
\frac{\partial}{ \partial x} \left( \int_{\delta}^{h_0} \begin{bmatrix}
P \\ 0 \end{bmatrix} dy \right) - 
\begin{bmatrix} h_0' \\ -1 \end{bmatrix} P |_{y=h_0} + 
\begin{bmatrix} \delta' \\ -1 \end{bmatrix} P |_{y=\delta}  - 
\mathbf{\mathbf{F}_{\mathrm{visc}}} \right\},
\end{aligned}
$}
\end{equation}
where the $\mathbf{\hat{x}}$ component, after substituting $Q_x$,
\begin{equation} \label{eq:MomCons2D_first}
\frac{\partial Q_x}{\partial t} +
\frac{\partial \mathcal{N}_x}{ \partial x}  =
-\frac{1}{\rho_f} \left\{ 
\frac{\partial}{ \partial x} \left( \int_{\delta}^{h_0} P dy \right) - 
h_0' P |_{y=h_0} + \frac{\partial \delta}{\partial x} P |_{y=\delta}  - 
F_{\mathrm{visc},x} \right\},
\end{equation}
\begin{equation} \label{eq:NonlinAdvectionDef}
\mathcal{N}_x = \int_{\delta}^{h_0} u^2 dy,
\end{equation}
and $F_{\mathrm{visc},x}$ is the net viscous stress acting on the walls.

To make further progress, we need a closure relation for $\mathcal{N}_x$ and $F_{\mathrm{visc,x}}$ in term of $Q_x$ and $\delta$, as well as a relation between the integrated and evaluated quantities of $P$ in equation \ref{eq:MomCons2D_first}.  
Such a closure can be rigorously determined for a narrow gap where both
$\frac{\bar{h}}{L} = \hat{h} \ll 1$ and $\hat{h}^2 Re_L \ll 1$ so that inertial terms and streamwise diffusion of momentum can be locally neglected, i.e. the lubrication limit \citep[p. 319]{Kundu2012}.  
This lubrication scaling of the N-S equations leads to the pressure being constant across the gap, $P = P(x,t)$, and the streamwise velocity becoming parabolic.
Using the definition in equation \ref{eq:FlowRatexdef} and integrating a parabolic $u$ 
leads to expressions for $Q_x$ in terms of $\delta$ and $P$.  $\mathcal{N}_x$ and $F_{\mathrm{visc,x}}$ can then be evaluated as
\begin{equation*}
\mathcal{N}_x  = \frac{6}{5} \frac{Q_x^2}{h_0 - \delta} 
\text{,  and  }
F_{\mathrm{visc,x}} =
-12 \mu_f \frac{ Q_x }{\left( h_0 - \delta \right)^2},
\end{equation*}
with the latter taking the form for a Newtonian fluid.

Even when $\hat{h}^2 Re_L$ is not small, a closure can be determined under the weaker assumption that the profile shape is unchanging with $x$ and $t$.  For a particular velocity profile $u$, we define a \emph{profile factor}, $\xi_x$  and Fanning \emph{friction factor}, $f$, so that the appropriate relations become
\begin{align} 
\mathcal{N}_x  & = \xi_x \frac{Q_x^2}{h_0 - \delta}, \label{eq:NonlinAdvectionGen} 
\\
F_{\mathrm{visc,x}} & =
- \frac{f(Q_x)}{4} \frac{Q_x^2}{\left(h_0 - \delta \right)^2},
\label{eq:FrictionTermGen}
\end{align}
respectively.  
Following \citep{Shimoyama1957,Inada1988,Nagakura1991}, we take
\begin{equation} \label{eq:FrictionTerm_Shimoyama}
f =
\begin{cases}
48 Re_h^{-1} & \text{if } Re_h < 1000\\
0.26 Re_h^{-0.24}& \text{if } Re_h \geq 1000
\end{cases},
\end{equation}  
whereas we model the profile factor as
\begin{equation} \label{eq:NonlinearTerm_general}
\mathcal{\xi}_x =
\begin{cases}
6/5 & \text{if } Re_h < 1000\\
1 & \text{if } Re_h \geq 1000
\end{cases},
\end{equation}
where the laminar value ($Re_h < 1000$) coincides with the aforementioned lubrication result, and the turbulent case follows from the blunted mean velocity profile in the outer region and neglects the thin inner region.

With this closure and $P = P(x,t)$, we need not consider the $\mathbf{\hat{y}}$ component of equation \ref{eq:MomConsInt_sub}. Equation \ref{eq:MomCons2D_final} can be simplified as
\begin{equation} \label{eq:MomCons2D_final}
\frac{\partial Q_x}{\partial t} + \xi_x 
\frac{\partial}{ \partial x} \left( \frac{Q_x^2}{h_0-\delta} \right) =  
-\frac{h_0 - \delta}{\rho_f} 
\frac{\partial P}{ \partial x} - \frac{f}{4} \frac{Q_x^2}{( h_0 - \delta)^2}.
\end{equation}

In order to solve equations  \ref{eq:MassCons2D_final} and \ref{eq:MomCons2D_final} uniquely, two pressure boundary conditions are required. A simple and common treatment is to use the extended Bernoulli equation with empirical loss coefficients associated with the specific geometry and flow conditions (including Reynolds number) near the inlet and outlet. 
We adopt here the approach used to treat leakage-flow instabilities \citep{Nagakura1991,Inada1988,Inada1990}, 

\begin{equation} \label{eq:PressureInOutBC}
P(t)|_{x=0}  = P_{\mathrm{in}} - 
\frac{\zeta_{\mathrm{in}}}{2} \rho_f \left[ \left( \frac{Q_x}{h_0 - \delta} \right)^2 \right]_{x=0}, \
P(t)|_{x=L}  = P_{\mathrm{out}} + 
\frac{\zeta_{\mathrm{out}}}{2} \rho_f \left[ \left( \frac{Q_x}{h_0 - \delta} \right)^2 \right]_{x=L}.
\end{equation}
where $\zeta_{\mathrm{in}} \geq 1 $ and $\zeta_{\mathrm{out}} \geq 0$ are loss coefficients, and the departure from equality represents non-isentropic processes. $P_{\mathrm{out}}$ and $P_{\mathrm{out}}$ are constants.

To summarize, equations \ref{eq:MassCons2D_final} and \ref{eq:MomCons2D_final}, together with the boundary conditions \ref{eq:PressureInOutBC} relate the pressure, flow rate, and deflection in either channel.  Both sets are closed with the geometrical constraint, equation \ref{eq:BeamShapeConstraint}, the common boundary conditions ($P_{\mathrm{in}}$ and $P_{\mathrm{out}}$), and the common equation of motion for the beam, equation \ref{eq:EulerBernoulliBeam2D}.  The top and bottom channels correspond to distinct channel shapes $h_0^{\mathrm{top}}$ and $h_0^{\mathrm{bot}}$, respectively.
These constitute the closed quasi-1D fluid-structure model.  

Before proceeding, we review the assumptions underpinning the model.  In addition to flow two-dimensionality, we require that the gap is thin $\hat{h} \ll 1$ and that axial variations in the channel gap are small: $h_0'^2 \ll 1$ and $\delta'^2 \ll 1$. In addition, we require the velocity profile to be slowly changing with both $x$ and $t$, so that $\xi_x$ can be regarded as a constant, as would strictly be the case when $\hat{h}^2 Re_L \ll 1$.  As we will show, the model also performs well in circumstances where  $\hat{h}^2 Re_L$ is not small.  Our goal, after linearizing the model, is to test the validity of these bounds by contrasting them with results from numerical simulations.  



\subsection{Linearized model} \label{sec:LinearizationPressure}

The primary goal of this paper is to predict the linear stability (flutter boundary) of an equilibrium beam shape $\delta_0(x)$, as a function of parameters on table \ref{tab:dimensionpars}.  We begin this process by expanding the dependent variables about their respective equilibrium values in a small parameter, $\varepsilon$, representing the amplitude of the beam displacement.  That is, we take
\begin{align*} 
\delta(x,t) & = \delta_0(x) + \varepsilon \delta_1(x,t) + \ldots \\
P(x,t) & = p_0(x) + \varepsilon p_1(x,t)  + \ldots \\
Q_x(x,t) & = q_{x0}(x) + \varepsilon q_{x1}(x,t)  + \ldots
\end{align*}
as well as the linearized friction factor 
\begin{align*}
 \label{eq:FrictionTermLinearExp}
f(Q_x) & \approx f(q_{x0}) + (Q_x - q_{x0})  \left[ \frac{\mathrm{d} f}{\mathrm{d} Q_x} \right]_{Q_x = q_{x0}} + \ldots \\
& \approx f_0 + \varepsilon \eta  q_{x1}(x,t) + \ldots,
\end{align*}
determined from laminar and turbulent relations in equation \ref{eq:FrictionTermGen} with the $Re_h = \hat{h} Re_{L}$.

At zeroth order, we obtain a differential equation describing the equilibrium beam shape
\begin{equation}
    EI_{\mathrm{2D}} \frac{\mathrm{d}^4}{\mathrm{d} x^4} \delta_0(x) = 
p_0^{\mathrm{bot}} - p_0^{\mathrm{top}}, \label{eq:Eps0BeamShape} 
\end{equation}
together with the homogeneous beam boundary conditions of equation \ref{eq:StructureBC_ClampedFree}.  Once again, the superscripts top and bot refer to parameters associated with $h_0^{\mathrm{top}}$ and $h_0^{\mathrm{bot}}$ as the channel shapes above and below the beam, respectively. The pressure distribution and flow rate in either channel (i.e. top and bottom) are given by
\begin{align}
p_0(x) & = {P_{\mathrm{in}}} - 
{\rho}_{f}\, q_{x0}^2\, \left( \frac{ {f}_{0}\, }{4} \int_{0}^{ {x}} \frac{d  {x_2}}{{h_e\!\left( {x_2}\right)}^3} \, - 
{\xi_x}\, \int_{h_e(0)}^{ h_e(x)} \frac{d h_e}{h_e^3}  + 
\frac{ {\zeta_{ {in}}}}{2\, {h_e\!\left(0\right)}^2} \right), \label{eq:Eps0Pressure} \\
q_{x0} & =  \left(
\frac{ {P_{\mathrm{in}}} -  {P_{\mathrm{out}}}\,}{   
{\rho}_{f}\, \left[
\frac{ \zeta_{\mathrm{out}}} {2 h_e\!\left(L\right)^2} + 
\frac{ \zeta_{\mathrm{in}}} {2 h_e\!\left(0\right)^2}  -
{\xi_x}\, \left(\int_{h_e(0)}^{h_e(L)} \frac{d h_e}{h_e^3} \right) +
\frac{f_{0}}{4}\, \left(\int_{0}^{L} \frac{d  {x_2}}{{h_e\!\left( {x_2}\right)}^3} \,\right) \right]
} \right)^\frac{1}{2},  \label{eq:Eps0FlowRatex}
\end{align}
where we have set $h_e(x) = h_0(x) - \delta_0 (x)$, the equilibrium channel height.  We note that at equilibrium, the flow rate in each channel is a constant independent of $x$. As an example, $p_0^{\mathrm{top}}$ in equation \ref{eq:Eps0BeamShape} refers to equations \ref{eq:Eps0Pressure} and \ref{eq:Eps0FlowRatex} where $h_e^{\mathrm{top}}$ is substituted for $h_e$, while $h_e^{\mathrm{bot}}$ substituted into $p_0^{\mathrm{bot}}$.  The same is true for the subsequent higher order terms in $\varepsilon$ (i.e. $p_1$ and $q_{x1}$ with top and bot superscripts).

The steady equilibrium equations here are similar to those obtained by \citeauthor{Inada1988} \citep{Inada1988}, assuming $\xi_x = 1$, and $\delta_0 = 0$. Equation \ref{eq:Eps0Pressure} has three distinct terms in the parenthesis, the first with $f_0$ as a factor represents the pressure drop due to viscous effects as a function of the integral equilibrium gap shape over $x$; the second integral, multiplied by $\xi_x$, comes from the inertia term and is only a function of the initial ($x=0$) and the current (at $x$) gap size, representing the pressure change due to the gap expansion or contraction; and the third is solely due to system inlet conditions.  If $\zeta_{\mathrm{in}} = 1$ (isentropic), then the only pressure drop comes from accelerating the flow to the average inlet velocity.    

Next we collect and equate coefficients to linear order in $\varepsilon$.  For the beam, we obtain
\begin{equation} \label{eq:Eps1BeamShape}
\rho_s h_b\frac{\partial^2 \delta_1}{\partial t^2}  + 
EI_{\mathrm{2D}}\frac{\partial^4 \delta_1}{\partial x^4}   =
p_1^{\mathrm{bot}} - p_1^{\mathrm{top}},
\end{equation}
together with its homogeneous boundary conditions.  For the pressure and flow rate, we have
\begin{equation} \label{eq:Eps1MassConsInt}
\frac{\partial  q_{x1} }{\partial t} =  \frac{\partial \delta_1}{\partial x} ,   
\end{equation}
\begin{equation} \label{eq:Eps1PressureGradient_1}
\resizebox{.9 \textwidth}{!}{$
\begin{aligned}
\frac{\partial p_1}{\partial x}  = 
\frac{ \rho_f} {h_e}\left\{
\left[\frac{ {\xi_x}\, q_{x0}^2\, }{h_e^2} \frac{\partial }{\partial x}   -  
\frac{3\, p_{0}^{\prime} }{\rho_f }  \right] \delta_{1} -  \left[
\frac{\partial }{\partial t}    + 
\frac{ 2 {\xi_x} q_{x0}\,}{h_e} \left( \frac{\partial}{\partial x}  - 
\frac{h_e^{\prime} }{h_e}  \right) + 
\frac{  q_{x0}\, }{2\, {h_e}^2} \left( {f}_{0}\, + 
\frac{ \eta q_{x0} }{2} \right) \right]  q_{x1} - \frac{\partial q_{x1} }{\partial t} \right\}. 
\end{aligned}
$} 
\end{equation} 
To obtain a closed form for $p_1$, equation \ref{eq:Eps1MassConsInt} is first integrated in $x$, solved for $q_{x1}$ in terms of $\delta_1$ and $q_{x1}(0,t)$, and substituted into equation \ref{eq:Eps1PressureGradient_1}.  
The result is, once again, integrated in $x$, and together with the boundary conditions
\begin{equation} \label{eq:Eps1PressureInBC}
p_1(0,t) =  
\frac{2  \left( {P_{\mathrm{in}}}\, - p_{0}\!\left(0\right)\, \right) }{h_e\!\left(0\right)} \delta_{1}\!\left(0,t\right) -
{\zeta_{\mathrm{in}}}\, \frac{{\rho}_{f}\, q_{0}\, }{{h_e\!\left(0\right)}^2} q_{x1}\!\left(0,t\right)
\end{equation}
and 
\begin{equation} \label{eq:Eps1PressureOutBC}
p_1(L,t) =
\frac{2\,  \left( {P_{\mathrm{out}}}\, -  p_{0}\!\left(L\right)\, \right)}{{h_e\!\left(L\right)}} \delta_{1}\!\left(L,t\right) +  
{\zeta_{\mathrm{out}}}\, \frac{ {\rho}_{f}\, q_{0}\, }{{h_e\!\left(L\right)}^2} q_{x1}\!\left(L,t\right),
\end{equation}
and with the beam equation, can be solved solely for $\delta_1$ and $q_{x1}$.

\subsection{Numerical solution of the perturbation equations} \label{sec:FSI_numsolve}

To numerically solve the linear system of PDEs given by equations \ref{eq:Eps1BeamShape} to \ref{eq:Eps1PressureGradient_1}, we expand the first-order beam displacement in a series of basis functions
\begin{equation} \label{eq:delta_exp}
    \delta_1(x,t) = \sum_{i=0}^{n} a_i(t) g_i(x) 
\end{equation}
where 
\begin{equation} \label{eq:ClampedFreeBC_BasisExpansion}
g_i(x) =
\begin{cases}
0 & \text{for } i = 0 \\
\phi_i(x) & \text{for } i = [1, n]
\end{cases}, 
\end{equation}
and $\phi_i(x)$ are solutions of the homogeneous (unforced) beam equation,
\begin{equation} \label{eq:biharmonicOperator}
\frac{\mathrm{d}^4 \phi }{\mathrm{d} x^4 }  + \beta^4 \phi = 0.
\end{equation}
in the domain $x \in [0,L]$. Its solution, $\phi_k$, $k \in \mathbb{Z}:[1,\infty]$, when subject to the clamped-free boundary conditions is
\begin{equation} \label{eq:ClampedFreeEigFun}
\phi_k(x)=\cosh\!\left(\beta_k\, x\right) - \cos\!\left(\beta_k\, x\right) + \left[ 
\frac{ \cos\!\left(\beta_k\,  {L}\right) + \cosh\!\left(\beta_k\,  {L}\right)}{\sin\!\left(\beta_k\,  {L}\right) + \sinh\!\left(\beta_k\,  {L}\right)} \right] \Big( \sin\!\left(\beta_k\, x\right)\, - \sinh\!\left(\beta_k\, x \right)\, \Big),
\end{equation}
with characteristic equation
\begin{equation} \label{eq:ClampedFreeCharEq}
\cosh \left(\beta_k L \right) \cos \left(\beta_k L \right) + 1 = 0.
\end{equation}
The first six values are listed in table \ref{tab:BLvecvalues}. 
\begin{table}[h] 
\centering
\caption{Table of solutions to the characteristic equation \ref{eq:ClampedFreeCharEq}.} \label{tab:BLvecvalues}
\begin{tabular}{| c  c  c  c  c c| }  \hline 
$\beta_1 L$ &$\beta_2 L$ & $\beta_3 L$ & $\beta_4 L$ & $\beta_5 L$ & $\beta_6 L$  \\ 
\hline
 1.8751 &    4.6941 &    7.8548 &   10.9955&   14.1372   & 17.2788 
  \\ \hline
\end{tabular} 
\end{table}

Equation \ref{eq:delta_exp} is substituted into the coupled beam-pressure equation (equation \ref{eq:Eps1BeamShape} and solution of \ref{eq:Eps1PressureGradient_1}), the result which is then projected onto the same basis functions (Galerkin projection).  By construction, the equation for the time evolution of $q_{x1}(0,t)$ is not a function of $x$, but rather of the boundary points at $x = 0$ and $x = L$.
The final total number of unknowns is $2 n + 2$.  These include the expansion coefficients $a_i(t)$, their time derivatives, ${\dot a}_i(t)$, and the remaining two unknowns are the entrance flow rate $q_{z1}(0,t)$ for the top and bottom channels.  We write the solution vector as
\begin{equation}
    {\bf x} = \begin{bmatrix} a_0 & a_1 & \ldots & a_n & {\dot a}_0 & {\dot a}_1 & \ldots & {\dot a}_n & q_{x1}^{\mathrm{bot}}(0,t) & q_{x1}^{\mathrm{top}}(0,t) \end{bmatrix}^T
\end{equation}
the resulting system of ODE are
\begin{equation} \label{eq:Eps1LinearEq}
    {\dot {\bf x}} = {\bf A} {\bf x}.
\end{equation}
The entries of ${\bf A}$ are given in the appendix.  The eigenvalues and eigenvectors of ${\bf A}$ are computed in section~\ref{sec:ConstChannelFlow} to determine the flutter boundary for the coupled FSI system.  We note here that the model, once nondimensionalized, involves four independent nondimensional groups given in table~\ref{tab:NDpars_clampedFree}: $U^*$, $M^*$, $\hat{h}$, and $\hat{h}^2 Re_L$.  The table also contains the beam frequency response parameters, including the Strouhal number $St$ as a function of the imaginary part of the eigenvalue $\lambda$ or the dimensional beam frequency response $f_s$.
    

\begin{table}[H] 
\centering
\caption{Table of clamped-free fluid-structure nondimensional parameters.} \label{tab:NDpars_clampedFree}
\begin{tabular}{| c  c  c  | }  \hline 
\textbf{Variable} & \textbf{Expression} &\textbf{Description}  \\ 
\hline
 $M^*$ & $\frac{ \rho_f L }{\rho_s h_b }$ & mass ratio \\  
 $U^*$ & $\frac{  q_{x0} L }{\bar{h}} 
 \left( \frac{\rho_s h_b }{E I_{\mathrm{2D}}} \right)^{\frac{1}{2}} $ & stiffness ratio \\  
 $\hat{h}$ & $\frac{\bar{h}}{L}$ & gap or throat ratio \\   
 $\hat{h}^2 Re_L$ & $ \left( \frac{\bar{h}}{L} \right)^2 \left( \frac{\rho_f q_{x0} L}{\bar{h} \mu_f} \right)$ & viscous parameter \\
 $St$ & $\frac{\text{Im} \left[ \lambda \right]}{2 \pi} = \frac{f_s L \bar{h}}{q_{x0}}$ & Strouhal number \\
 $f_s^*$ & $U^* St$ & nondimensional frequency
 \\ \hline
\end{tabular} 
\end{table}

\section{Immersed-Boundary Direct Numerical Simulation and Data Processing} \label{sec:DNSFSIAlgorithm}

In order to validate aspects of the quasi-1D model developed in section \ref{sec:lubeclosure}, we employ a two-dimensional fluid-structure algorithm \citep{Goza2017} that utilizes the immersed boundary (IB) projection method \citep{Taira2007,Colonius2008} along with Newton-Raphson approach to solve the strongly-coupled fluid-structure system.  Strong-coupling ensures that the nonlinear constraint between the fluid and the structure is enforced at each time step, and is necessary for accurate computation of large structural deformations.

The current implementation uses a co-rotational formulation of the structural equations for the beam, where strains are assumed small within constituent material equations in the beam-local frame \citep{Crisfield1991}.  The beam-fluid nondimensional parameters that govern the dynamics are $Re_L$, $M^*$, and $U^*$ and the same as those defined in the quasi-1 dimensional model in table \ref{tab:NDpars_clampedFree}. 

The internal flow passage is created by immersed boundaries, including one spanning the inlet where the streamwise velocity is set to a given  profile and the normal velocity is set to zero.

The initial discrete delta function used in the IB method is constructed as follows. The 3-point kernel of \citeauthor{Yang2009} \citep{Yang2009} is smoothed three times using their recursive formula such that its total support is 6 points. This kernel was chosen as a compromise between computational efficiency and accuracy and smoothness of the local stresses \citep{Goza2016}.

The algorithm has been verified extensively in \citep{Goza2018thesis} for external flows, where regimes of the standard and inverted flags were explored and compared to results from other strongly-coupled fluid-structure solvers.  It has also been verified \citep{Tosi2018thesis} for internal flows using the suggested benchmark of an elastic member in an internal, incompressible, laminar flow and compared to \citep{Turek2006,Tian2014,Bhardwaj2012,Shoele2014}.  Results including frequency and amplitude response, along with the drag coefficient, were consistent with the other compared schemes for the two cases at $Re_L = 350$ and 700. 

In order to evaluate the linear stability of the FSI system, we employ an empirical approach based on the Dynamic Mode Decomposition (DMD) to calculate growth/decay rates, frequencies, and mode shapes directly from the (nonlinear) simulation data. DMD is a data decomposition technique that approximates the eigenmodes of the linear operator that best describes the dynamics of the system, i.e. from one time instance to the next \citep{Schmid2010,Chen2012,Tu2013,Goza2018Modal}.  We take the exact-DMD algorithm and augment our data matrix, comprising of beam $\mathbf{\hat{y}}$ positions along its Lagrangian IB coordinate, with the eigensystem realization algorithm (ERA) to ensure robustness of our eigenvalue calculations \citep{Tu2013}.  DMD eigenvalues and modes for the time series near the beam equilibrum are used to compare with quasi-1D model eigenvalues and modes.  



\subsection{Cantilever in Constant Channel} \label{sec:ConstChannelFlow}

Although the quasi-1D model is formulated for arbitrarily shaped channels, flow in a constant, symmetric channel presents a simple geometry that can be simulated and compared to existing inviscid models \citep{Alben2015, Shoele2015}, and provides a relevant problem configuration to many of the aforementioned applications.

\begin{figure}[H]
    \centering
   \includegraphics[trim={.5cm 7cm .5cm 5cm},clip,width=.75\textwidth]{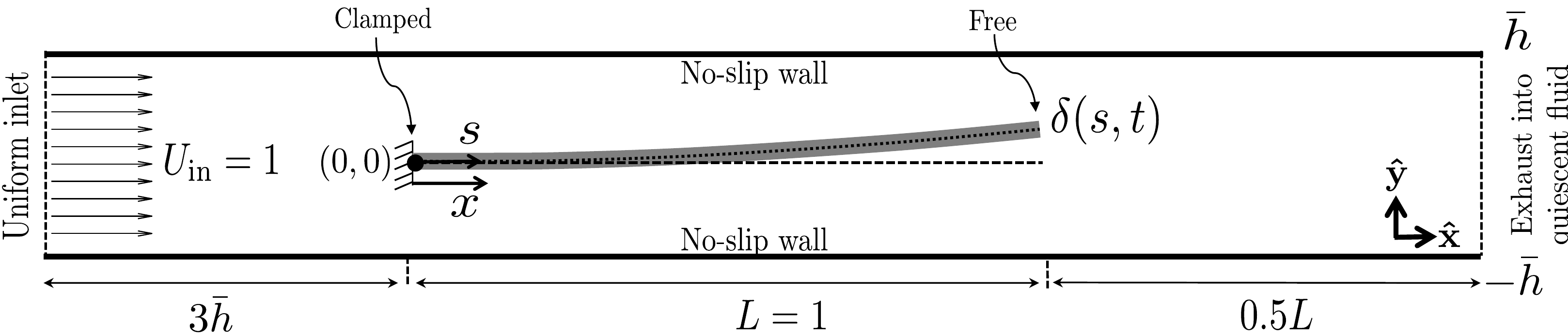}
        \caption{Illustration for fluid-structure constant channel domain and boundary conditions.}  
     \label{fig:constchannelgeo}
\end{figure}

The FSI DNS computational domain is illustrated in figure \ref{fig:constchannelgeo}.  
The coordinate system has its origin at the beam leading edge, and the channel is defined at a constant half width, $\bar{h}$.  
The initial beam position $\delta(s,0) = 0$, and its velocity is $\frac{\partial}{\partial t} \delta(s,0) = 0$, where $s$ is the Lagrangian coordinate that describes the beam parameterized by its arc length. 
The boundary conditions are clamped and free for leading and trailing edges of the beam, respectively.  A uniform velocity profile is specified at the inlet as $U_{\mathrm{in}}$ and used as the reference parameter for non-dimensionalization of other quantities.  
The flow is impulsively started and the beam is perturbed by a small body force at the initial time step in order to break symmetry.

Figure \ref{fig:constchannelgeo} is contrasted with the domain in figure \ref{fig:NozzleDiffuserGeometryl}, where quasi-1D model parameters are illustrated ($U_c = \frac{q_{x0}}{\bar{h}}$).  
First, the steady flow rate, $q_{x0}$, specified for the quasi-1D model, constitutes half the integrated $\mathbf{\hat{x}}$ flow velocity (in the channel) over $\mathbf{\hat{y}}$ from $-\bar{h}$ to $\bar{h}$ immediately downstream of the inlet. The beam boundary conditions for the quasi-1D model are the same as the DNS. 
Results also do not depend on inlet and outlet distances except through loss coefficients $\zeta_{\mathrm{in}}$ and $\zeta_{\mathrm{out}}$ defined in equations \ref{eq:PressureInOutBC}.  We assume that stagnation pressure loss at either end is negligible and take the no loss coefficients values ($\zeta_{\mathrm{in}} = 1$ and $\zeta_{\mathrm{out}} = 0$). 

Lastly, we are interested in the dynamics near the DNS initial conditions, which, apart from the small body force to deflect the beam, represents a (potentially unstable) equilibrium for both the DNS and the quasi-1D model.
In this limit the coordinate $x$, shown in both figures \ref{fig:NozzleDiffuserGeometryl} and \ref{fig:constchannelgeo}, coincides with the Lagrangian coordinate $s$ in figure \ref{fig:constchannelgeo}. For small perturbations, the Lagrangian beam shape then becomes the beam $\mathbf{\hat{y}}$ displacement, $\delta(x,t)$.  

We will use DMD to find the least-stable mode of the DNS results using the procedure discussed.  These results are directly comparable with the eigenvalues and eigenmodes of the quasi-1D linear operator $\mathbf{A}$ in equation \ref{eq:Eps1LinearEq}. In order to compare to DMD growth rate, the dynamically significant eigenvalues of the quasi-1D will be shown as well (the least, and sometimes second least, stable eigenvalues for the quasi-1D model). If this eigenvalue has an imaginary part (and complex conjugate pair since the data matrix is real), we will track the positive frequency counterpart.  
Given the parameters in table \ref{tab:NDpars_clampedFree}, the model eigenvalues and DMD spectrum, represented as $\lambda$, are scaled with the inverse of nondimensional convective time units. 

\subsection{FSI DNS Discretization} \label{sec:ConstChannelDiscretization}

The FSI DNS Eulerian mesh is uniform in $\mathbf{\hat{x}}$ and $\mathbf{\hat{y}}$ with grid spacing $\Delta x^* = \frac{\Delta x}{L}$.  Because the parametric study spans a wide range of $Re_L$ and $\hat{h}$ values, with over 4000 simulations carried out, the $\Delta x^*$ is automatically determined by the most restrictive of three conditions: the grid Reynolds number $Re_{\Delta x} \leq 2$; the minimum number of grid elements in $\hat{h}$ is 20; the minimum number of elements on the beam surface is 160,
\begin{equation} \label{eq:gridconditions}
\Delta x^* = \min\left\{\frac{2}{Re_L}, \frac{\hat{h}}{20}, \frac{1}{160} \right\}.
\end{equation}
The Lagrangian grid spacing is always $\Delta s^* = 2 \Delta x^*$, as suggested in \citep{Goza2016}.  The time step size $\Delta t^*$ is determined by holding the $CFL = \frac{\Delta t^*}{\Delta x^*} = 0.2$ for the $\Delta x^*$ that satisfies the criteria \ref{eq:gridconditions}. These conditions were determined by trial and error to capture at least 200 time steps per beam oscillating cycle for all results. 
The grid Reynolds number chosen captures fluid advection and diffusion terms well. The resulting $\Delta x^*$, $\Delta t^*$ combination produces results within acceptable wall-time for the number of simulations ran in this study.  We explore the effect of grid refinement from criteria \ref{eq:gridconditions} in our results next.

\subsection{Grid Convergence and Effective Beam Thickness} \label{sec:GridConv}

The convergence of the FSI DNS beam dynamics as a function of the spatial discretization is explored through a grid refinement study.  
The parameters are $\hat{h} = 0.05$, $\hat{h}^2 Re_L = 0.5, \ M^* = 0.01$, and the corresponding $\Delta x = 0.0025$.  
This parameter set is chosen specifically because it insures that the closure approximations introduced in section \ref{sec:lubeclosure} are satisfied; thus the quasi-1D model is expected to accurately represent the fully-coupled system. $U^*$ is varied as the bifurcation parameter, while $M^*$, $\hat{h}$, and $\hat{h}^2 Re_L$ are held constant. DMD eigenvalues are calculated for $\Delta x^*$, $\frac{\Delta x^*}{2}$, and $\frac{\Delta x^*}{3}$, where $\Delta x^*$ satisfies criteria \ref{eq:gridconditions}.  The DMD procedure is applied to the beam displacement results from simulations for each refined $\Delta x^*$ value.  

\begin{figure}[H]
    \centering
    \begin{subfigure}[t]{0.5\textwidth}
        \centering
        \captionsetup{width=.8\linewidth}
   		\includegraphics[width=1\textwidth]{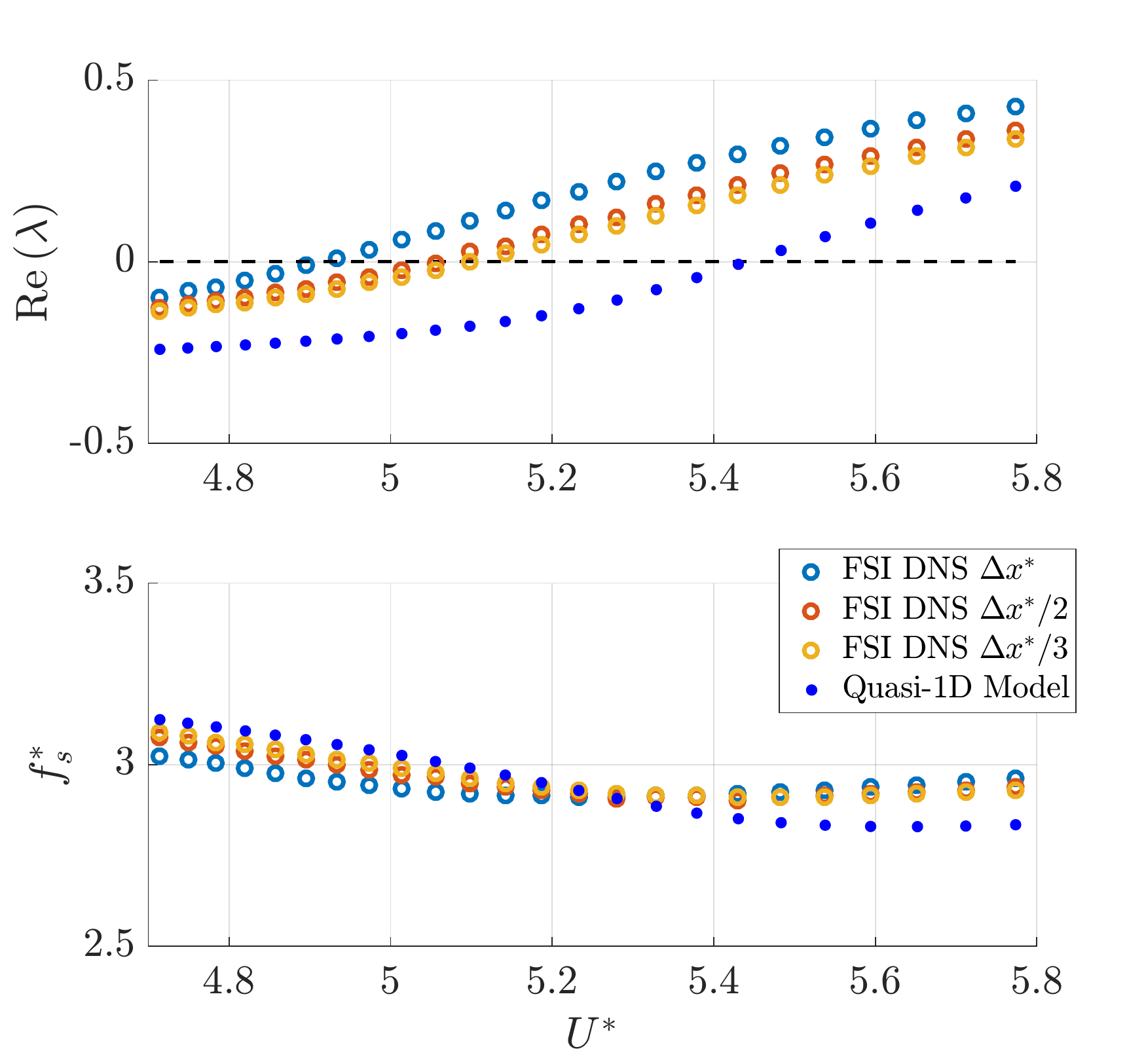}
        \caption{DMD spectrum and quasi-1D model eigenvalues in DNS grid refinement study.} 
     	\label{fig:GridRefinementDNS}  
    \end{subfigure}%
    ~ 
    \begin{subfigure}[t]{0.5\textwidth}
        \centering
        \captionsetup{width=.8\linewidth}
   \includegraphics[width=1\textwidth]{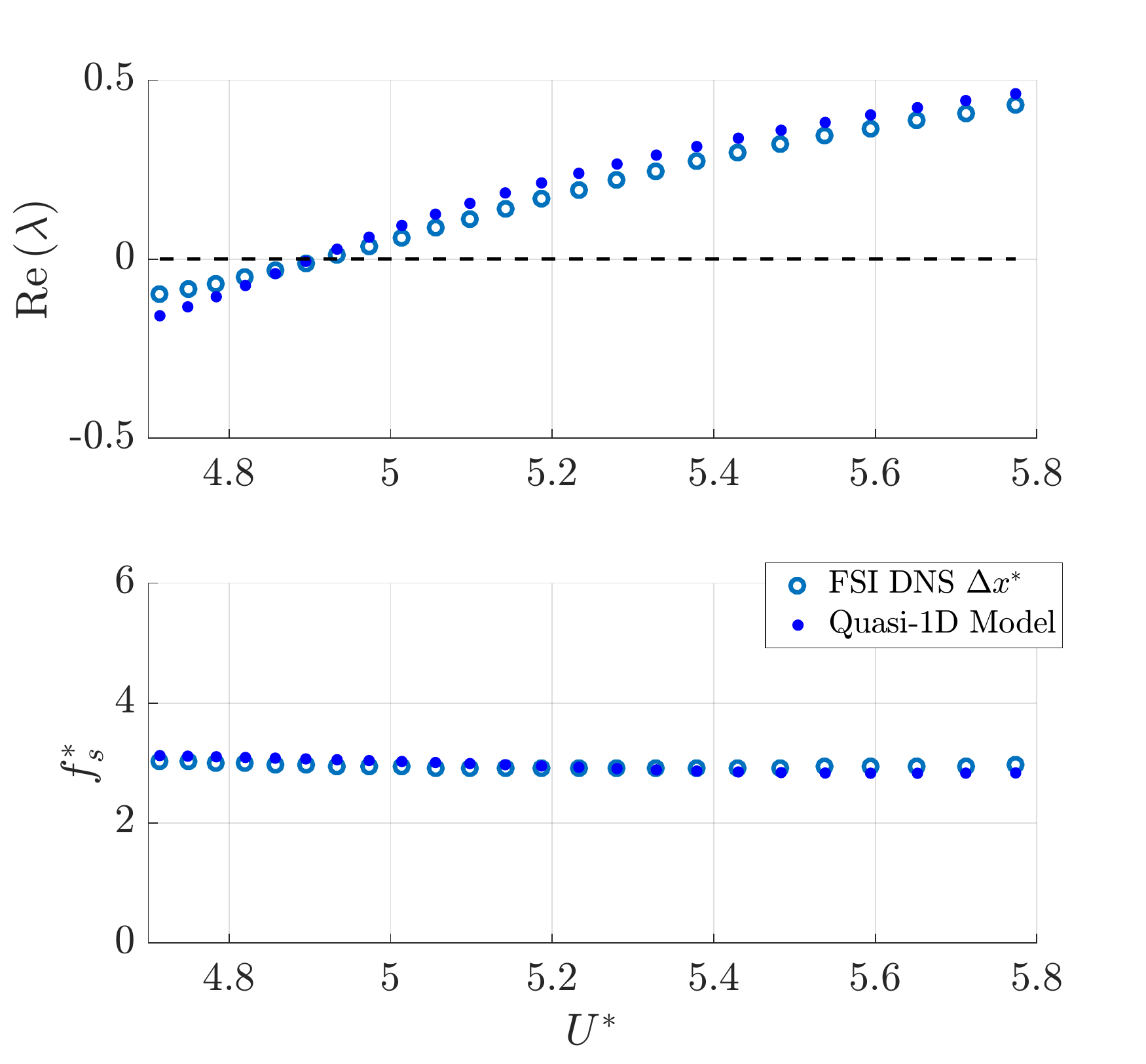}
        \caption{DMD spectrum and quasi-1D model eigenvalues for corrected channel width $\hat{h}_{\mathrm{corr}}$.} 
     \label{fig:Hhatcorrection}  
    \end{subfigure}
    \caption{DNS FSI Grid convergence results comparison with quasi-1D model with $\hat{h}$ (left) and $\hat{h}_{\mathrm{corr}}$ (right). }  
\end{figure}

Figure \ref{fig:GridRefinementDNS} shows the real and imaginary parts of the least-damped eigenvalue  for all $U^*$ and $\Delta x^*$ grid values. The leading quasi-1D model eigenvalues are also shown.  As the FSI DNS grid is refined, the DMD spectrum appears to be converging to the model eigenvalues: the real part of the DMD spectrum are monotonically moving toward the real part of the model eigenvalues; yet most notably, the imaginary part of the DMD spectrum is moving down for points where $U^*<5.27$ and up for $U^*>5.27$, approximating the quasi-1D curve shape.    

To better understand the slow $\lambda$ convergence, we consider the beam thickness in light of the immersed boundary projection method.  
If the Eulerian mesh does not resolve the physical thickness of the beam, the immersed boundary produces an ``effective'' beam thickness that is proportional to $\Delta x^*$. This phenomenon is caused by the discrete kernel of the delta function being always positive, in addition to the unidirectional flow conditions on both sides of the beam (i.e. moving in the $+\mathbf{\hat{x}}$ direction). 
The IB projection method ensures that the no-slip condition is \textit{exactly} enforced at each Lagrangian IB point. 
Once the discrete delta function smears it onto the flow, the Navier-Stokes equations are altered by the wall forcing for Eulerian points within the delta function kernel support. 
The flow outside its support, however, behaves as if the no-slip condition had been applied at the IB point locations.  
This effective thickness is not necessarily a physical quantity and cannot be systematically measured.  An upper bound for this effective thickness is the number of Eulerian support points of delta function kernel (6 in our case) and expected to vary with $\Delta x^*$.  
By refining $\Delta x^*$, we are also decreasing the beam's effective thickness, and slightly increasing the channel size.
This effect would be more pronounced as $\hat{h}$ decreases, when the dynamics become a stronger function of $\hat{h}$ \citep{Guo2000,Shoele2014,Shoele2015,Cisonni2017}.

Running the full parametric study with the finest grid in figure \ref{fig:GridRefinementDNS} presents issues due to computational resource restrictions and time constraints\footnote{$\Delta x^*$ grid results require approximately 32 hours of wall-time computation, while $\frac{\Delta x^*}{3}$ grid requires approximately 13 days.}.  
These problems would be significantly amplified should the actual beam thickness need to be resolved by the Eulerian mesh.  

Hence, we explore instead altering the channel thickness within the quasi-1D  model to understand whether the slight change in channel width could explain the slow convergence. Figure \ref{fig:Hhatcorrection} shows the comparison of the DMD spectrum for $\Delta x^*$ and the quasi-1D model with a corrected channel width, 
\begin{equation} \label{eq:hhatcorr}
\hat{h}_{\mathrm{corr}} = \hat{h} - 1.7 \Delta x^*, 
\end{equation}
which gives an effective beam thickness slightly smaller than 3.5 Eulerian cells (or slightly larger than 1.5 Lagrangian cells).  This result replicates the DMD spectrum extremely well, both in shape and in the instability boundary at $U^*_{\mathrm{cr}} = 4.90$; it also confirms the sensitivity of the dynamics to $\hat{h}$, and lends credibility to this convergence hypothesis.  
Hence, we apply the grid criteria \ref{eq:gridconditions} in subsequent results presented in section \ref{sec:DNSQ1DComp}.  We assume that the effective beam thickness is captured by equation \ref{eq:hhatcorr} when comparing between FSI DNS DMD spetra and quasi-1D eigenvalues for all grids in this study.  

\section{Results}

\subsection{Comparison of FSI DNS and Quasi-1D Model Results} \label{sec:DNSQ1DComp}
We begin by assessing the validity of the quasi-1D model for parameter values compatible with the closure assumptions enumerated in section \ref{sec:lubeclosure}.  We consider the specific cases listed in table~
  \ref{tab:ConstChannelMstCases}.  Similar to section \ref{sec:ConstChannelDiscretization}, FSI DNS simulations are run for each set of parameters, with $U^*$ treated as the bifurcation parameter. For each parameter trio $\left[ \hat{h}, \ \hat{h}^2Re_L, \ M^* \right]$, 
we find the pair $U^*_{\mathrm{cr}}$ and Im$\left[ \lambda \right]_{\mathrm{cr}}$ as the critical values for the flutter boundary by linearly interpolating between the two nearest parametric mesh points between stable and unstable results. All cases in table \ref{tab:ConstChannelMstCases} initially explored a $U^*$ range that spanned at least two orders of magnitude, with refined cases near the bifurcation point that yielded appropriate results for linear interpolation to find $U^*_{\mathrm{cr}}$ and $f_{s \mathrm{cr}}^*$ values. For all cases, $Re_L$ values were chosen such that $\hat{h}^2 Re_L$ remains constant across the different $\hat{h}$ values tested, consistent with the fact that $Re_L$ only appears in the model equations through the group $\hat{h}^2 Re_L$.  All cases in table \ref{tab:ConstChannelMstCases} fall within the \emph{laminar} flow regime (i.e. parabolic flow profile), consistent with the definition in equations \ref{eq:FrictionTerm_Shimoyama} and \ref{eq:NonlinearTerm_general} for the linearized $f_0$, where $\hat{h} Re_L < 1000$.

\begin{table}[hbt!] 
\centering
\caption{Table of cases for constant channel flow simulations with $U^*$ as the bifurcation parameter and varying $M^*$ and $\hat{h}^2 Re_L$.} \label{tab:ConstChannelMstCases}
\begin{tabular}{| c c c c | }  \hline 
Case $\#$	&	$\hat{h}$ &	$ \hat{h}^2 Re_L$ & $M^*$  \\ \hline
1 &	0.025&	0.500&	[$0.01-0.50$] \\
2 &	0.050&	0.500&	[$0.01-1.00$] \\
3&	0.050&	1.250&	[$0.01-1.00$] \\
4&	0.050&	2.500&	[$0.01-1.00$] \\
5&	0.125&	0.500&	[$0.01-1.00$] \\
6&	0.125&	1.250&	[$0.01-1.00$] \\ 
7&  0.050	&	[0.10-4.50]	& 0.01   \\
8&	0.125	&	[0.10-9.50]	& 0.01  \\
\hline 
\end{tabular} 
\end{table}

We first focus on cases 1 to 6, where $U^*_{\mathrm{cr}}$ is tracked as the mass ratio is finely incremented.  The results are presented in figures~\ref{fig:hhat_025_Reh2_p5_All} to \ref{fig:hhat_125_Reh2_1p25_All}.  In addition to the flutter boundary, each figure shows the corresponding frequency of the unstable mode and beam mode shapes for selected mass ratios.  Figure \ref{fig:hhat_025_Reh2_p5_All} shows the narrowest channel at $\hat{h} = 0.025$.
The quasi-1D model predicts the flutter boundary exceptionally well for the range of $M^*$ simulated.  The corresponding beam mode shapes are given in figures \ref{fig:hhat_025_Reh2_p5_Modes_p01} and \ref{fig:hhat_025_Reh2_p5_Modes_p3}.  For the modes, the corresponding $U^*$ values were chosen from the nearest supercritical value from the available results for both the quasi-1D and FSI DNS computations.  The beam shapes are qualitatively similar in the DMD and quasi-1D results. We attribute small oscillations superimposed onto the primary mode shape to the DMD data matrix having components from the impulsive start and body force perturbation at $t = 0$.  Mode switching is evident as $M^*$ increases from 0.01 to 0.3: not only is there an abrupt jump in frequency, but an additional effective node appears in both the real and imaginary parts of the modes shown. 
Compared to the orthogonal beam modes in vacuum, described in equation \ref{eq:ClampedFreeEigFun}  we see a strong resemblance to beam mode two at $M^*=0.01$ and the third mode at $M^*=0.3$.  The mode numbers refer to the eigenvalue index on table \ref{tab:BLvecvalues}.  

Similar results are obtained when the channel width is doubled to $\hat{h}=0.05$ holding $\hat{h}^2 Re_L$ constant (case 2, figure \ref{fig:hhat_05_Reh2_p5_All}).   In particular, we confirm that quasi-1D model replicates the flutter boundary well for all $M^*$ values simulated. Quasi-1D results show multiple bifurcations at a single value of $M^*$.  This occurs when, at a given $M^*$, a second eigenvalue crosses the stability boundary in the stable-to unstable direction as $U^*$ increases.  

Figure \ref{fig:hhat_05_Reh2_1p25_All} shows similar results for case 3 where $\hat{h}^2 Re_L$ is raised to 1.5, holding $\hat{h} = 0.05$. The flutter boundary has moved to higher values of $U^*$, indicating stabilization with increasing $Re_L$.  
This trend is reversed, however, when Reynolds numbers are high enough and inertial forces begin to dominate the dynamics. Quasi-1D and FSI DNS modes still mirror each other, but the mode switching inflection point has moved to a higher $M^*$ (relative to cases 1 and 2), so that $M^* = 0.01$ and $M^* = 0.3$ are primarily composed of beam mode two.  

A further increase $\hat{h}^2 Re_L = 2.5$ yields results in figure \ref{fig:hhat_05_Reh2_2p5_All} (case 4). The quasi-1D model marginally under-predicts the FSI DNS boundary, with the discrepancy increasing with $M^*$.  Consequently, the mode switching $M^*$ point and higher mode boundary are under-predicted, yet $f^*_{s\mathrm{cr}}$ remains in close agreement.  The mode shapes at $M^* = 0.01$ also show some disagreement, with the DMD results appearing to have a larger contribution resembling orthogonal beam mode three.  

Further increasing the channel width to $\hat{h}=0.125$ causes an increasing discrepancy between the DNS and quasi-1D model prediction, particularly at higher $M^*$ values.  For case 5 (figure \ref{fig:hhat_125_Reh2_p5_All}) the model accurately predicts flutter properties for $M^* < 0.2$, but underestimates the higher mode boundary and critical frequencies for $M^* > 0.4$.  This is also evident in the mode shapes, as the $M^* = 0.01$ modes are in agreement, but the $M^* = 0.3$ modes are not. 
For case 6 (figure~\ref{fig:hhat_125_Reh2_1p25_All}), $\hat{h}^2 Re_L$ is increased to 1.25, where we also begin to see larger differences in the lower $M^*$ range.  
The quasi-1D model over-estimates $U^*_{\mathrm{cr}}$ relative to the FSI DNS simulations for $M^* < 0.2$ but under-estimates it for $M^* > 0.4$, also missing the mode switching $M^*$ point. $f^*_{s\mathrm{cr}}$ values remain well predicted through all values of $M^*$, however, as long as the system is in the correct branch.  

Overall, the level of agreement between the quasi-1D model and the simulation results is consistent with the assumptions underpinning the model.  When $\hat{h}$ and $\hat{h}^2 Re_L \le 0.5$ are sufficiently low, the quasi-1D model predicts critical flutter values well even for higher $M^*$ values, where condition $\delta'^2 \ll 1$ is less apt
as the beam is oscillating in higher modes.  Results where $\hat{h}^2 Re_L = 0.5$ is kept constant and $\hat{h}$ increases from 0.025 to 0.125 illustrate that, indeed, as $\hat{h}$ approaches $O(1)$, the quasi-1D boundary predictions worsen. Their results miss the mode switching $M^*$ along with the critical properties for $M^* \geq 0.4$.  Yet even at $\hat{h} = 0.125$, critical values for $M^*<0.2$ remain well approximated by the model.  
This indicates that as long as $\delta'^2 \ll 1$ holds, with only the lowest mode is considered, the model remains accurate.  Also of note is that at lower $M^*$ values, inertial terms associated with the beam, i.e. moving channel walls, tend to dominate over the fluid inertia.  The former is well captured with our closure, while the latter may not be due to the dependence of $\xi_x$ in $t$ and $x$.       
Cases where we hold $\hat{h} = 0.05$ constant and increase $\hat{h}^2 Re_L$ from 0.5 to 2.5 ($Re_L$ from 200 to 1000), show the quasi-1D model is not restricted to $\hat{h}^2 Re_L \ll 1$ for accurate predictions.  This is true even as the beam is excited at higher modes through increasing $M^*$.  However, considering case 6, we see the $\hat{h}$ effect as $\delta'^2$ increases through mode switching (i.e. increasing $M^*$).  In summary, condition  $\hat{h} \ll 1$  appears to be the most stringent restriction: as long as $\hat{h}$ remains small (i.e. $\hat{h} \leq 0.05$), the quasi-1D model predicts the flutter boundary for a wide range of $M^*$ and $\hat{h}^2Re_L$.



\newcommand{\hhatvar}{0.025 } 
\newcommand{\Rehtvar}{0.5 } 
\newcommand{\casenumvar}{1 }

\begin{figure}[h]

\sbox0{\begin{subfigure}{0.25\textwidth}
		\captionsetup{width=1\linewidth,font={footnotesize}}
        \includegraphics[width=1\textwidth]{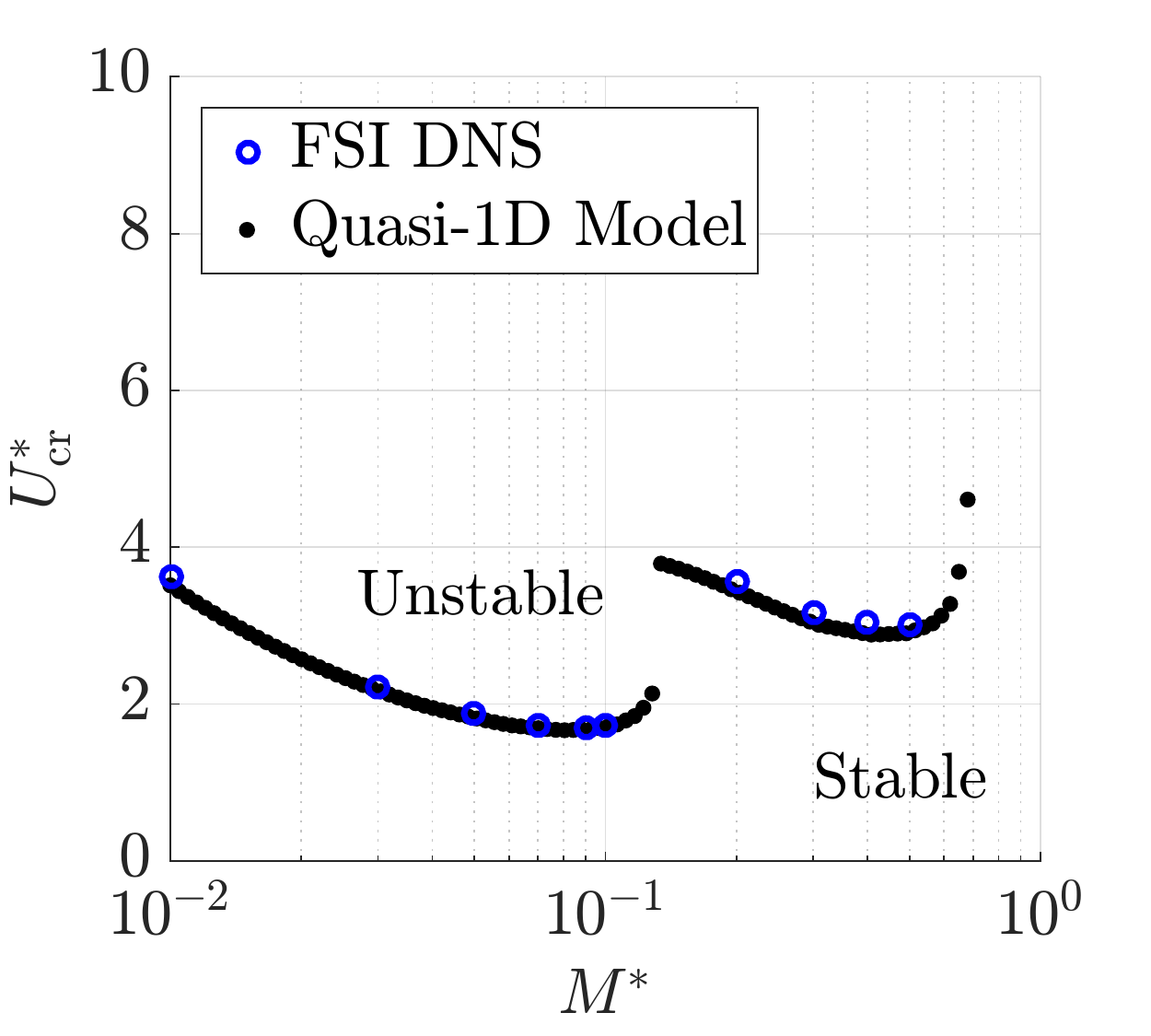}
    	\caption{$U^*_{\mathrm{cr}}$ as a function of $M^*$.} 
        \label{fig:hhat_025_Reh2_p5_CritVals_Ucr}
      \end{subfigure}}
\sbox1{\begin{subfigure}{0.25\textwidth}
		\captionsetup{width=1\linewidth,font={footnotesize}}
        \includegraphics[width=1\textwidth]{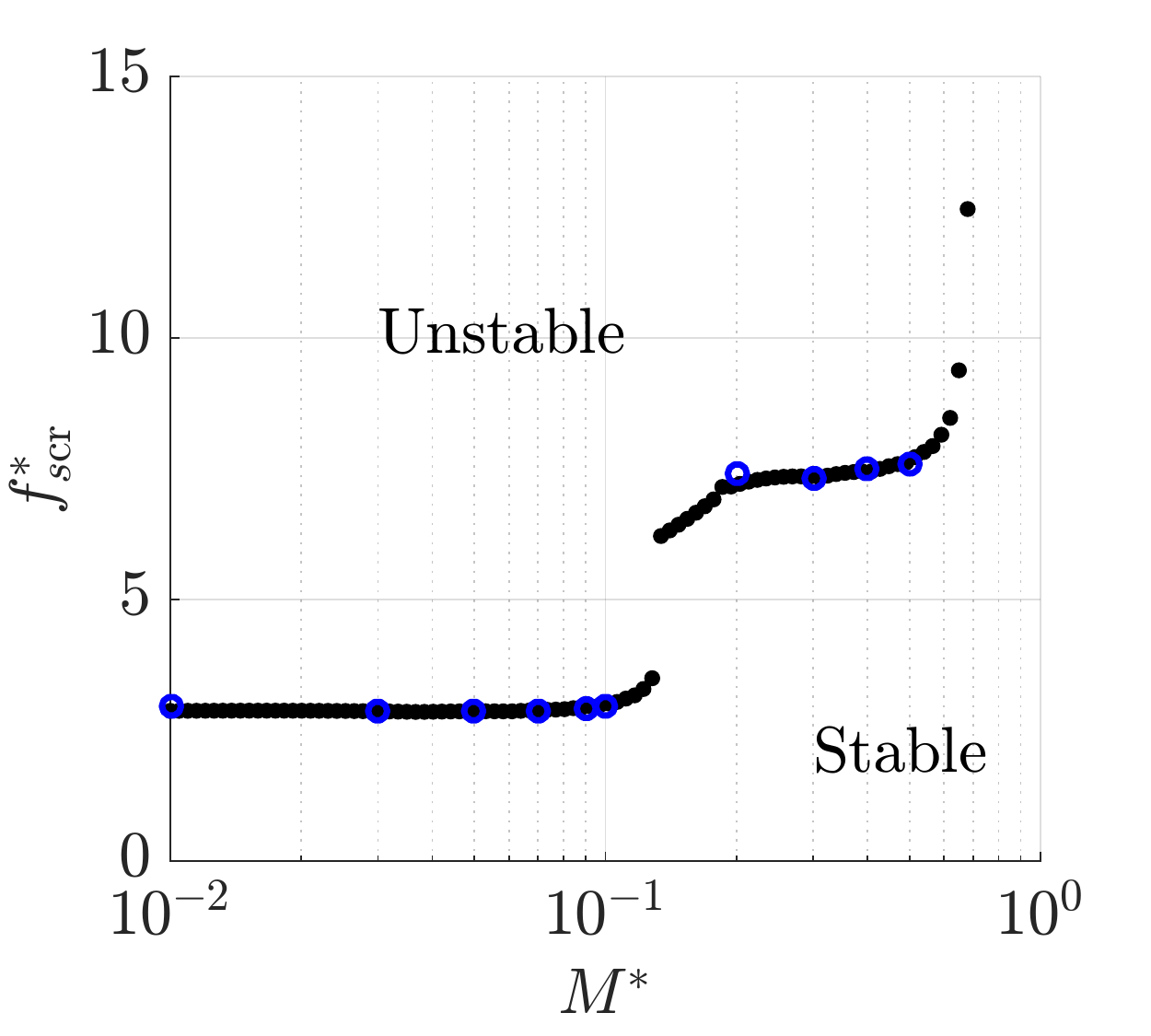}
    	\caption{$f^*_{s\mathrm{cr}}$ as a function of $M^*$.} 
        \label{fig:hhat_025_Reh2_p5_CritVals_fcr}
      \end{subfigure}}
\sbox2{\begin{subfigure}{0.25\textwidth}
		\captionsetup{width=1\linewidth,font={footnotesize}}
        \includegraphics[
        width=1\textwidth]{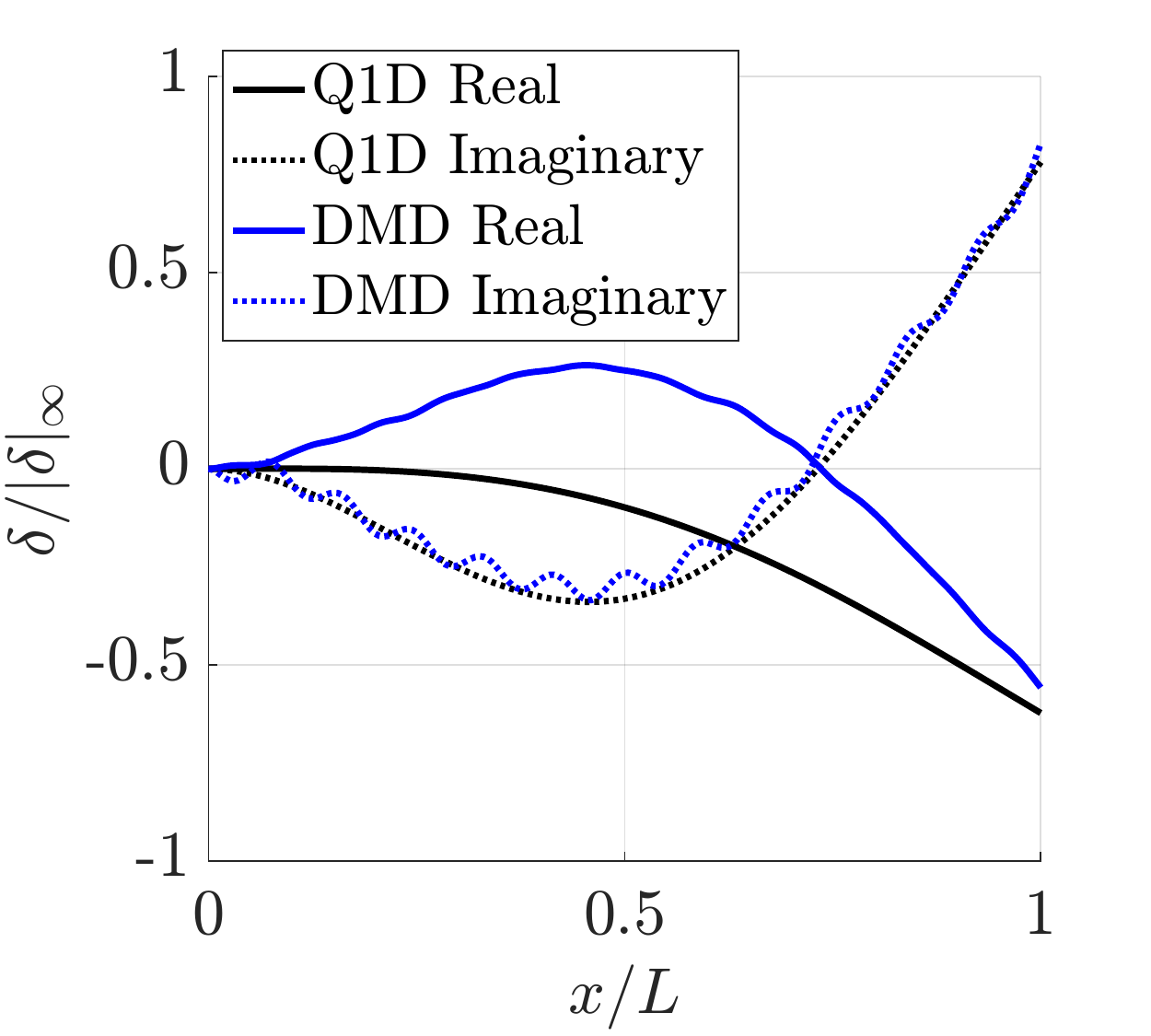}%
    	\caption{Mode shapes at $M^*=0.01$.} 
        \label{fig:hhat_025_Reh2_p5_Modes_p01}
      \end{subfigure}}
\sbox3{\begin{subfigure}{0.25\textwidth}
		\captionsetup{width=1\linewidth,font={footnotesize}}
        \includegraphics[
        width=1\textwidth]{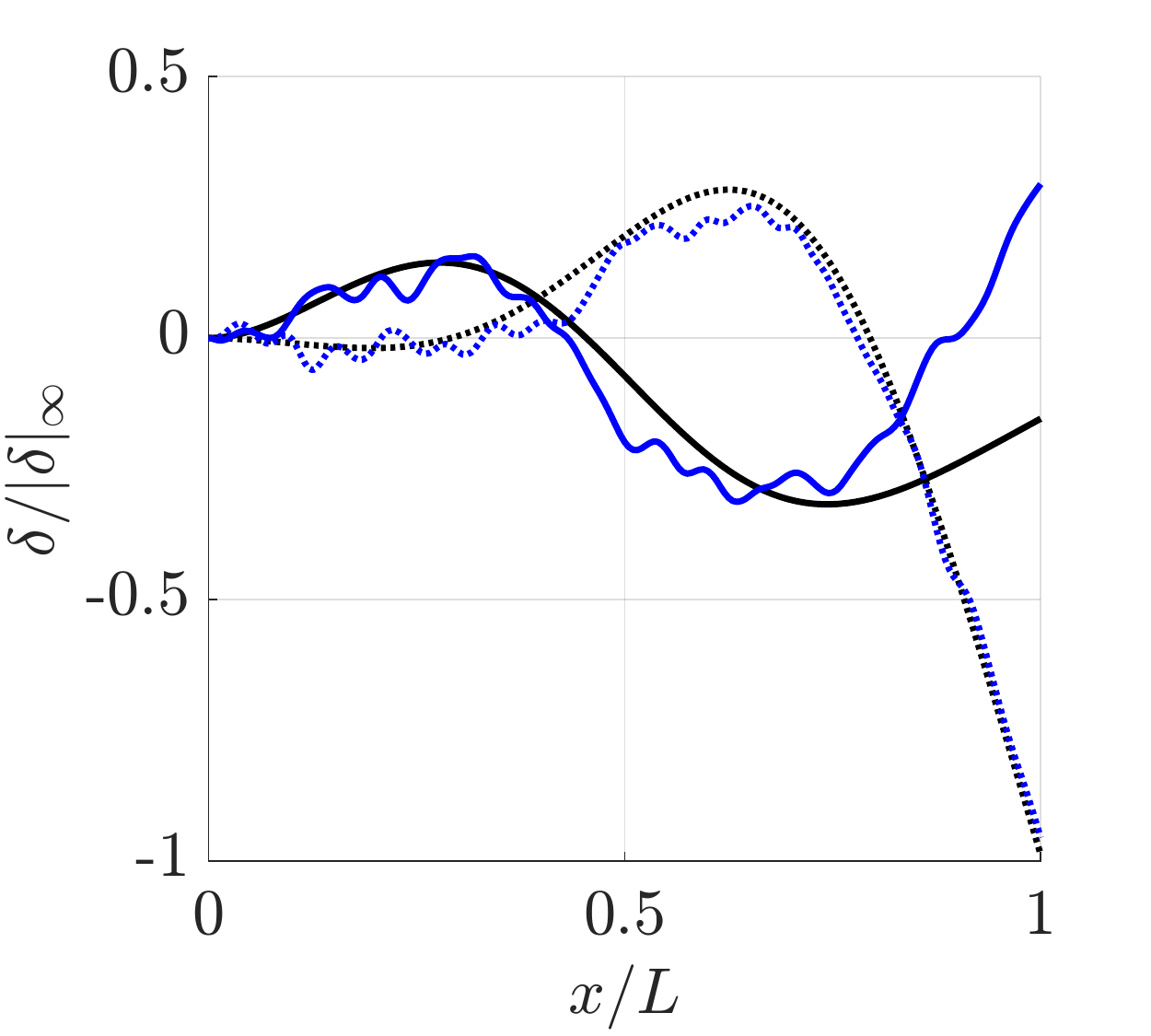}%
    	\caption{Mode shapes at $M^*=0.3$.} 
        \label{fig:hhat_025_Reh2_p5_Modes_p3}
      \end{subfigure}}
\centering
  \usebox0\hfil \usebox1\hfil
  \usebox2\hfil \usebox3\par
	\caption{Comparison of FSI DNS and quasi-1D model for case \casenumvar ($\hat{h} = \hhatvar$,$\hat{h}^2 Re_L =\Rehtvar$).}
    \label{fig:hhat_025_Reh2_p5_All}
\end{figure}

\renewcommand{\hhatvar}{0.05 } 
\renewcommand{\Rehtvar}{0.5 } 
\renewcommand{\casenumvar}{2 } 

\begin{figure}[H]

\sbox0{\begin{subfigure}{0.25\textwidth}
		\captionsetup{width=1\linewidth,font={footnotesize}}
        \includegraphics[width=1\textwidth]{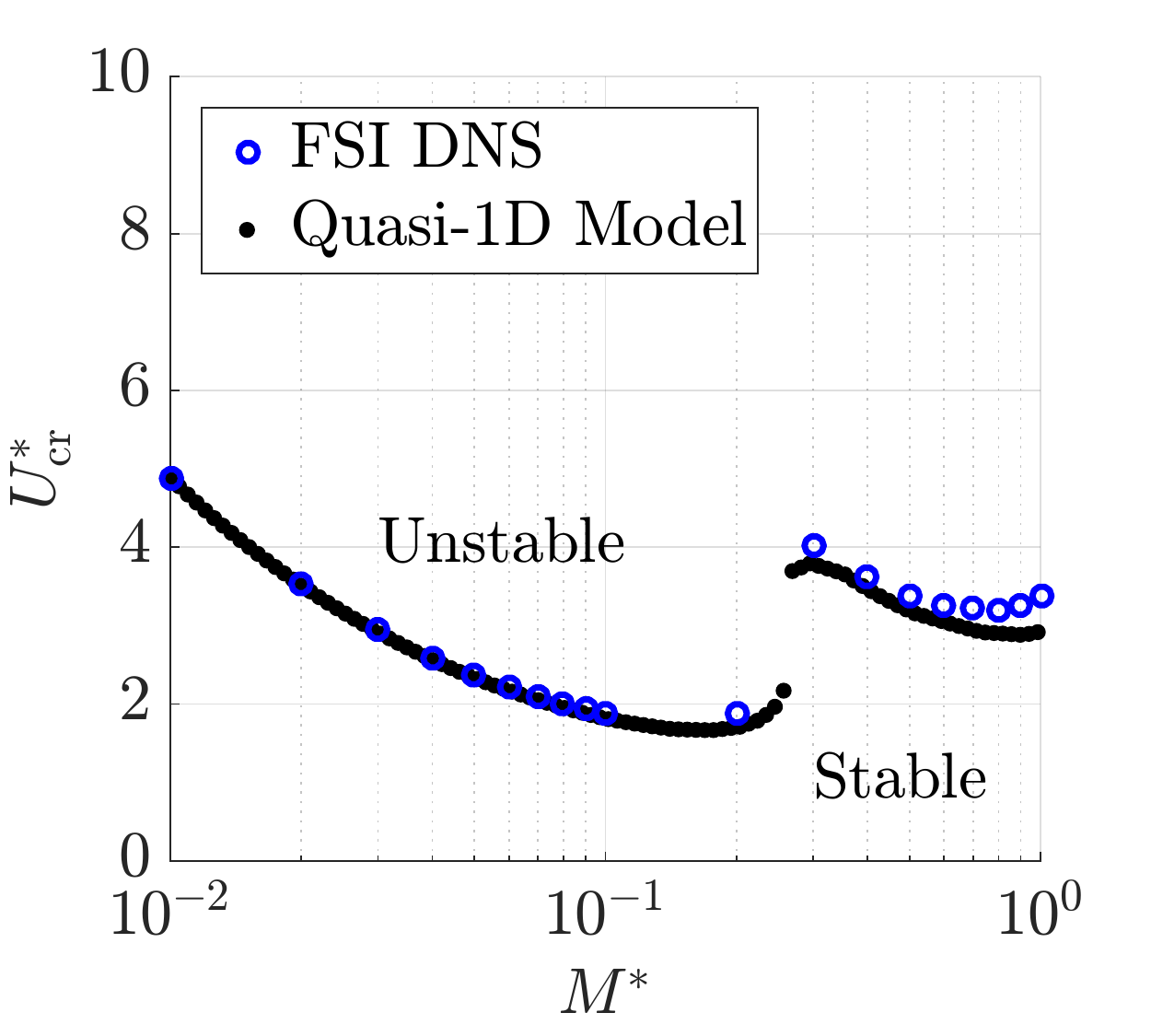}
        \caption{$U^*_{\mathrm{cr}}$ as a function of $M^*$.} 
        \label{fig:hhat_05_Reh2_p5_CritVals_Ucr}
      \end{subfigure}}
\sbox1{\begin{subfigure}{0.25\textwidth}
		\captionsetup{width=1\linewidth,font={footnotesize}}
        \includegraphics[width=1\textwidth]{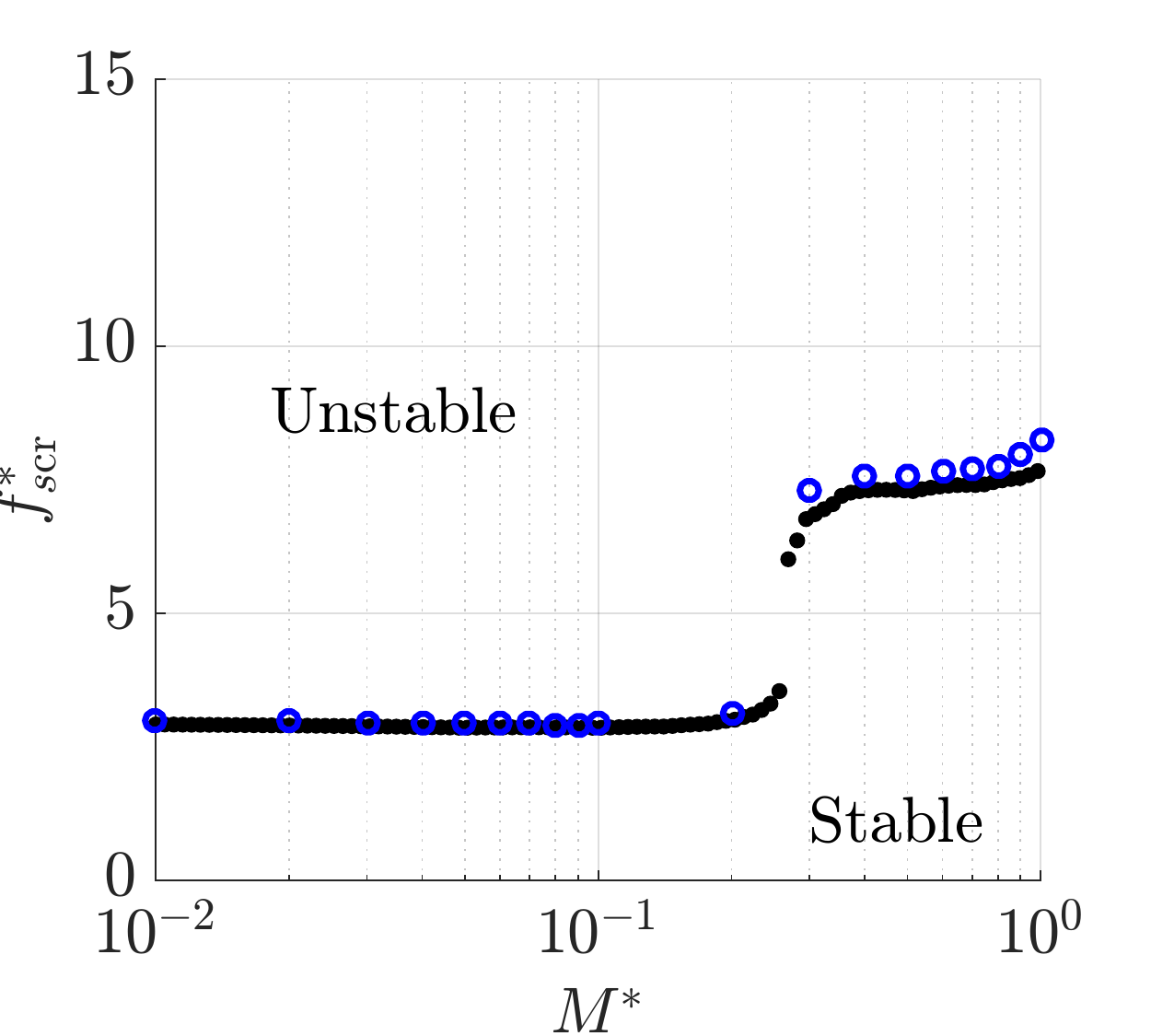}
        \caption{$f^*_{s\mathrm{cr}}$ as a function of $M^*$.} 
        \label{fig:hhat_05_Reh2_p5_CritVals_fcr}
      \end{subfigure}}
\sbox2{\begin{subfigure}{0.25\textwidth}
		\captionsetup{width=1\linewidth,font={footnotesize}}
        \includegraphics[
        width=\textwidth]{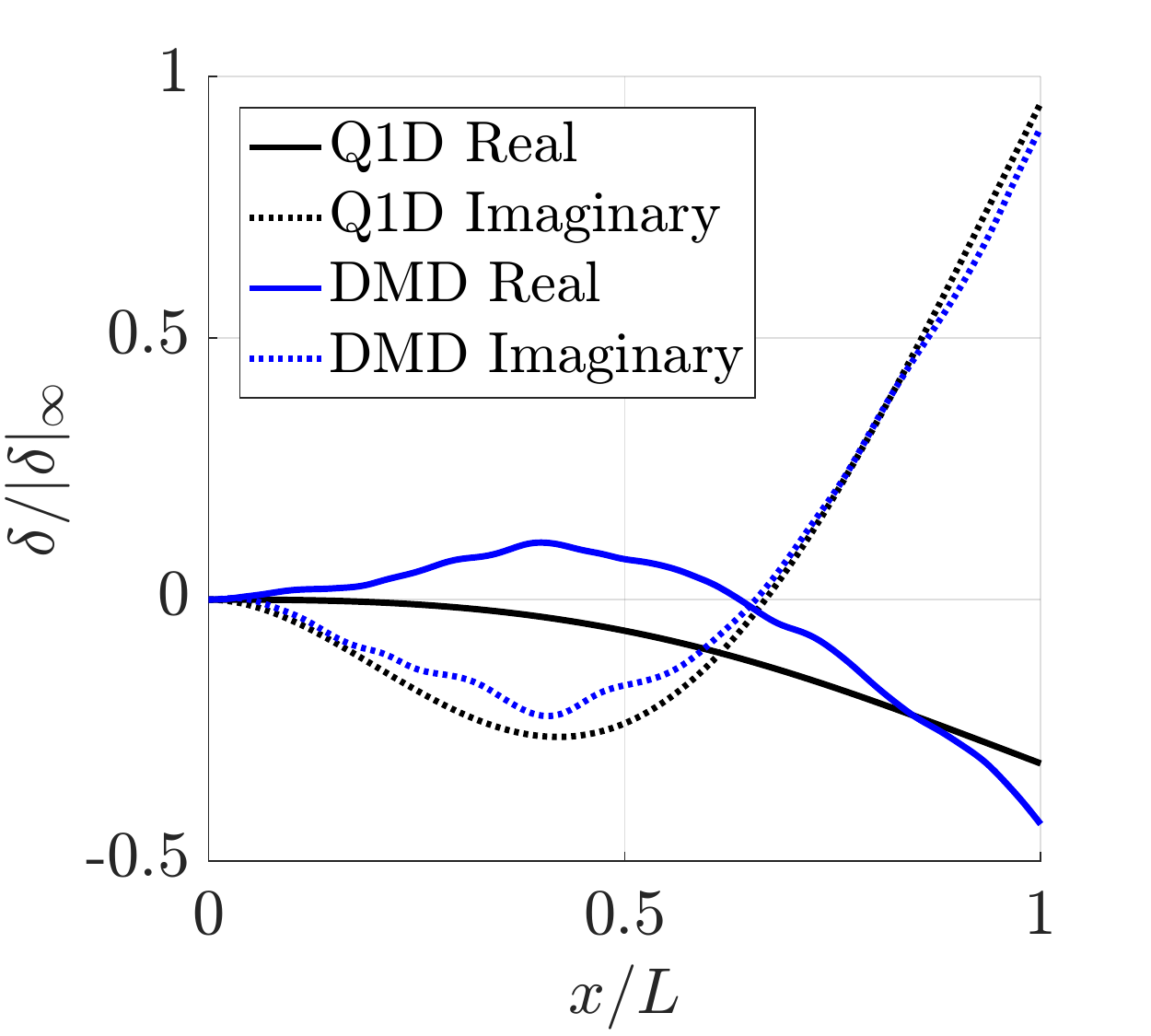}%
    	\caption{Mode shapes at $M^*=0.01$.} 
        \label{fig:hhat_05_Reh2_p5_Modes_p01}
      \end{subfigure}}
\sbox3{\begin{subfigure}{0.25\textwidth}
		\captionsetup{width=1\linewidth,font={footnotesize}}
        \includegraphics[
        width=\textwidth]{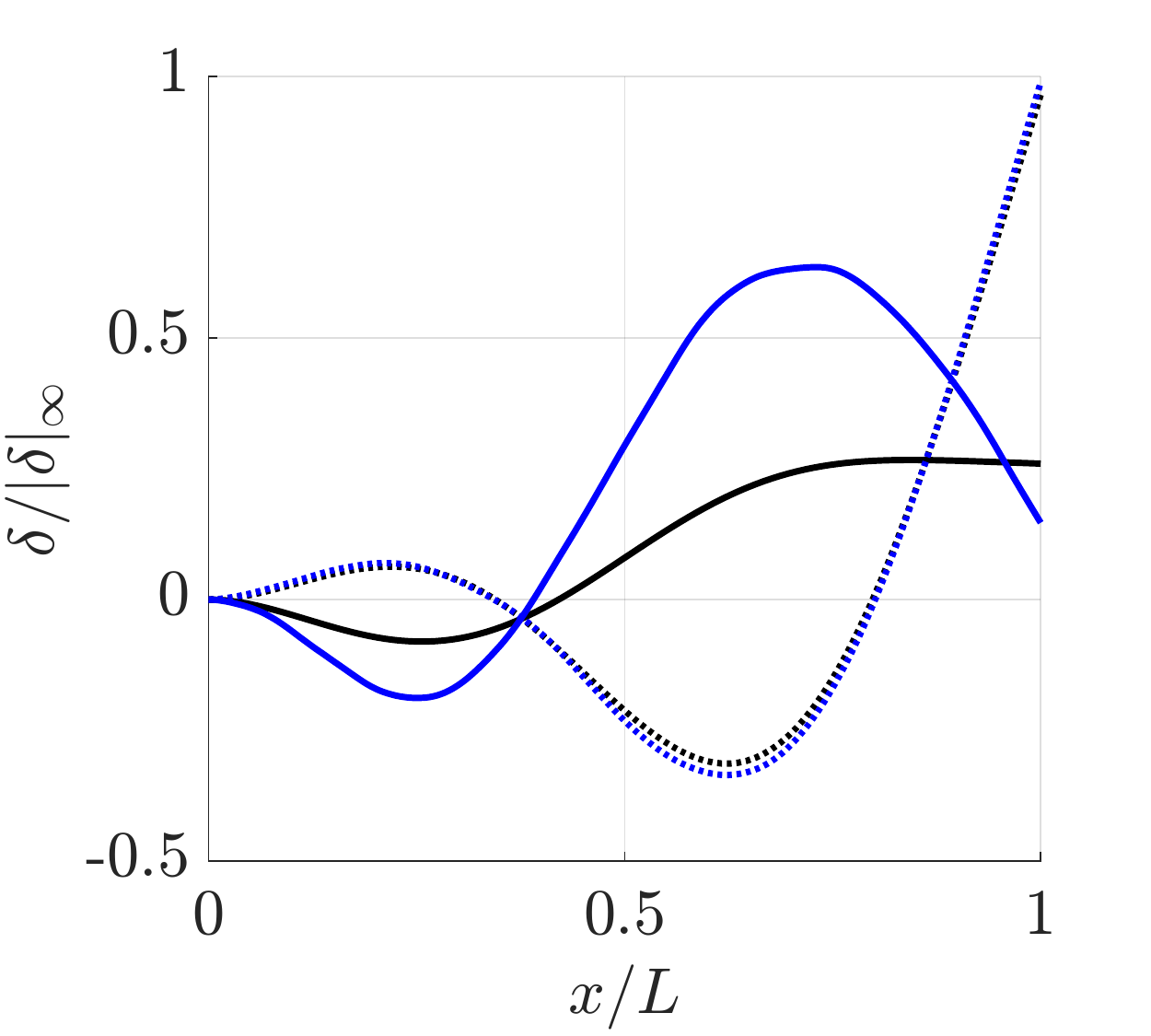}%
    	\caption{Mode shapes at $M^*=0.3$.}
        \label{fig:hhat_05_Reh2_p5_Modes_p3}
      \end{subfigure}}
\centering
  \usebox0\hfil \usebox1\hfil 
  \usebox2\hfil \usebox3\par
	\caption{Comparison of FSI DNS and quasi-1D model for case \casenumvar ($\hat{h} = \hhatvar$,$\hat{h}^2 Re_L =\Rehtvar$).}
    \label{fig:hhat_05_Reh2_p5_All}
\end{figure}

\renewcommand{\hhatvar}{0.05 } 
\renewcommand{\Rehtvar}{1.25 } 
\renewcommand{\casenumvar}{3 } 

\begin{figure}[H]

\sbox0{\begin{subfigure}{0.25\textwidth}
		\captionsetup{width=1\linewidth,font={footnotesize}}
        \includegraphics[width=1\textwidth]{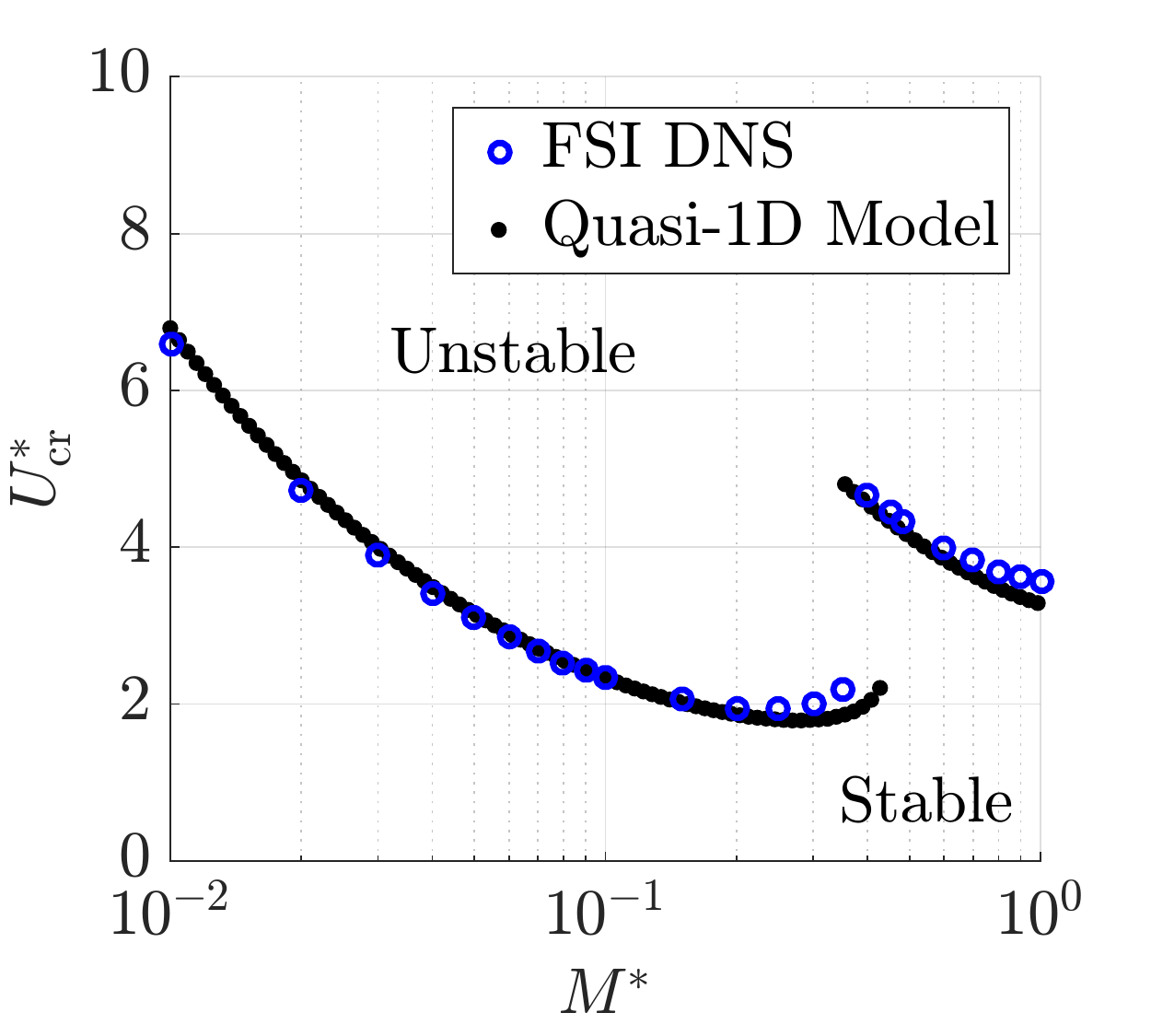}
        \caption{$U^*_{\mathrm{cr}}$ as a function of $M^*$.} 
        \label{fig:hhat_05_Reh2_1p25_CritVals_Ucr}
      \end{subfigure}}
\sbox1{\begin{subfigure}{0.25\textwidth}
		\captionsetup{width=1\linewidth,font={footnotesize}}
        \includegraphics[width=1\textwidth]{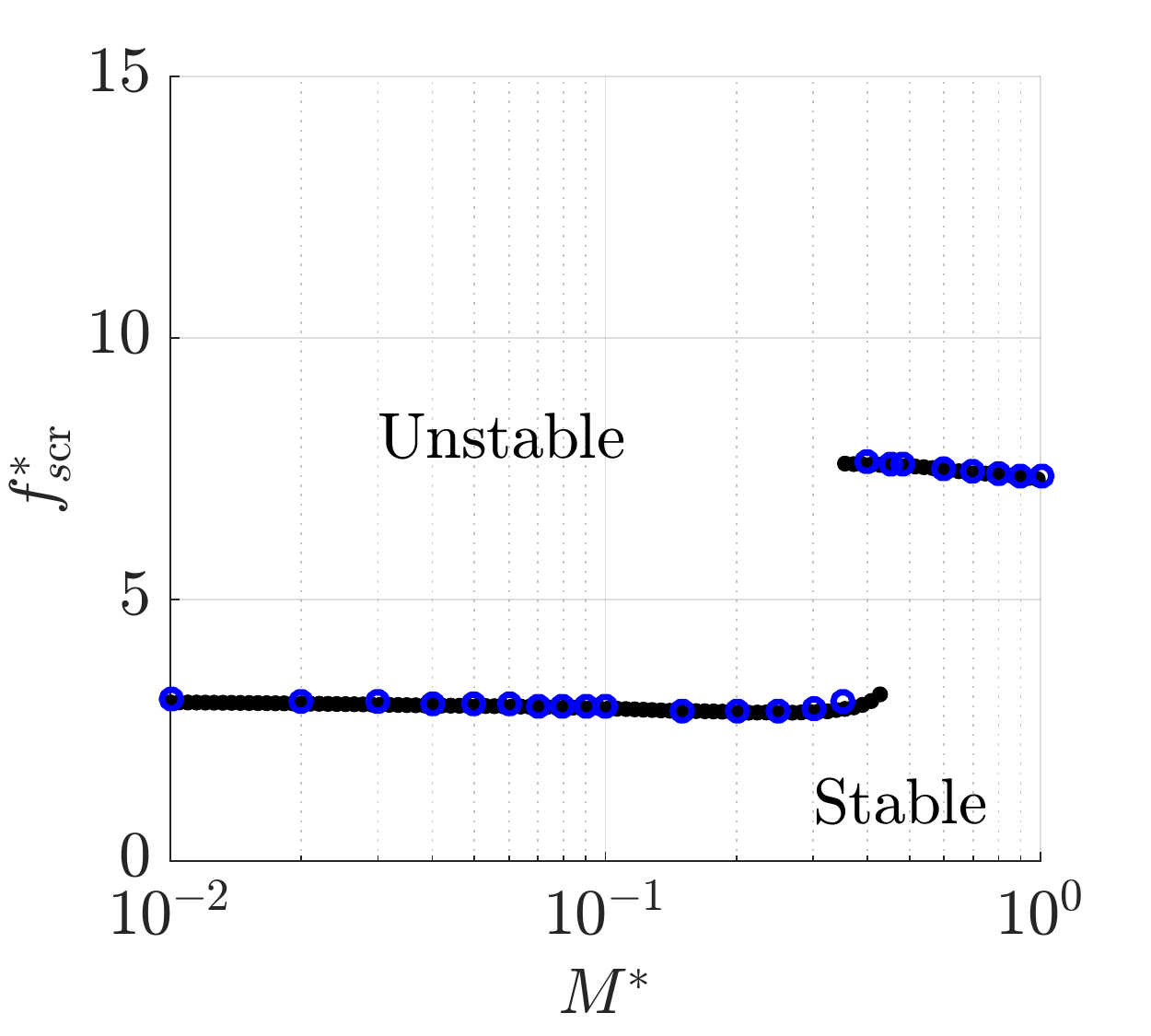}
        \caption{$f^*_{s\mathrm{cr}}$ as a function of $M^*$.} 
        \label{fig:hhat_05_Reh2_1p25_CritVals_fcr}
      \end{subfigure}}
\sbox2{\begin{subfigure}{0.25\textwidth}
		\captionsetup{width=1\linewidth,font={footnotesize}}
        \includegraphics[
        width=\textwidth]{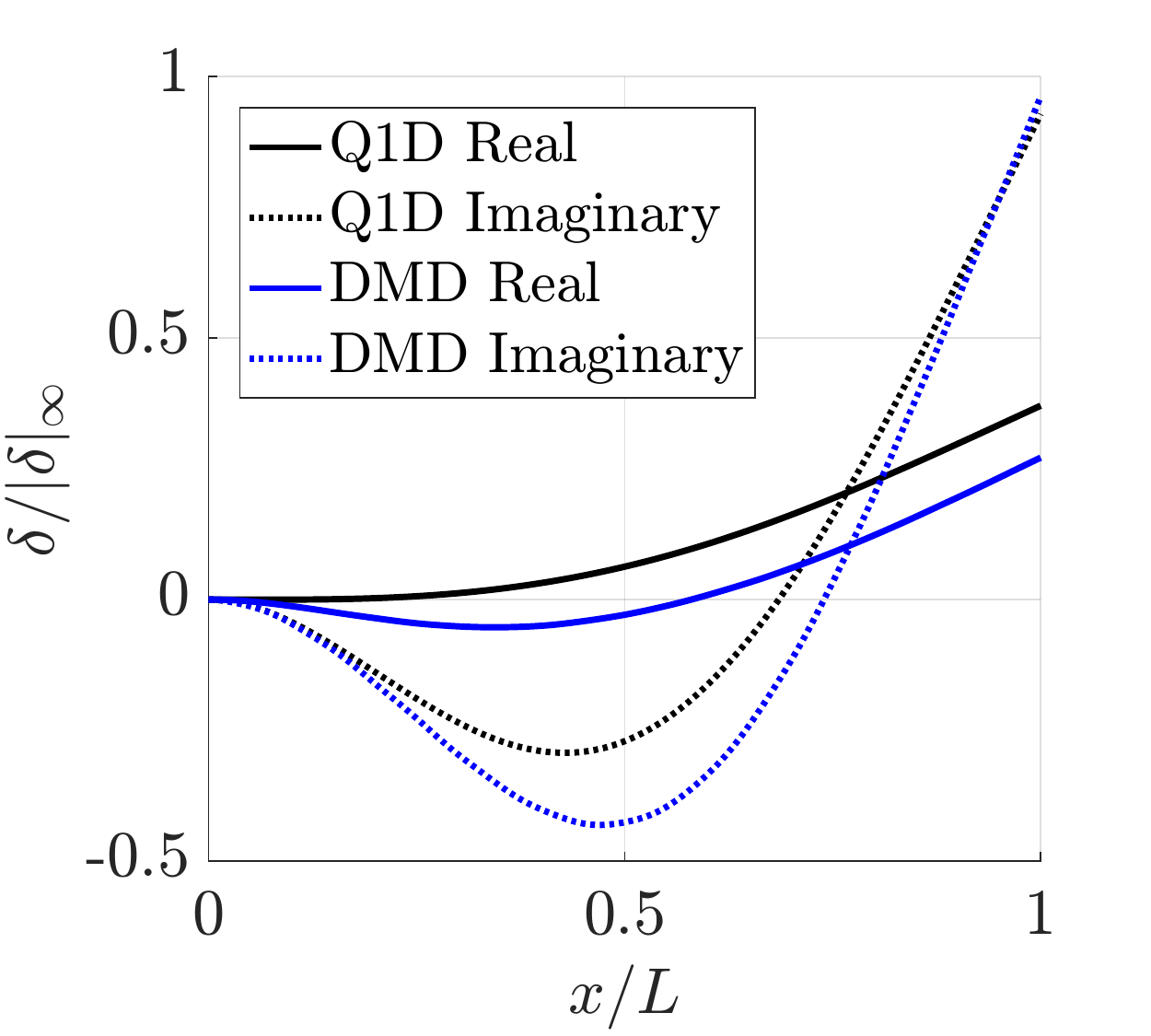}
    	\caption{Mode shapes at $M^*=0.01$.} 
        \label{fig:hhat_05_Reh2_1p25_Modes_p01}
      \end{subfigure}}
\sbox3{\begin{subfigure}{0.25\textwidth}
		\captionsetup{width=1\linewidth,font={footnotesize}}
        \includegraphics[
        width=\textwidth]{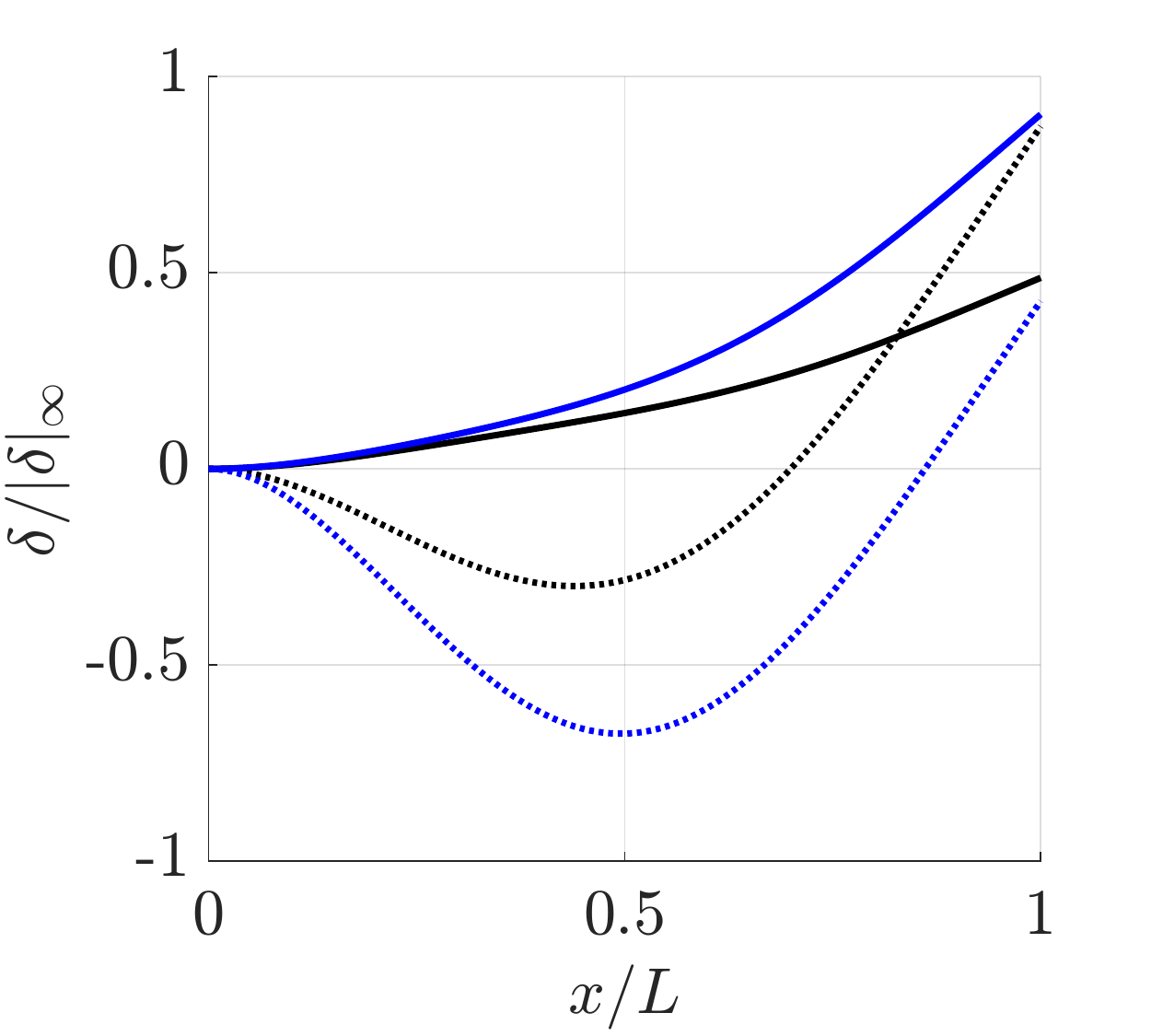}
    	\caption{Mode shapes at $M^*=0.3$.}
        \label{fig:hhat_05_Reh2_1p25_Modes_p3}
      \end{subfigure}}
\centering
  \usebox0\hfil \usebox1\hfil 
  \usebox2\hfil \usebox3\par
	\caption{Comparison of FSI DNS and quasi-1D model for case \casenumvar ($\hat{h} = \hhatvar$,$\hat{h}^2 Re_L =\Rehtvar$).}
    \label{fig:hhat_05_Reh2_1p25_All}
\end{figure}

\renewcommand{\hhatvar}{0.05 } 
\renewcommand{\Rehtvar}{2.5 } 
\renewcommand{\casenumvar}{4 }

\begin{figure}[H]

\sbox0{\begin{subfigure}{0.25\textwidth}
   		\captionsetup{width=1\linewidth,font={footnotesize}}
        \includegraphics[width=1\textwidth]{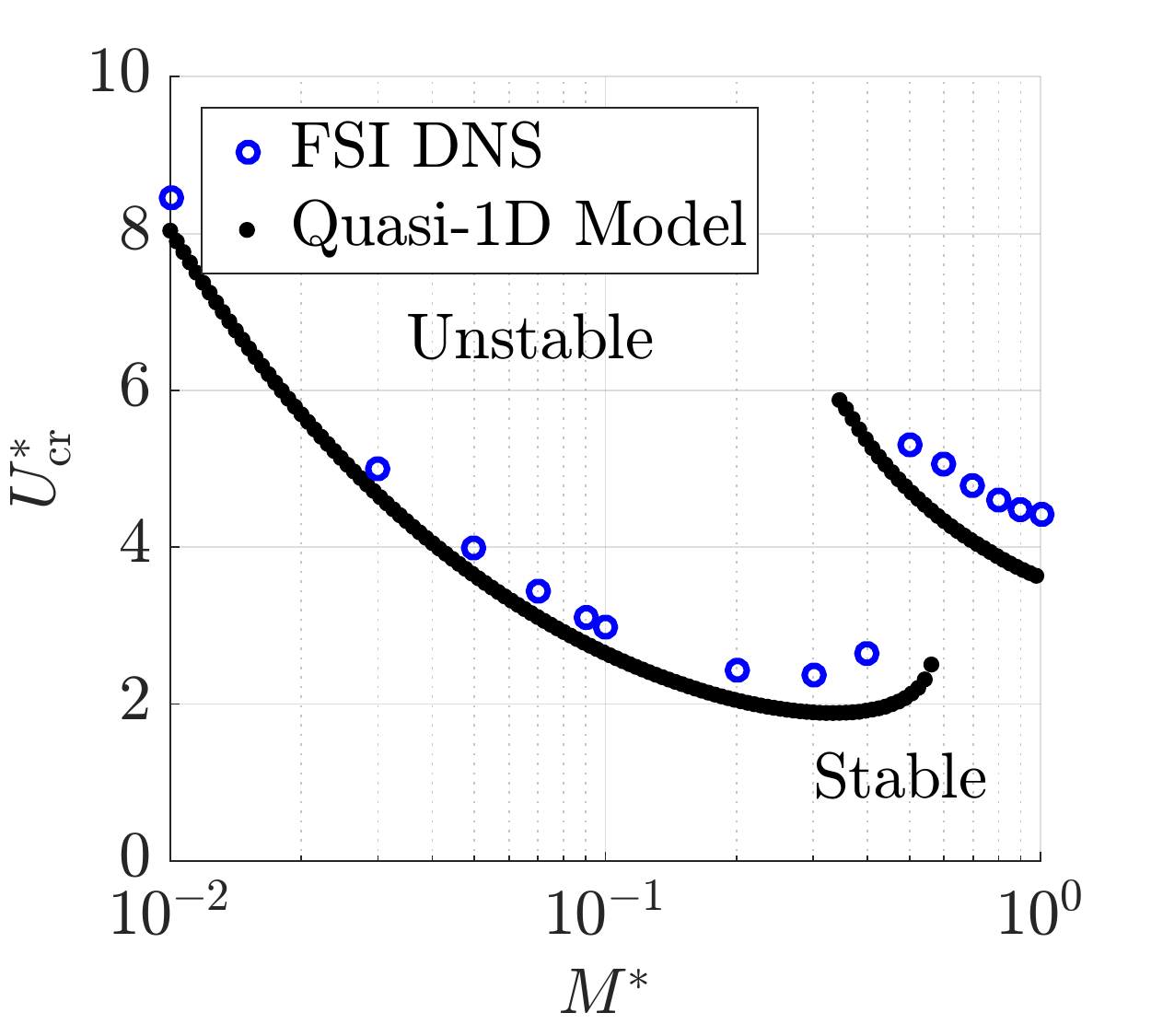}
        \caption{$U^*_{\mathrm{cr}}$ as a function of $M^*$.} 
        \label{fig:hhat_05_Reh2_2p5_CritVals_Ucr}
      \end{subfigure}}
\sbox1{\begin{subfigure}{0.25\textwidth}
		\captionsetup{width=1\linewidth,font={footnotesize}}
        \includegraphics[width=1\textwidth]{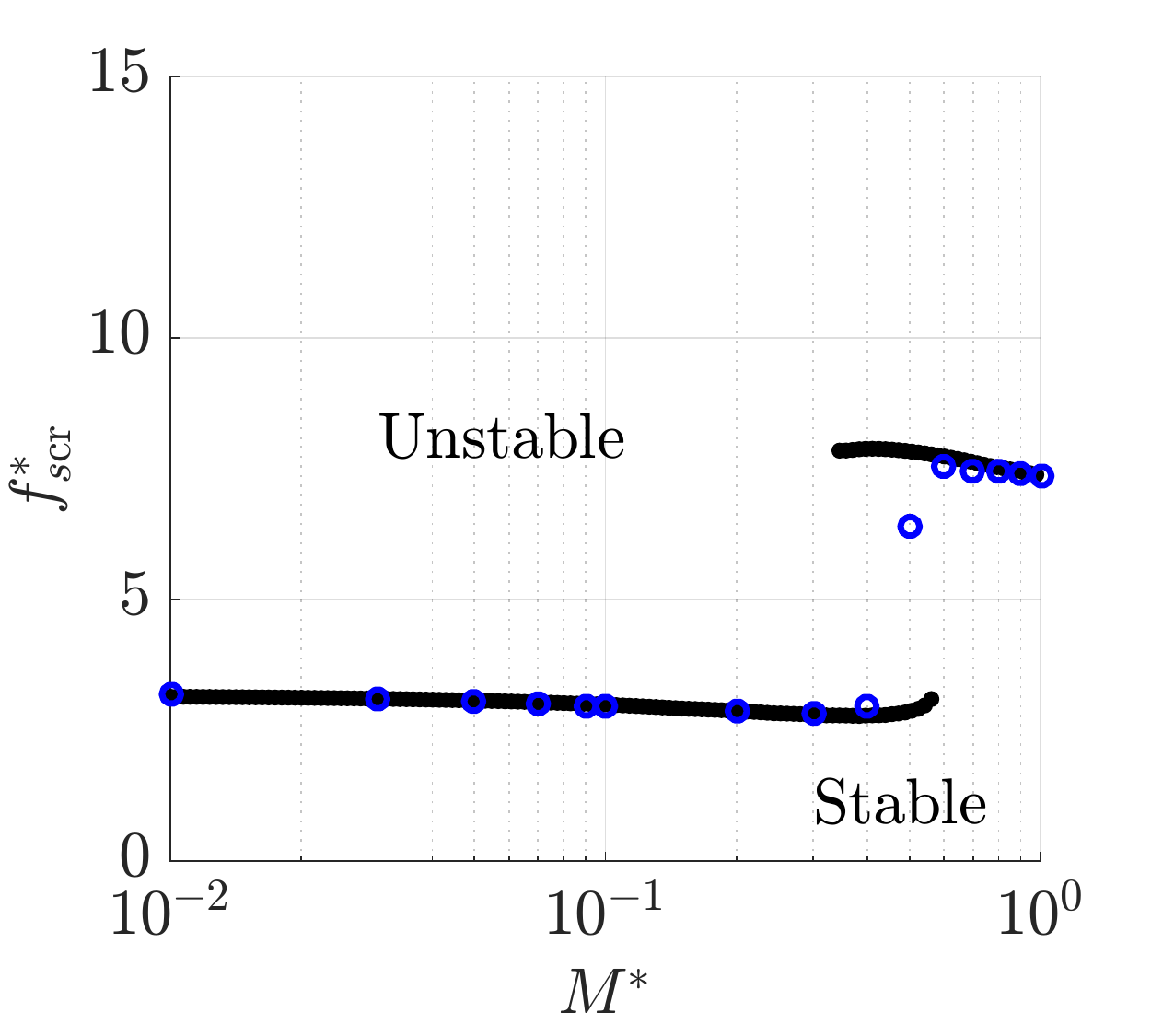}
        \caption{$f^*_{s\mathrm{cr}}$ as a function of $M^*$.} 
        \label{fig:hhat_05_Reh2_2p5_CritVals_fcr}
      \end{subfigure}}
\sbox2{\begin{subfigure}{0.25\textwidth}
        \captionsetup{width=1\linewidth,font={footnotesize}}
    	\includegraphics[
    	width=\textwidth]{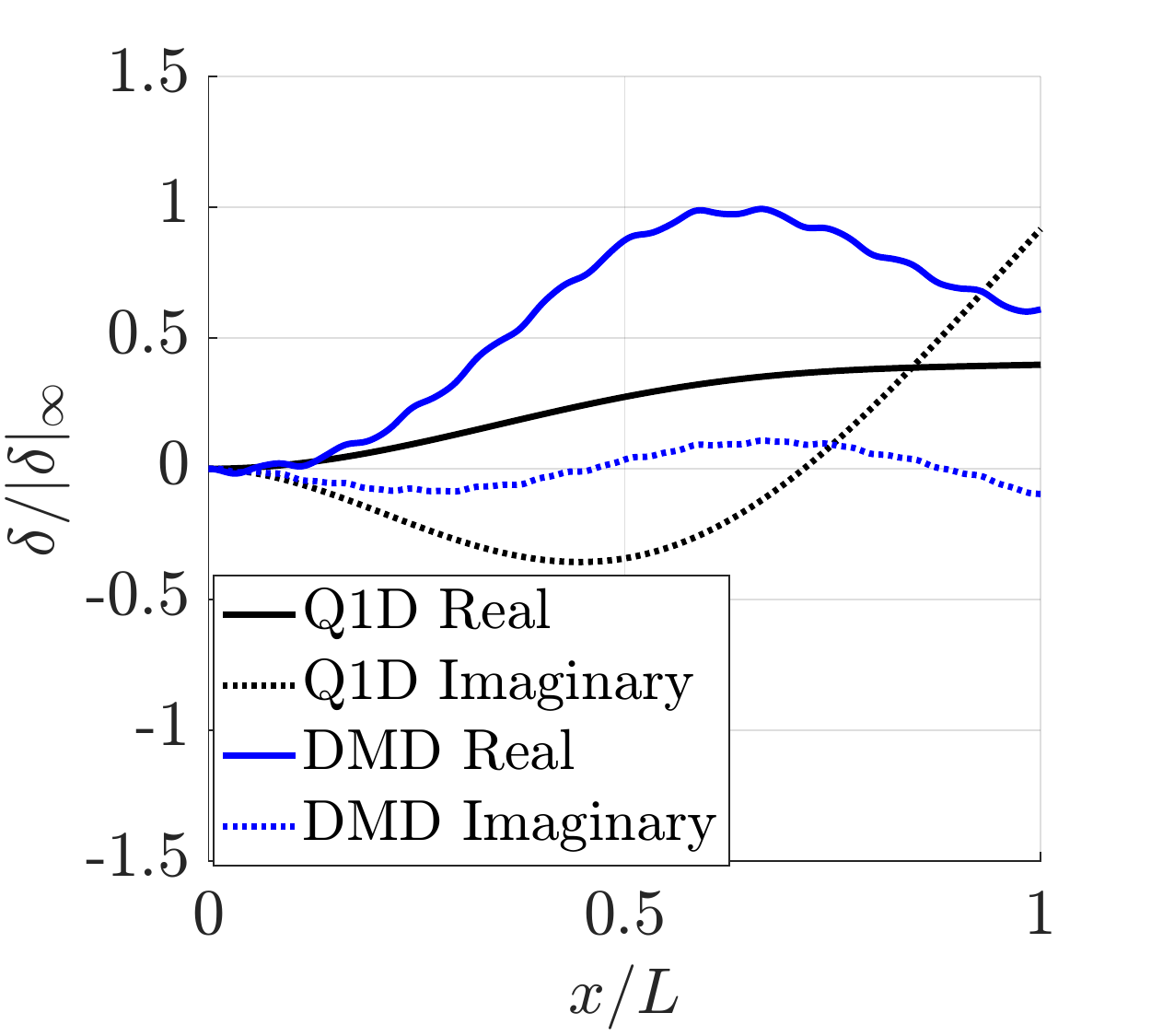}%
    	\caption{Mode shapes at $M^*=0.01$.} 
        \label{fig:hhat_05_Reh2_2p5_Modes_p01}
      \end{subfigure}}
\sbox3{\begin{subfigure}{0.25\textwidth}
        \captionsetup{width=1\linewidth,font={footnotesize}}
    	\includegraphics[
    	width=\textwidth]{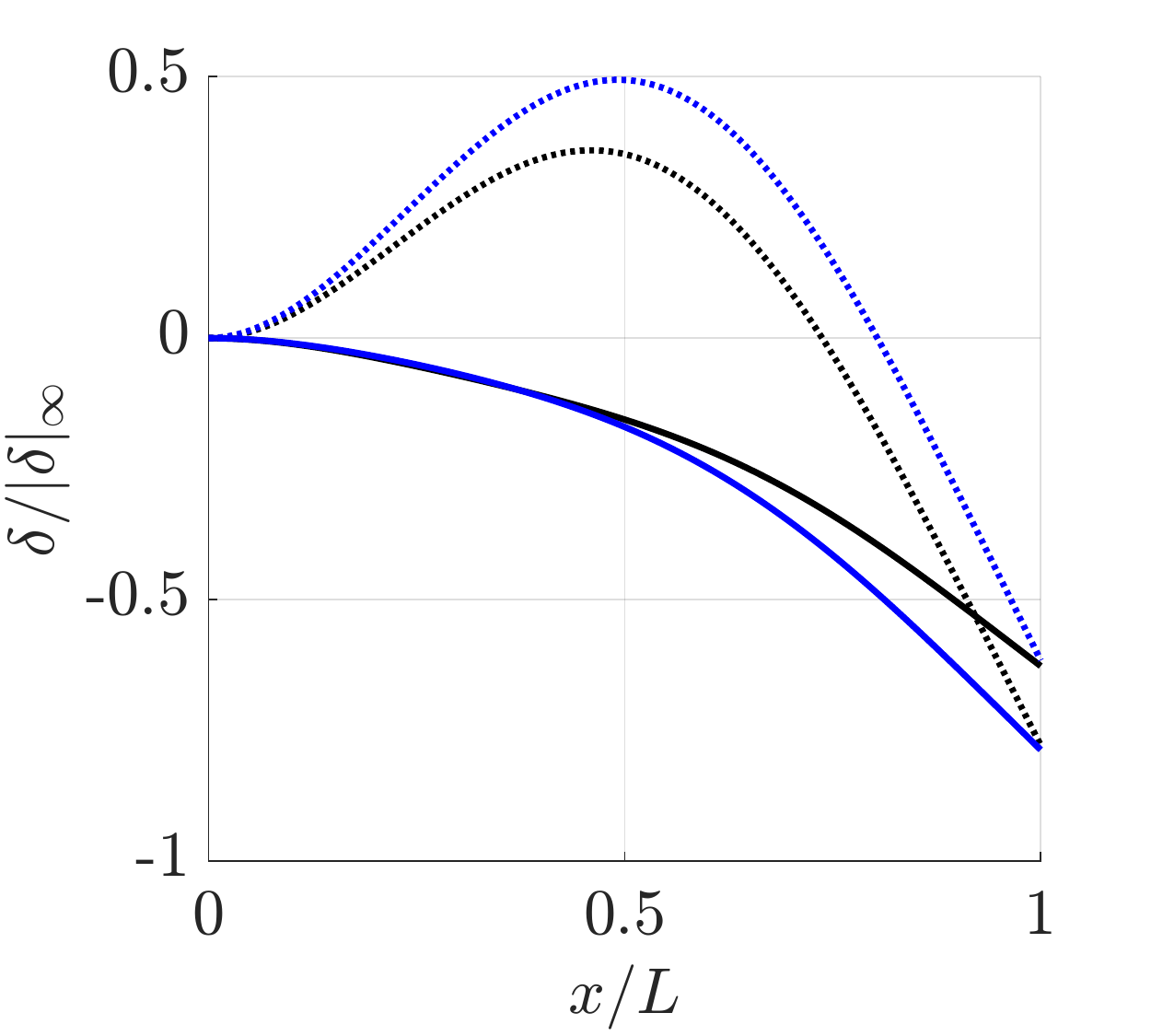}%
    	\caption{Mode shapes at $M^*=0.3$.}
        \label{fig:hhat_05_Reh2_2p5_Modes_p3}
      \end{subfigure}}
\centering
    \usebox0\hfil \usebox1\hfil 
    \usebox2\hfil \usebox3\par
	\caption{Comparison of FSI DNS and quasi-1D model for case \casenumvar ($\hat{h} = \hhatvar$,$\hat{h}^2 Re_L =\Rehtvar$).}
    \label{fig:hhat_05_Reh2_2p5_All}
\end{figure}

\renewcommand{\hhatvar}{0.125 } 
\renewcommand{\Rehtvar}{0.5 } 
\renewcommand{\casenumvar}{5 } 

\begin{figure}[H]

\sbox0{\begin{subfigure}{0.25\textwidth}
        \captionsetup{width=1\linewidth,font={footnotesize}}
   		\includegraphics[width=1\textwidth]{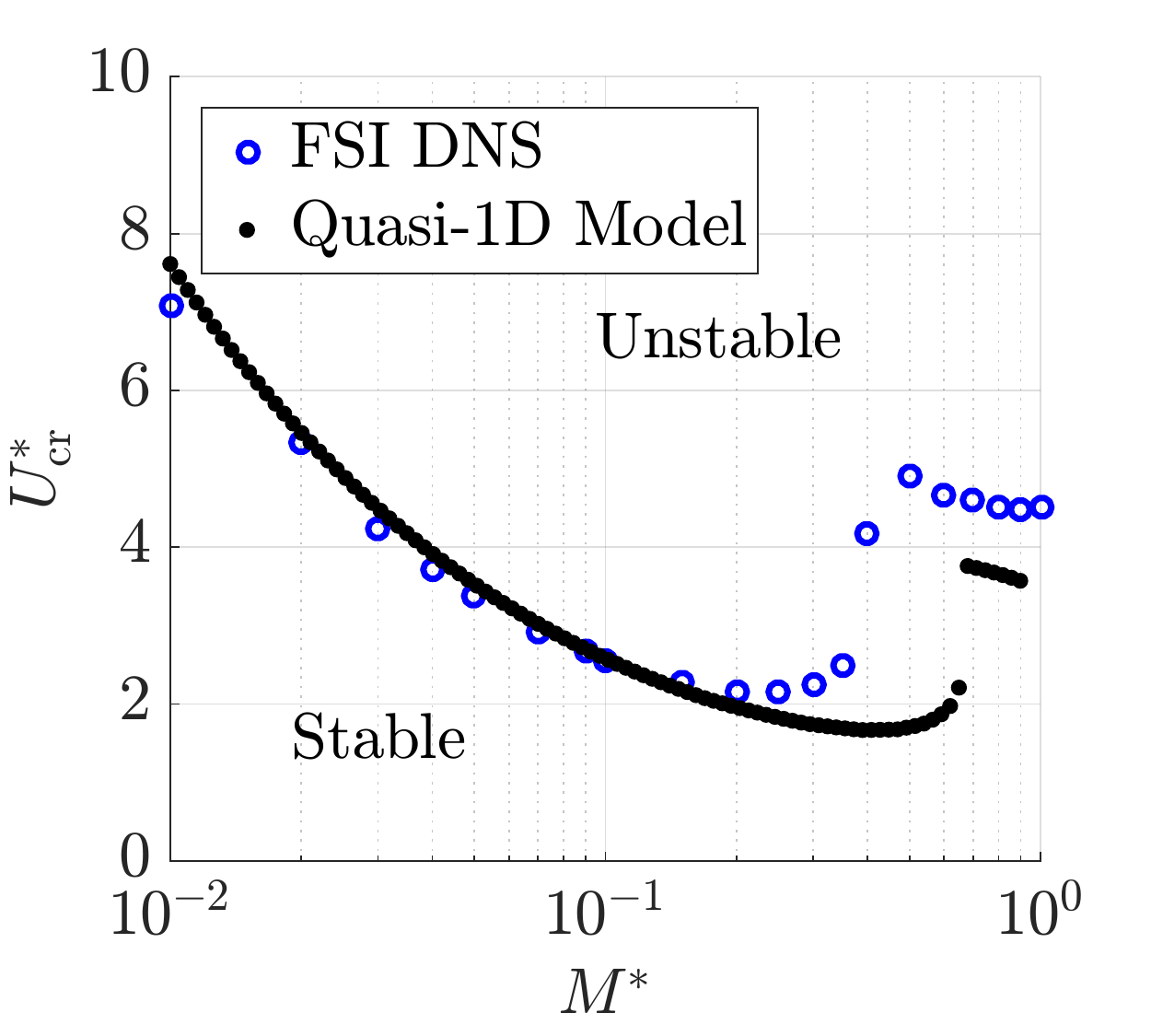}
        \caption{$U^*_{\mathrm{cr}}$ as a function of $M^*$.} 
        \label{fig:hhat_125_Reh2_p5_CritVals_Ucr}
      \end{subfigure}}
\sbox1{\begin{subfigure}{0.25\textwidth}
        \captionsetup{width=1\linewidth,font={footnotesize}}
   		\includegraphics[width=1\textwidth]{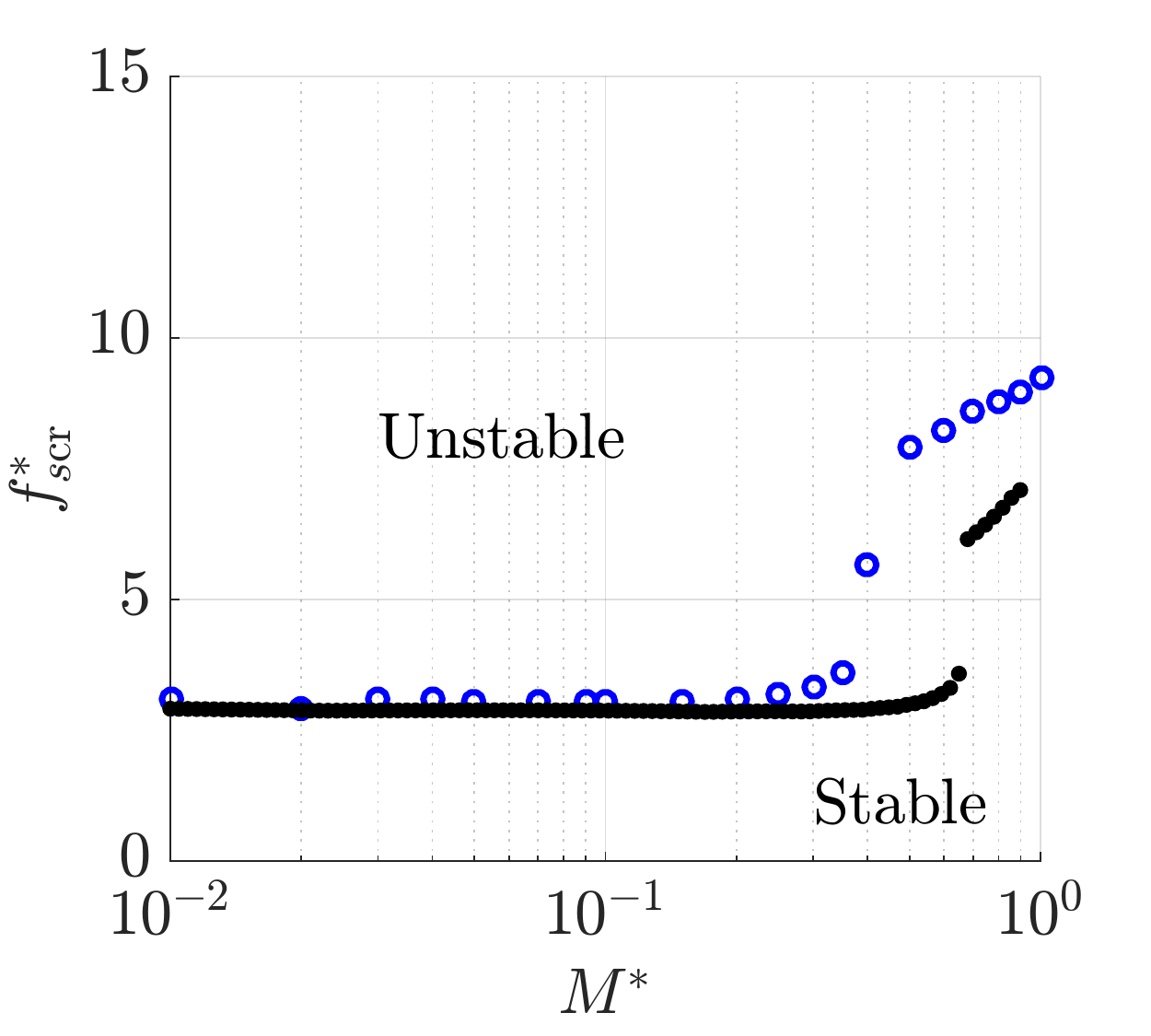}
        \caption{$f^*_{s\mathrm{cr}}$ as a function of $M^*$.} 
        \label{fig:hhat_125_Reh2_p5_CritVals_fcr}
      \end{subfigure}}
\sbox2{\begin{subfigure}{0.25\textwidth}
        \captionsetup{width=1\linewidth,font={footnotesize}}
    	\includegraphics[
    	width=\textwidth]{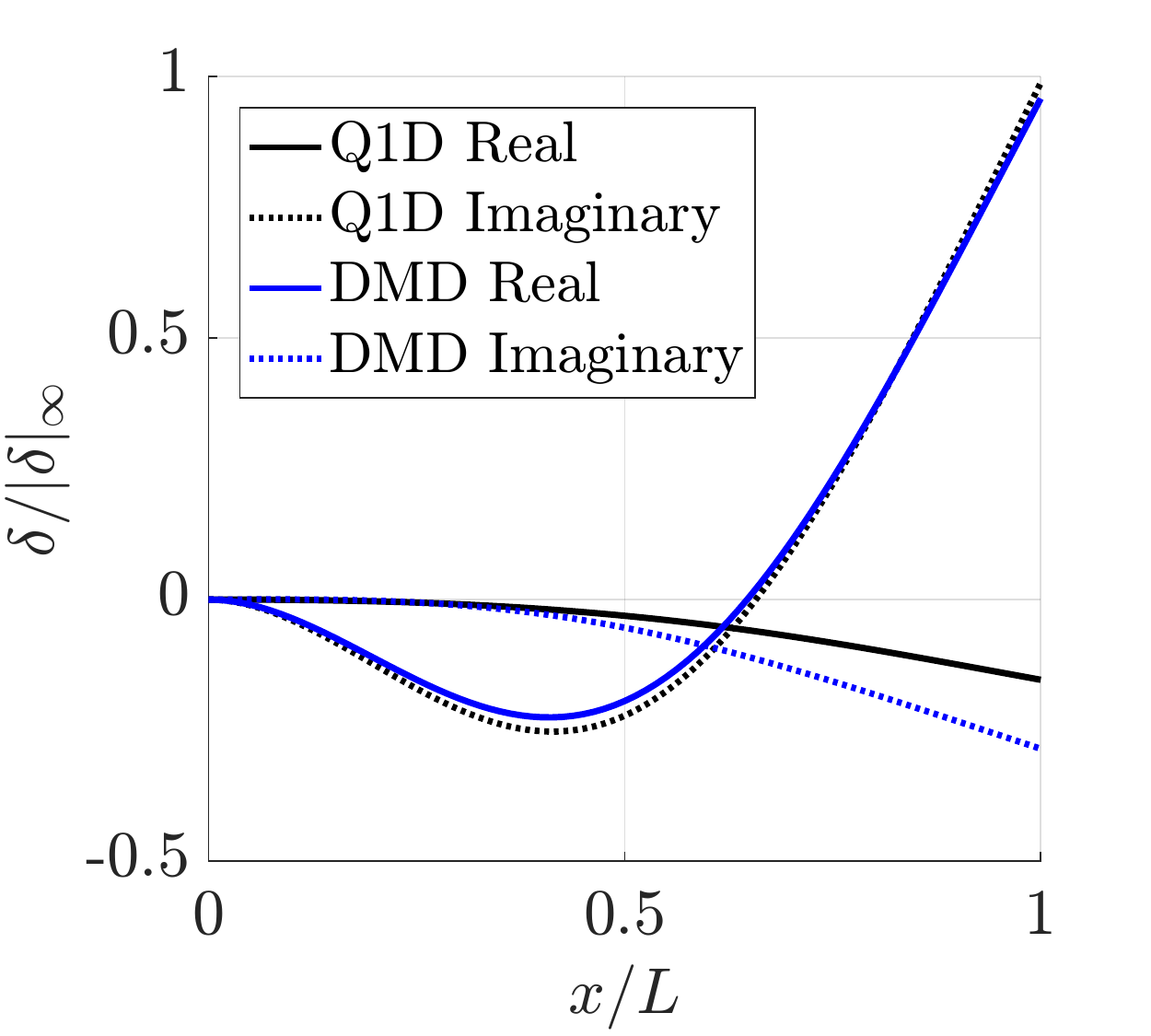}
    	\caption{Mode shapes at $M^*=0.01$.} 
        \label{fig:hhat_125_Reh2_p5_Modes_p01}
      \end{subfigure}}
\sbox3{\begin{subfigure}{0.25\textwidth}
        \captionsetup{width=1\linewidth,font={footnotesize}}
    	\includegraphics[
    	width=\textwidth]{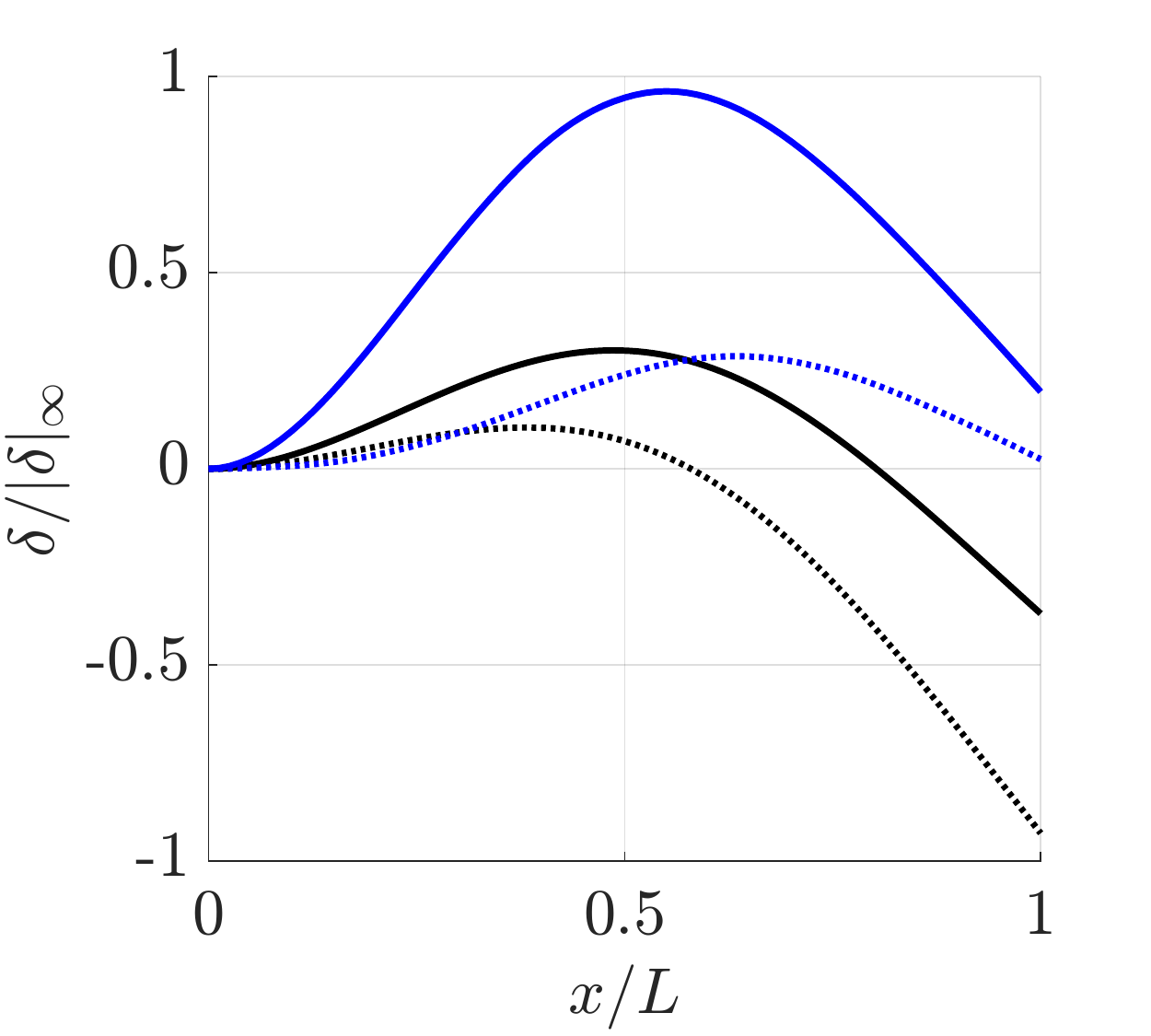}%
    	\caption{Mode shapes at $M^*=0.3$.}
        \label{fig:hhat_125_Reh2_p5_Modes_p3}
      \end{subfigure}}
\centering
    \usebox0\hfil \usebox1\hfil 
    \usebox2\hfil \usebox3\par
	\caption{Comparison of FSI DNS and quasi-1D model for case \casenumvar ($\hat{h} = \hhatvar$,$\hat{h}^2 Re_L =\Rehtvar$).}
    \label{fig:hhat_125_Reh2_p5_All}
\end{figure}

\renewcommand{\hhatvar}{0.125 } 
\renewcommand{\Rehtvar}{1.25 } 
\renewcommand{\casenumvar}{6 }

\begin{figure}[H]

\sbox0{\begin{subfigure}{0.25\textwidth}
        \captionsetup{width=1\linewidth,font={footnotesize}}
   		\includegraphics[width=1\textwidth]{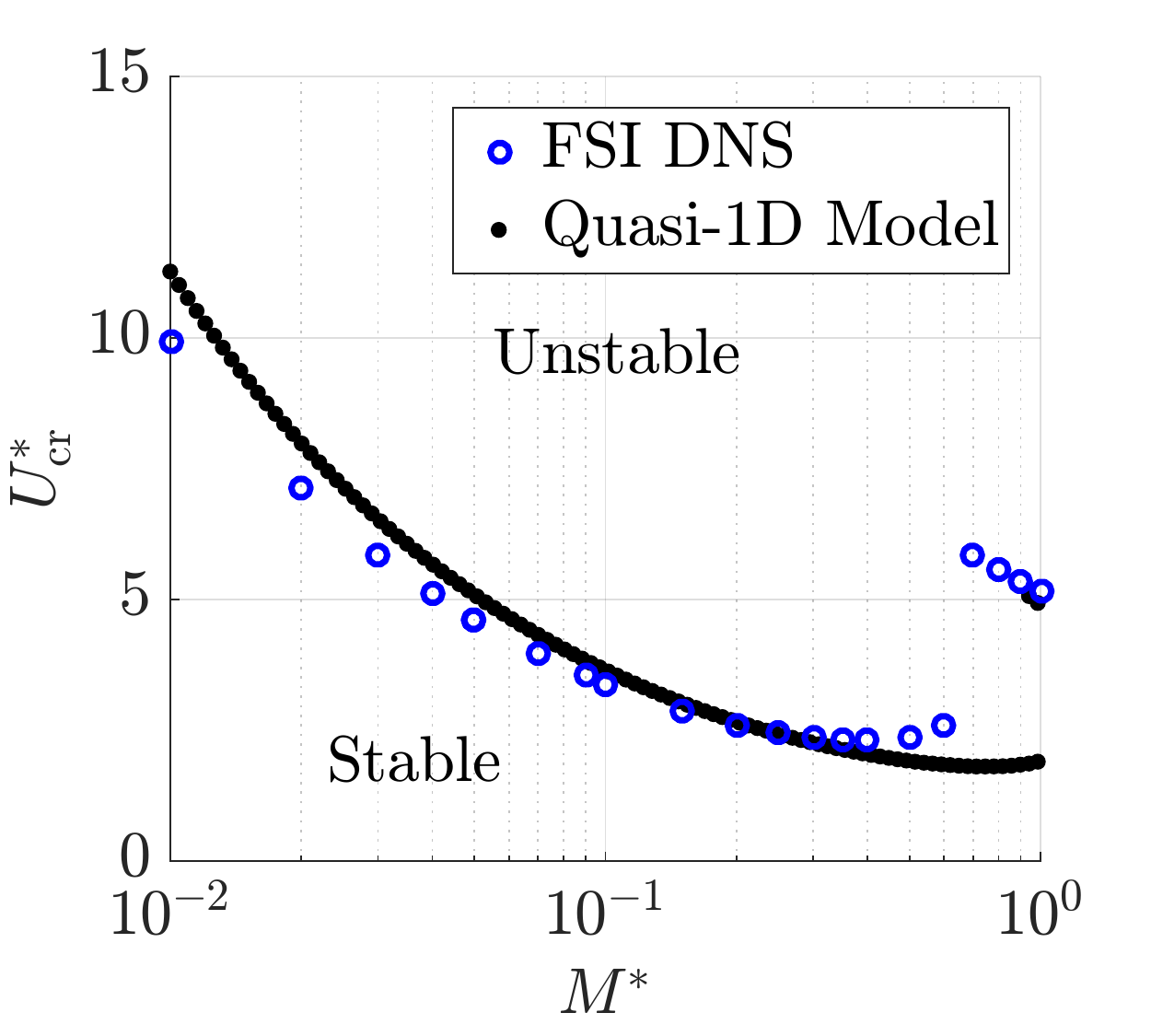}
        \caption{$U^*_{\mathrm{cr}}$ as a function of $M^*$.} 
        \label{fig:hhat_125_Reh2_1p25_CritVals_Ucr}
      \end{subfigure}}
\sbox1{\begin{subfigure}{0.25\textwidth}
        \captionsetup{width=1\linewidth,font={footnotesize}}
   		\includegraphics[width=1\textwidth]{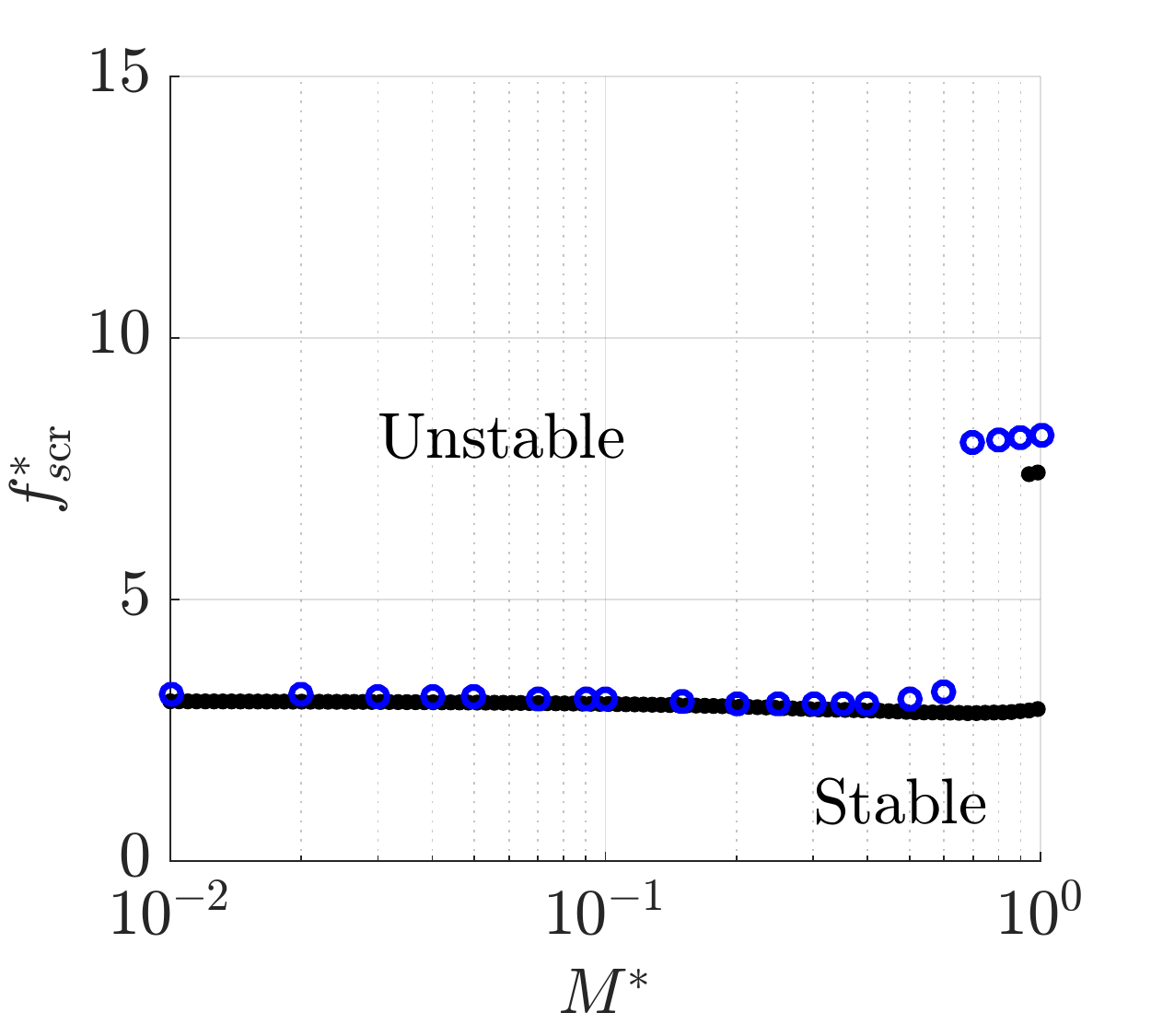}
        \caption{$f^*_{s\mathrm{cr}}$ as a function of $M^*$.} 
        \label{fig:hhat_125_Reh2_1p25_CritVals_fcr}
      \end{subfigure}}
\sbox2{\begin{subfigure}{0.25\textwidth}
        \captionsetup{width=1\linewidth,font={footnotesize}}
    	\includegraphics[
    	width=\textwidth]{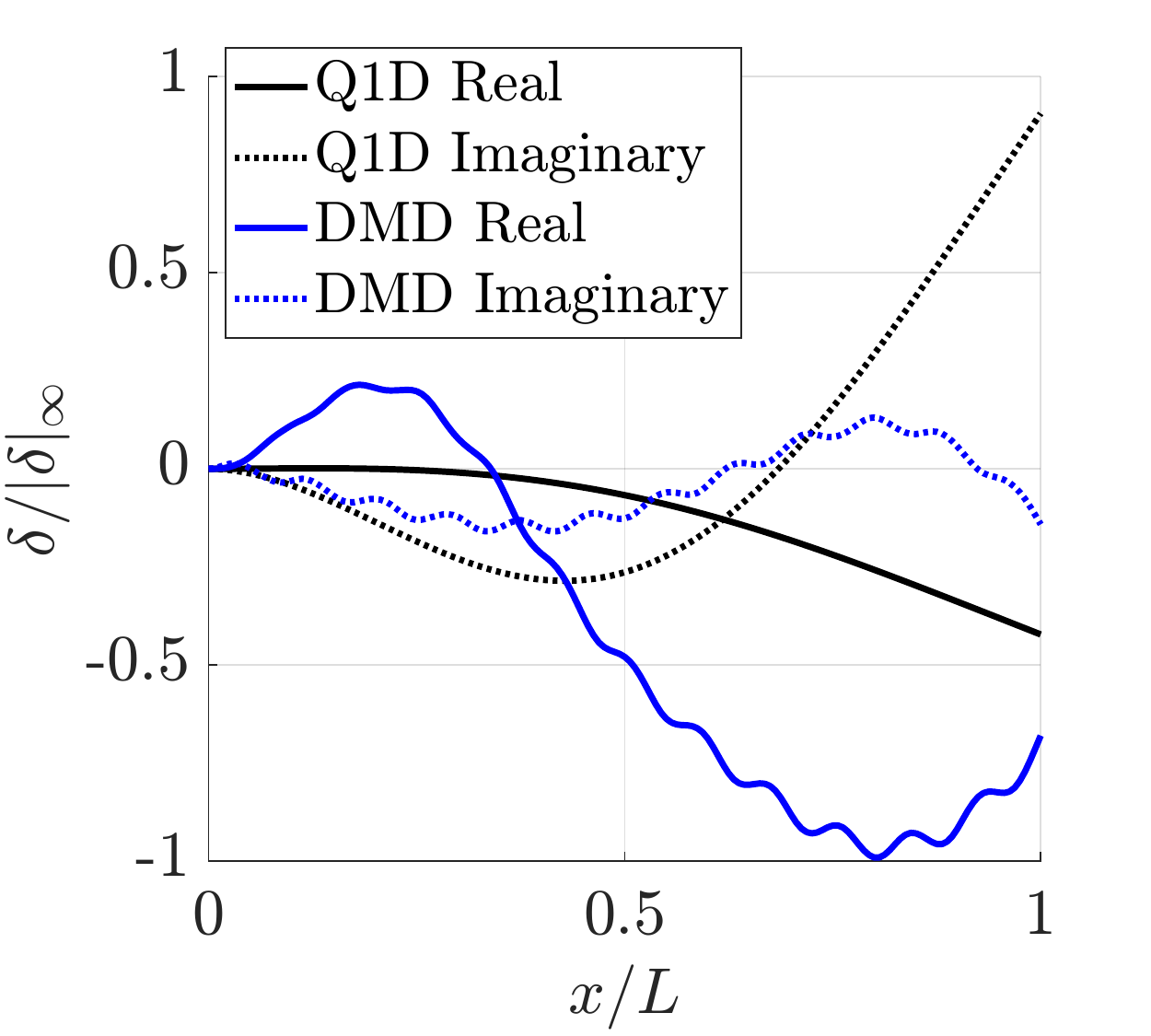}
    	\caption{Mode shapes at $M^*=0.01$.} 
        \label{fig:hhat_125_Reh2_1p25_Modes_p01}
      \end{subfigure}}
\sbox3{\begin{subfigure}{0.25\textwidth}
        \captionsetup{width=1\linewidth,font={footnotesize}}
    	\includegraphics[
    	width=\textwidth]{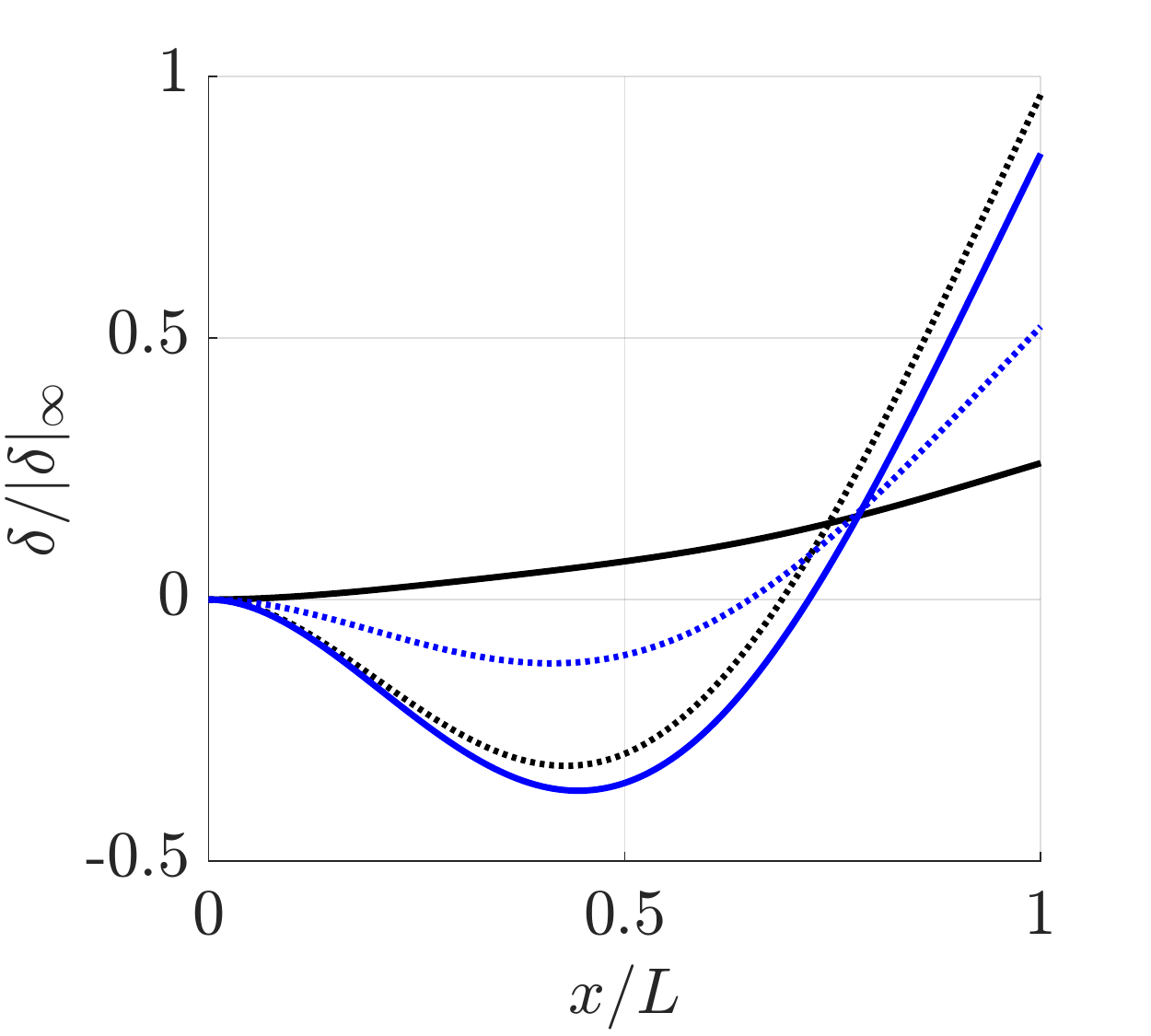}%
    	\caption{Mode shapes at $M^*=0.3$.}
        \label{fig:hhat_125_Reh2_1p25_Modes_p3}
      \end{subfigure}}
\centering
    \usebox0\hfil \usebox1\hfil 
    \usebox2\hfil \usebox3\par
	\caption{Comparison of FSI DNS and quasi-1D model for case \casenumvar ($\hat{h} = \hhatvar$,$\hat{h}^2 Re_L =\Rehtvar$).}
    \label{fig:hhat_125_Reh2_1p25_All}
\end{figure}

Cases 7 and 8 explore a larger range of ${\hat h}^2 Re_L$ for a low mass ratio $M^* = 0.01$ (heavy beam) where only the lowest beam mode is excited.  
Figure \ref{fig:hhat_p05_Reh2_3_Mst_p01_MstSims_All} presents results for case 7, where $\hat{h} = 0.05$, with the flutter boundary now depicted as a function of $\hat{h}^2Re_L$, and representative mode shapes depicted for $\hat{h}^2Re_L = 3$.  At the upper limit of $\hat{h}^2Re_L$, the equivalent $Re_L$ reaches 1800.  
The quasi-1D model closely replicates the simulation results, with a small discrepancy in $U^*_{\mathrm{cr}}$ apparent when $\hat{h}^2 Re_L > 3$. Representative modes shapes also agree.  In case 8 (figure \ref{fig:hhat_p125_Reh2_3_Mst_p01_MstSims_All}) $\hat{h}$ is increased to 0.125 and  $\hat{h}^2Re_L$ ranges from 0.1 to 9.5, with an equivalent $Re_L = 608$ at the upper limit. 
Similar to case 7, quasi-1D results agree reasonably well over the range of $\hat{h}^2Re_L$, with a slight deviation near $\hat{h}^2Re_L = 2$. Most notable, however, is the agreement at $\hat{h}^2Re_L > 5$, as both results appear to asymptote to $U^*_{\mathrm{cr}} \sim \left( \hat{h}^2 Re_L \right)^{1/2}$.  
Modes are shown in figure \ref{fig:hhat_p125_Reh2_3_Mst_p01_MstSims_Modes}, with FSI DNS mode again showing some a larger component of the third orthogonal beam mode than the quasi-1D model predicts.

In summary, results show that the quasi-1D model well predicts the flutter boundary up to $\hat{h} \le 0.125$ and $\hat{h}^2Re_L \le 10$ for the heavy beam.  The large values of $\hat{h}^2 Re_L$ suggest that $\hat{h}^2 Re_L \ll 1$ need not hold strongly in the limit $M^* \ll 1$.  
The impact of the viscous term on the flutter boundary though non-negligible, is well captured by the parabolic profile description of $\mathcal{N}_x$ and $F_{\mathrm{visc}x}$.  
This suggests that for low $M^*$ values, the inertia associated with the walls largely dominates over that associated with changing the velocity profile shape, i.e. $u$ largely remains locally parabolic as $\delta$ evolves in $t$ and $x$.   

\newcommand{\Mstvar}{0.01 } 

\renewcommand{\hhatvar}{0.05 } 
\renewcommand{\casenumvar}{7 }

\begin{figure}[h]

\sbox0{\begin{subfigure}{0.33\textwidth}
        \captionsetup{width=.8\linewidth,font={footnotesize}}
   		\includegraphics[width=1\textwidth]{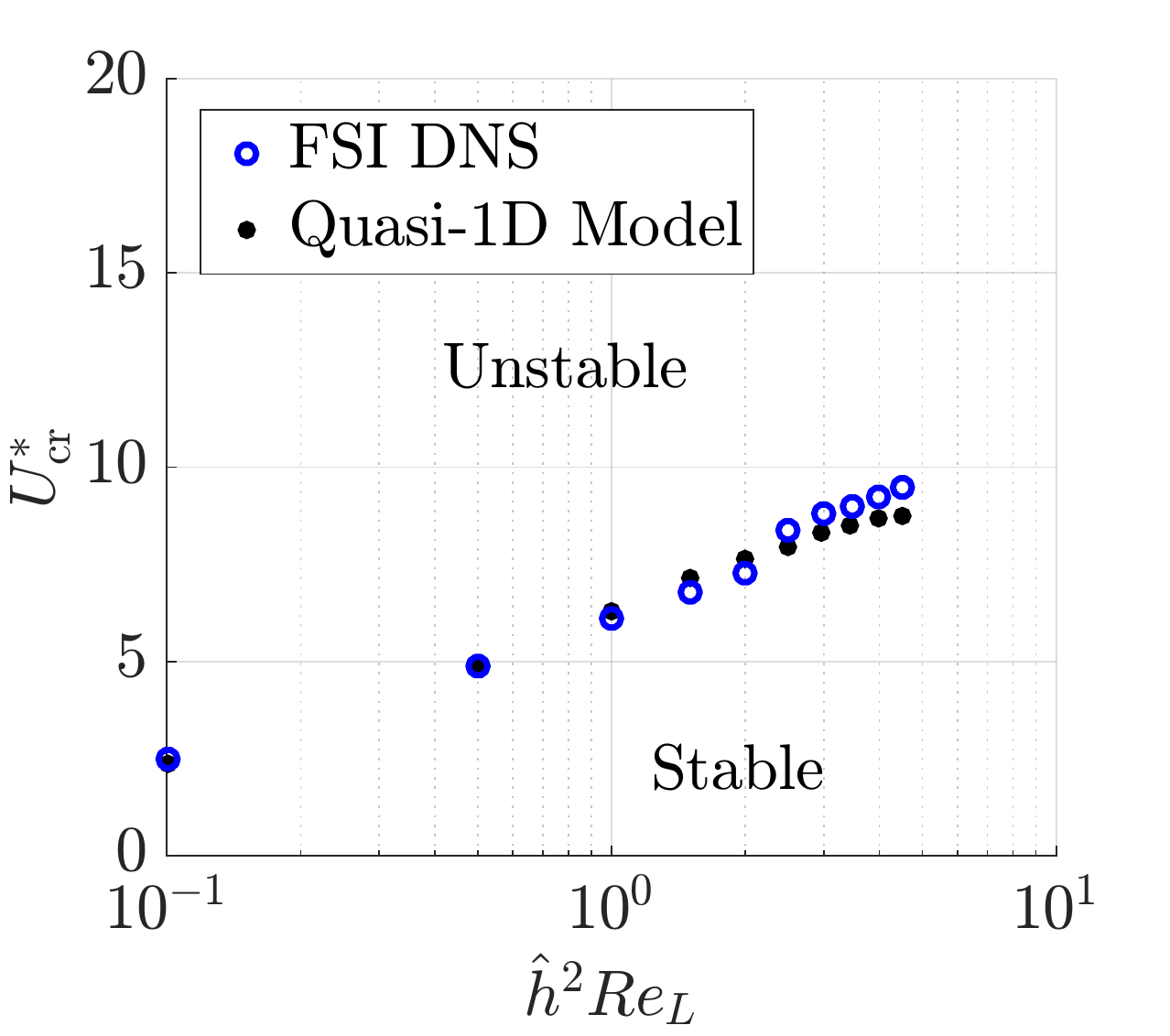}
		\caption{$U^*_{\mathrm{cr}}$ versus $\hat{h}^2 Re_L$.}
		\label{fig:hhat_p05_Reh2_3_Mst_p01_MstSims_CritVals_Ucr}
      \end{subfigure}}
\sbox1{\begin{subfigure}{0.33\textwidth}
        \captionsetup{width=.8\linewidth,font={footnotesize}}
   		\includegraphics[width=1\textwidth]{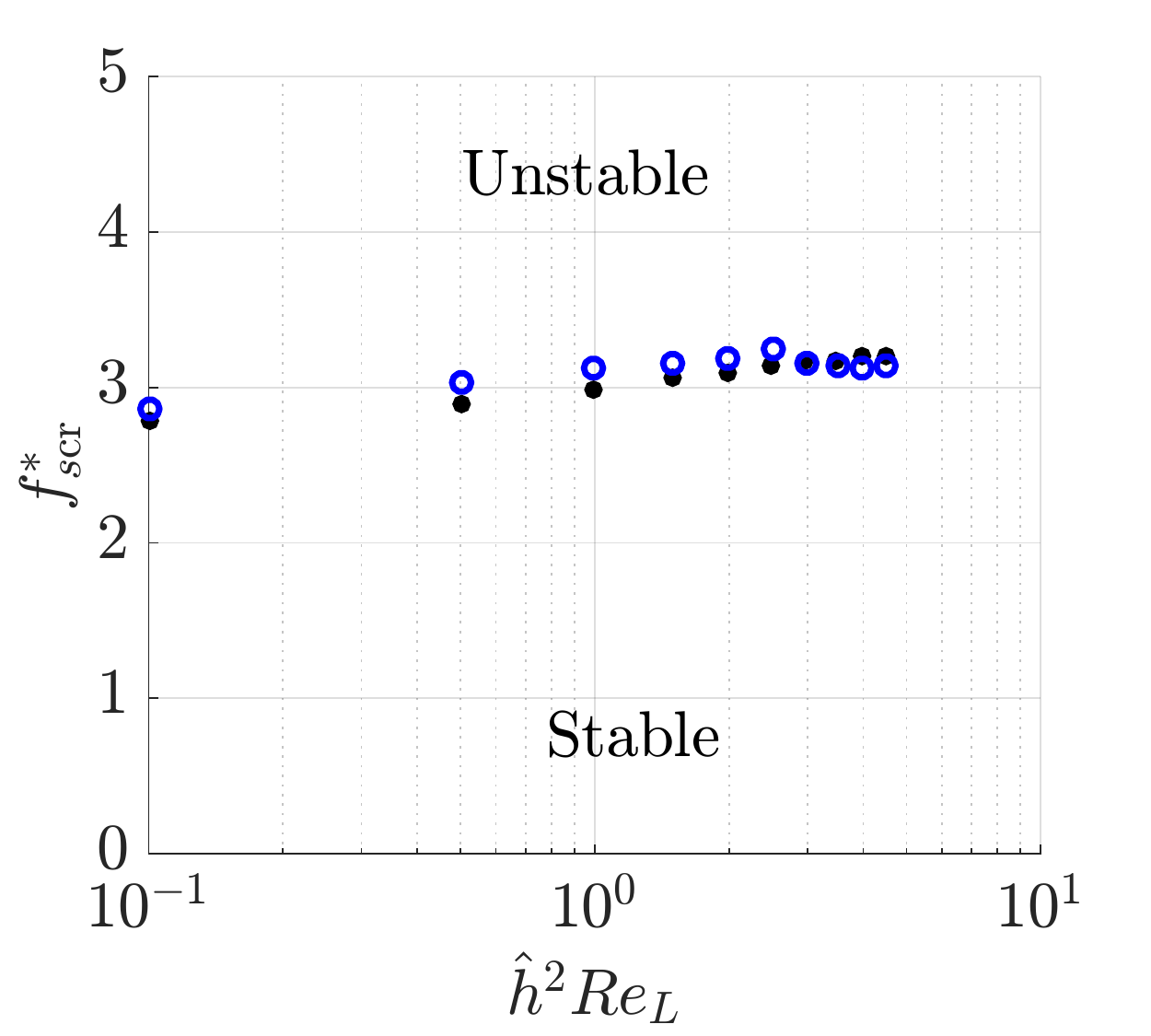}
        \caption{$f^*_{s\mathrm{cr}}$ versus $\hat{h}^2 Re_L$.}
        \label{fig:hhat_p05_Reh2_3_Mst_p01_MstSims_CritVals_fcr}
      \end{subfigure}}
\sbox2{\begin{subfigure}{0.33\textwidth}
        \captionsetup{width=.8\linewidth,font={footnotesize}}
    	\includegraphics[
    	width=\textwidth]{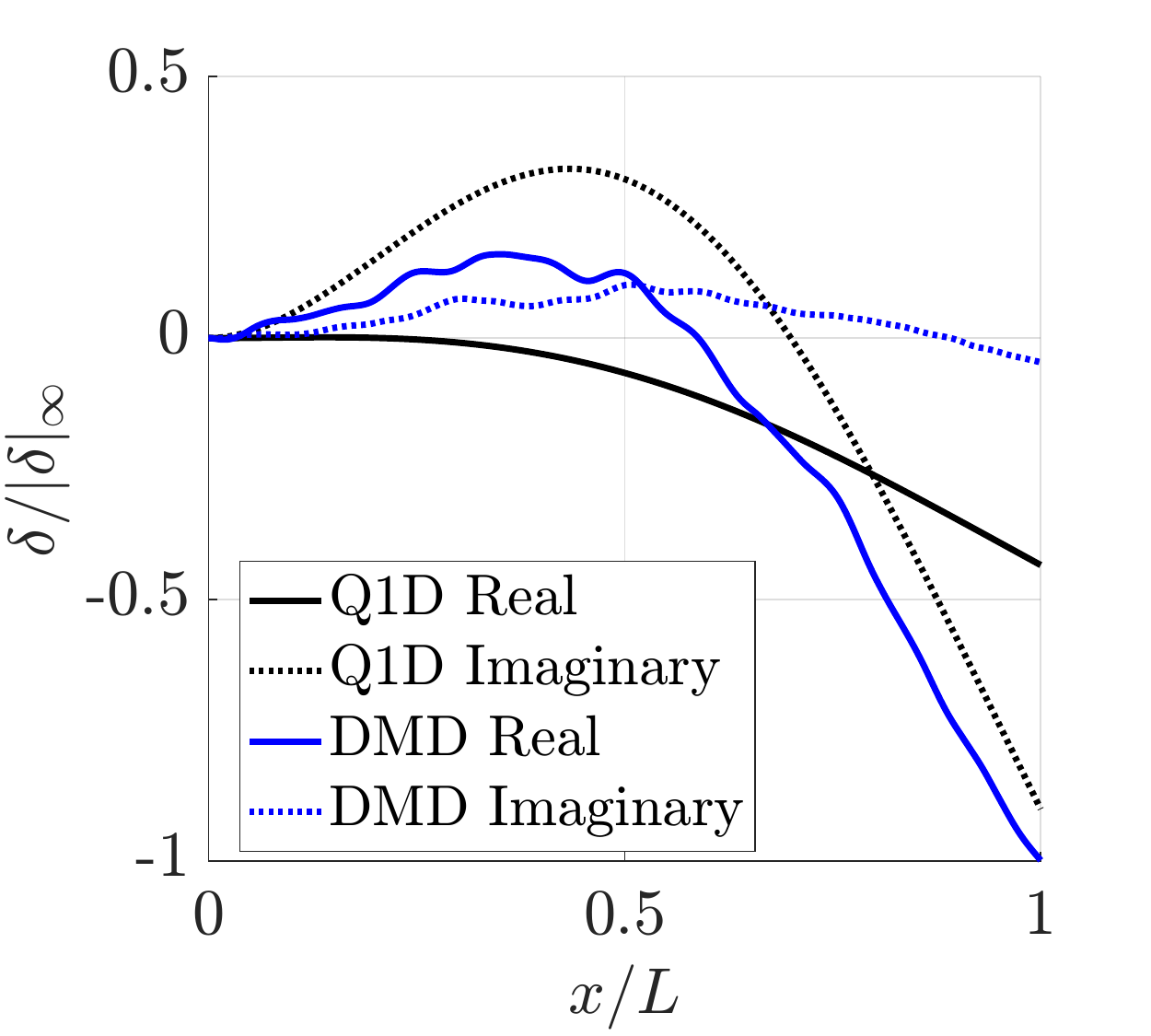}
    	\caption{Mode shapes at $\hat{h}^2 Re_L = 3$.} 
        \label{fig:hhat_p05_Reh2_3_Mst_p01_MstSims_Modes}
      \end{subfigure}}
  \usebox0\hfil \usebox1\hfil  \usebox2\par
	\caption{Comparison of FSI DNS and quasi-1D model for case \casenumvar ($\hat{h} = \hhatvar$,$M^* =\Mstvar$).}
\label{fig:hhat_p05_Reh2_3_Mst_p01_MstSims_CritVals} 
    \label{fig:hhat_p05_Reh2_3_Mst_p01_MstSims_All}
\end{figure}

\renewcommand{\hhatvar}{0.125 } 
\renewcommand{\casenumvar}{8 } 

\begin{figure}[h]

\sbox0{\begin{subfigure}{0.33\textwidth}
        \captionsetup{width=.8\linewidth,font={footnotesize}}
   		\includegraphics[width=1\textwidth]{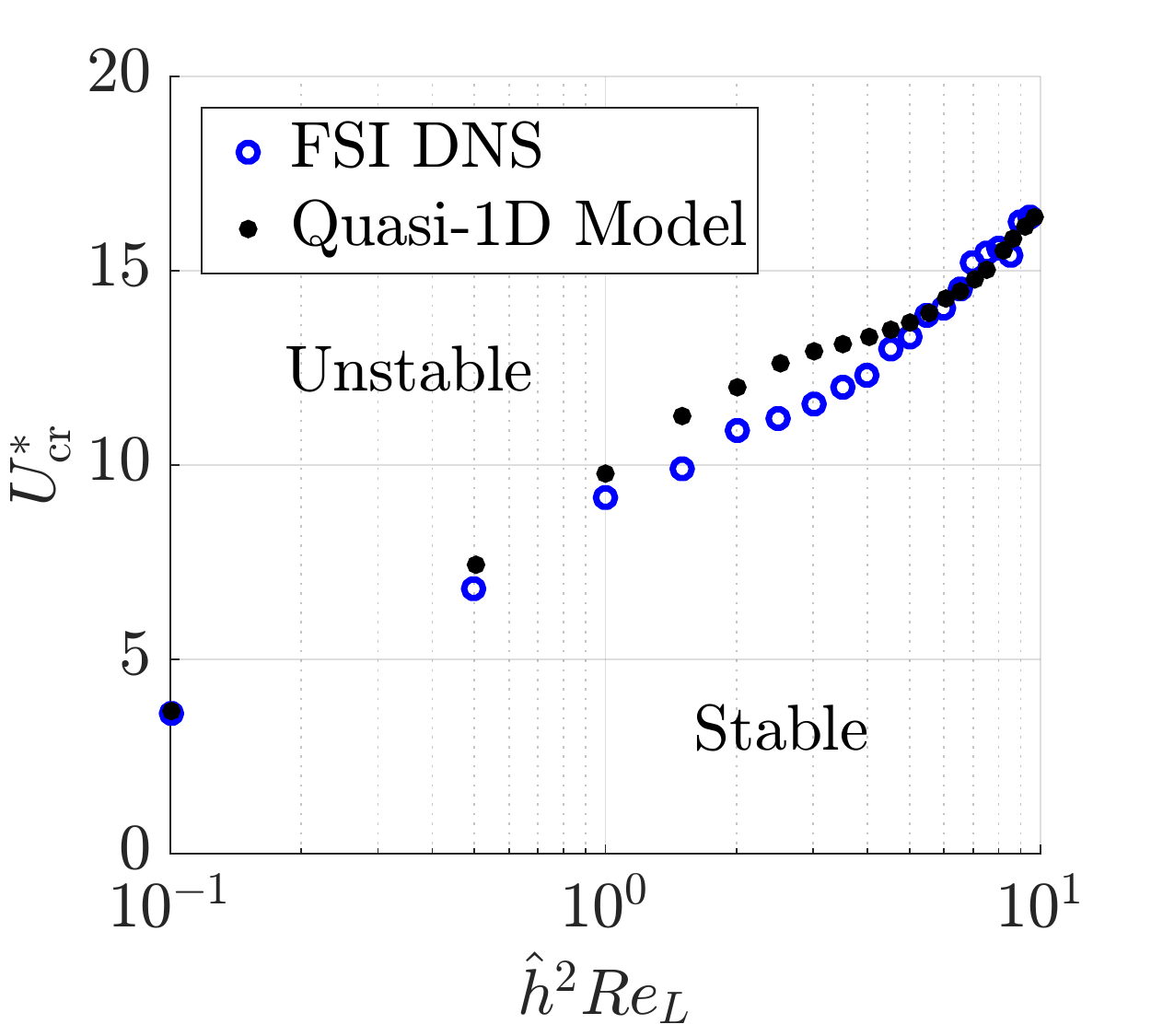}
        \caption{$U^*_{\mathrm{cr}}$ versus $\hat{h}^2 Re_L$.} 
        \label{fig:hhat_p125_Reh2_3_Mst_p01_MstSims_CritVals_Ucr}
      \end{subfigure}}
\sbox1{\begin{subfigure}{0.33\textwidth}
        \captionsetup{width=.8\linewidth,font={footnotesize}}
   		\includegraphics[width=1\textwidth]{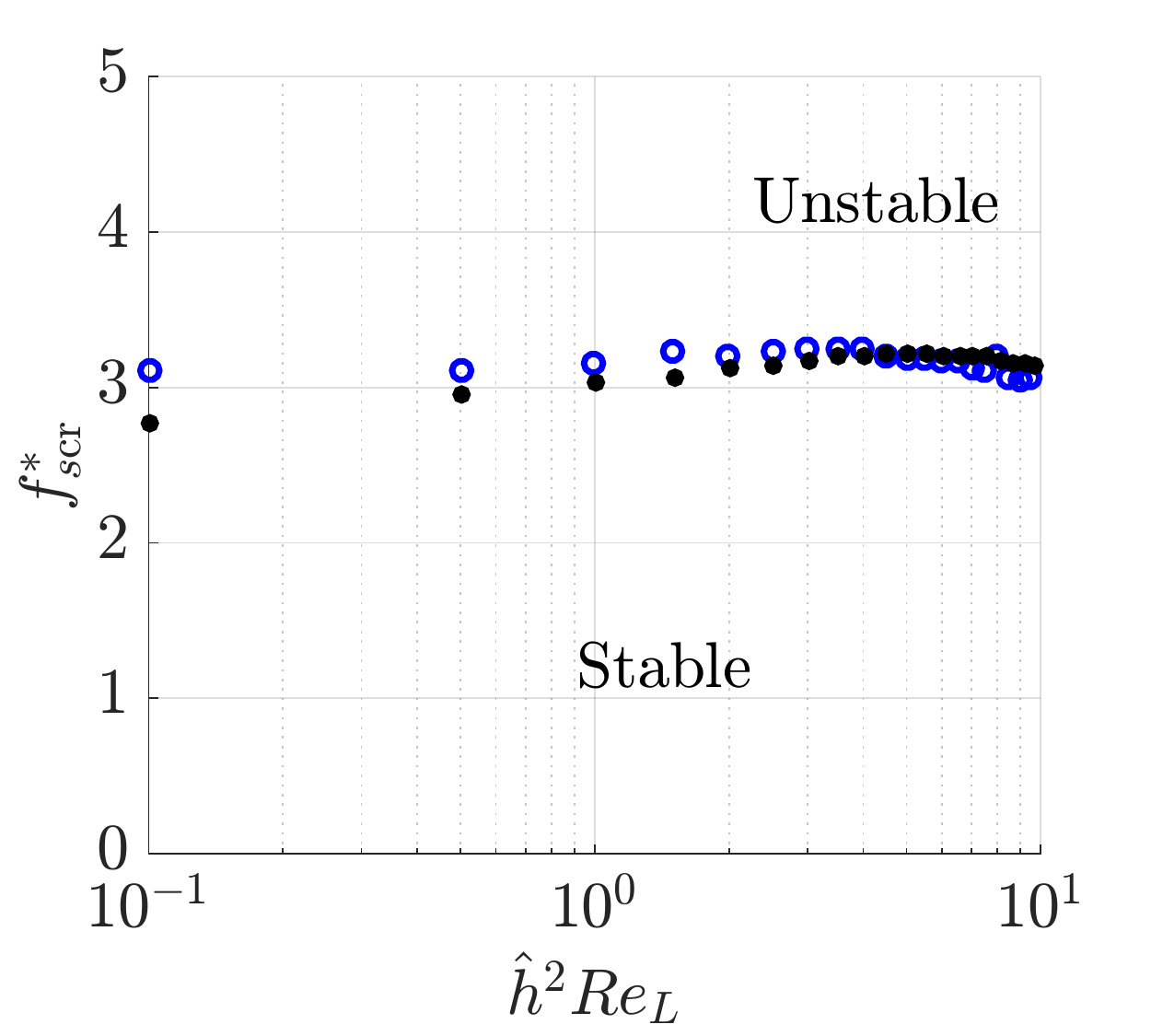}
        \caption{$f^*_{s\mathrm{cr}}$ versus $\hat{h}^2 Re_L$.} 
        \label{fig:hhat_p125_Reh2_3_Mst_p01_MstSims_CritVals_fcr}
      \end{subfigure}}
\sbox2{\begin{subfigure}{0.33\textwidth}
        \captionsetup{width=.8\linewidth,font={footnotesize}}
    	\includegraphics[
    	width=\textwidth]{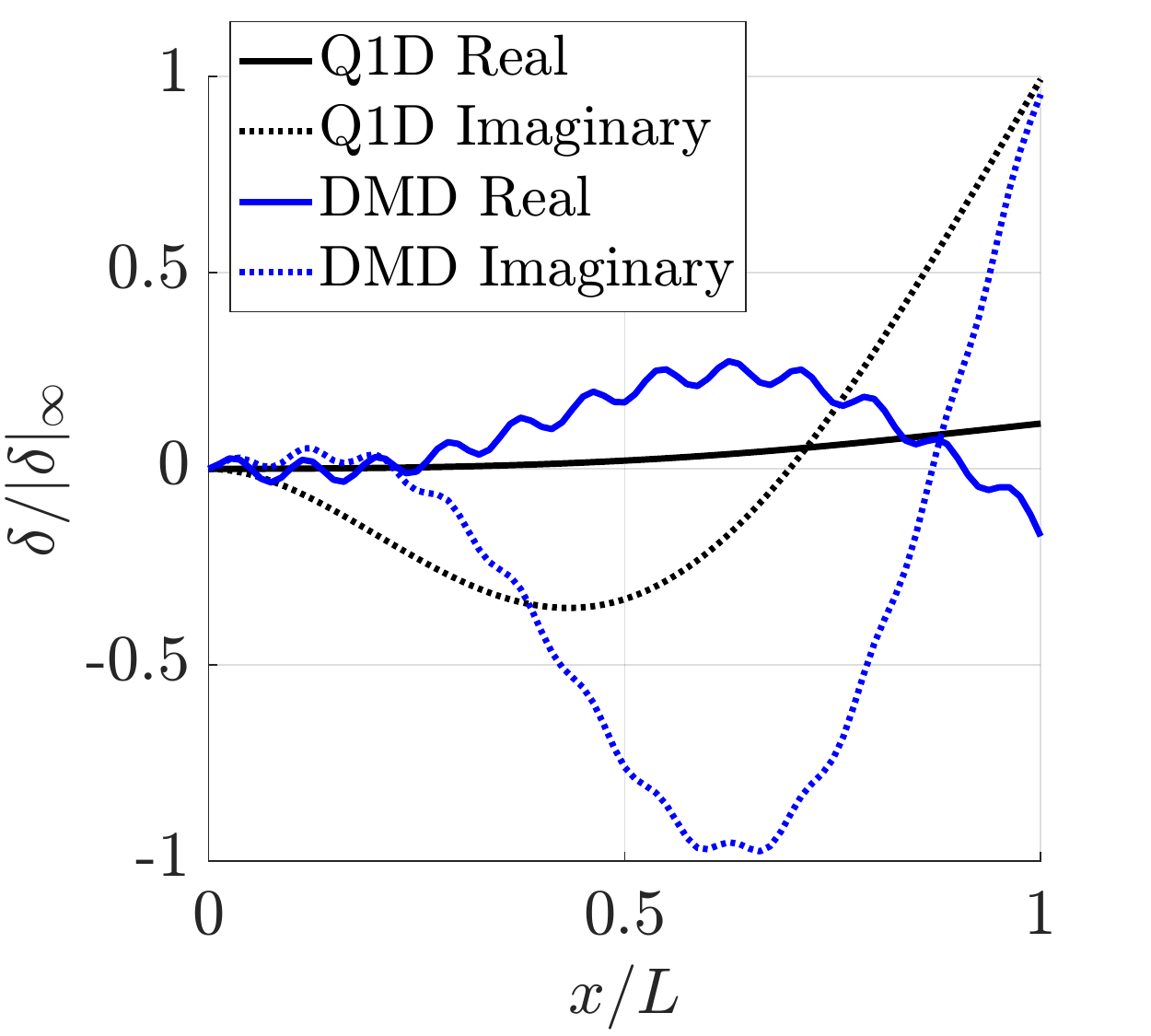}
    	\caption{Mode shapes at $\hat{h}^2 Re_L = 3$.} 
        \label{fig:hhat_p125_Reh2_3_Mst_p01_MstSims_Modes}
      \end{subfigure}}
  \usebox0\hfil \usebox1\hfil  \usebox2\par
	\caption{Comparison of FSI DNS and quasi-1D model for case \casenumvar ($\hat{h} = \hhatvar$,$M^* =\Mstvar$).}
    \label{fig:hhat_p125_Reh2_3_Mst_p01_MstSims_CritVals} 
    \label{fig:hhat_p125_Reh2_3_Mst_p01_MstSims_All}
\end{figure}


\subsection{Quasi-1D Flutter Boundary Comparison to Inviscid Model} \label{sec:InviscidComp}

\begin{figure}[H]   
    \centering
    \includegraphics[width=.55\textwidth]{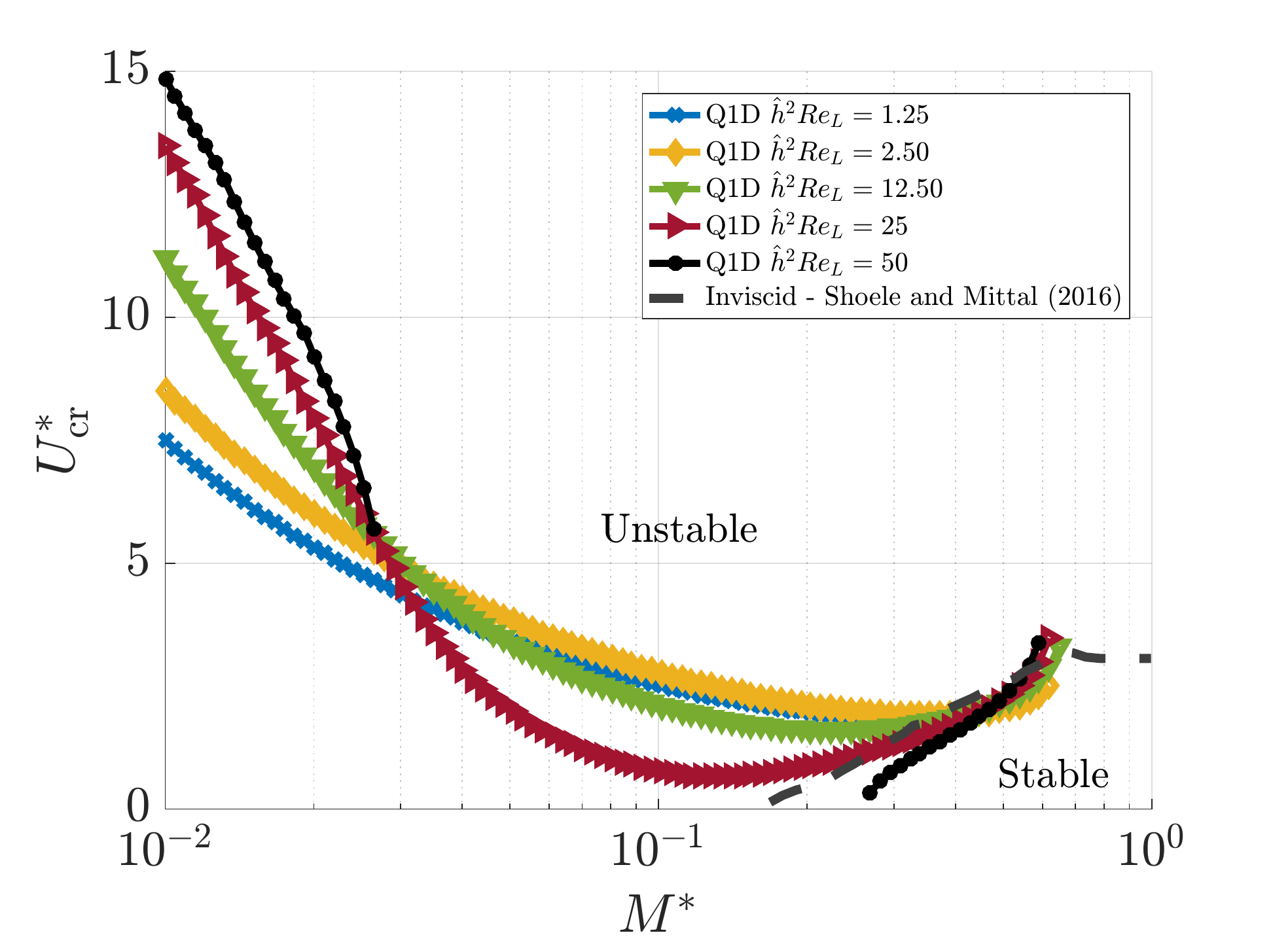}%
  \caption{Comparison of flutter boundary for lowest frequency mode between different quasi-1D model (Q1D) $\hat{h}^2Re_L$ values and inviscid model by \citeauthor{Shoele2015} \citep{Shoele2015} at $\hat{h} = 0.05$. }  
  \label{fig:InvComp}
\end{figure}

The ability of the quasi-1D model to predict the flutter stability behavior even when $\hat{h}^2 Re_L$ is not small appears promising.  We would like to compare quasi-1D predictions in cases where $\hat{h}^2 Re_L \gg 1$ while $\hat{h} \ll 1$ to an inviscid flow solution. 
The inviscid model in \citep{Shoele2015} presents the flutter stability boundary as a function of $\hat{h}$, with $\hat{h} = 0.05$ as the lowest provided value.
Thus, taking the case for $\hat{h} = 0.05$ and $\hat{h}^2Re_L = 1.25$ as the starting point, we compare $U^*_{\mathrm{cr}}$ as a function of $M^*$ for $\hat{h}^2Re_L = [1.25 - 50]$ ($Re_L = [500 - 2\times 10^{4}]$) for the lowest frequency mode branch.  Results are shown in figure \ref{fig:InvComp}.  
Notable trends appear as $\hat{h}^2 Re_L$ increases: first, as $\hat{h}^2Re_L$ increases from 1.25 to 2.50, the system is stabilized as the stability boundary shifts upwards.   Yet as $\hat{h}^2Re_L$ is further increased to 12.50 and thereafter the system is destabilized for $M^* > 0.03$, with the boundary eventually disappearing for $0.03 < M^* < 0.2$ at $\hat{h}^2Re_L = 50$.  This means that no matter the stiffness of the system, the first mode is unstable if the beam is heavy enough, but not too heavy.  The original stabilization trend for increasing $\hat{h}^2Re_L$ remains true, however, for $M^* < 0.03$.   
Furthermore, as  $\hat{h}^2Re_L$ becomes large, the quasi-1D flutter boundary appears to near the inviscid results acquired from \citep{Shoele2015} for $0.1< M^* < 1$, with the mode switching inflection nearly matching over all $\hat{h}^2Re_L$ boundaries shown.  


\citeauthor{Shoele2015}  \citep{Shoele2015} conjectured based on earlier DNS studies \citep{Shoele2014} that $Re_L \approx 200$ was sufficient to observe inviscid behavior for the  $\hat{h}$ values in their study.  Though this may be true for $\hat{h} > 0.125$, beyond which the quasi-1D model's predictions cannot be trusted, the inviscid behavior boundary for $\hat{h} = 0.05$ is predicted at about $Re_L \approx 2 \times 10^{4}$ (figure \ref{fig:InvComp}).  Predictions of the quasi-1D model indicate that the choice of $Re_L$ for inviscid treatment is a strong function of $\hat{h}$.  

In light of these results and those in section \ref{sec:DNSQ1DComp}, we utilize the quasi-1D model to produce the complete flutter boundary for $M^*=0.01$ in figure \ref{fig:Mst_p01_hhatRehhat2var_CritVals} and $M^*=0.1$ in figure \ref{fig:Mst_p1_hhatRehhat2var_CritVals}. Both figures only show the flutter boundary for the lowest frequency mode branch. Three trends become apparent from these plots:
first, as $M^*$ increases, the lowest frequency mode is destabilized significantly;
as $\hat{h}^2 Re_L$ increases, the mode is stabilized. This stabilization is accelerated at higher $\hat{h}$, for values for $\hat{h}^2 Re_L < 10$.  Figure \ref{fig:InvComp} shows that opposite is true as $\hat{h}^2 Re_L$ is increased further. 
Lastly, as $\hat{h}$ increases, the mode is stabilized, as has been pointed out previously \citep{Shoele2014}.

Though $f^*_{s\mathrm{cr}}$ remains within a narrow range for the lowest frequency mode, an interesting pattern arises as $\hat{h}^2 Re_L$ and $\hat{h}$ are varied.  Bands of lower frequency appear alternating with higher frequency states in both $M^*$ values.   This indicates that these parameters have an effect on the frequency response, but it is much less pronounced than their effects on the stability boundary as judged by $U^*_{\mathrm{cr}}$, where it should be noted that over the entire range of $\hat{h}$ and $\hat{h}^2 Re_L$ considered, there is only about 10\% variation in $f^*_{s\mathrm{cr}}$.

\renewcommand{\casenumvar}{2 } 
\renewcommand{\Mstvar}{0.01 } 

\begin{figure}[H]
    \centering
    \begin{subfigure}[t]{0.5\textwidth}
        \centering
   	\includegraphics[width=1\textwidth]{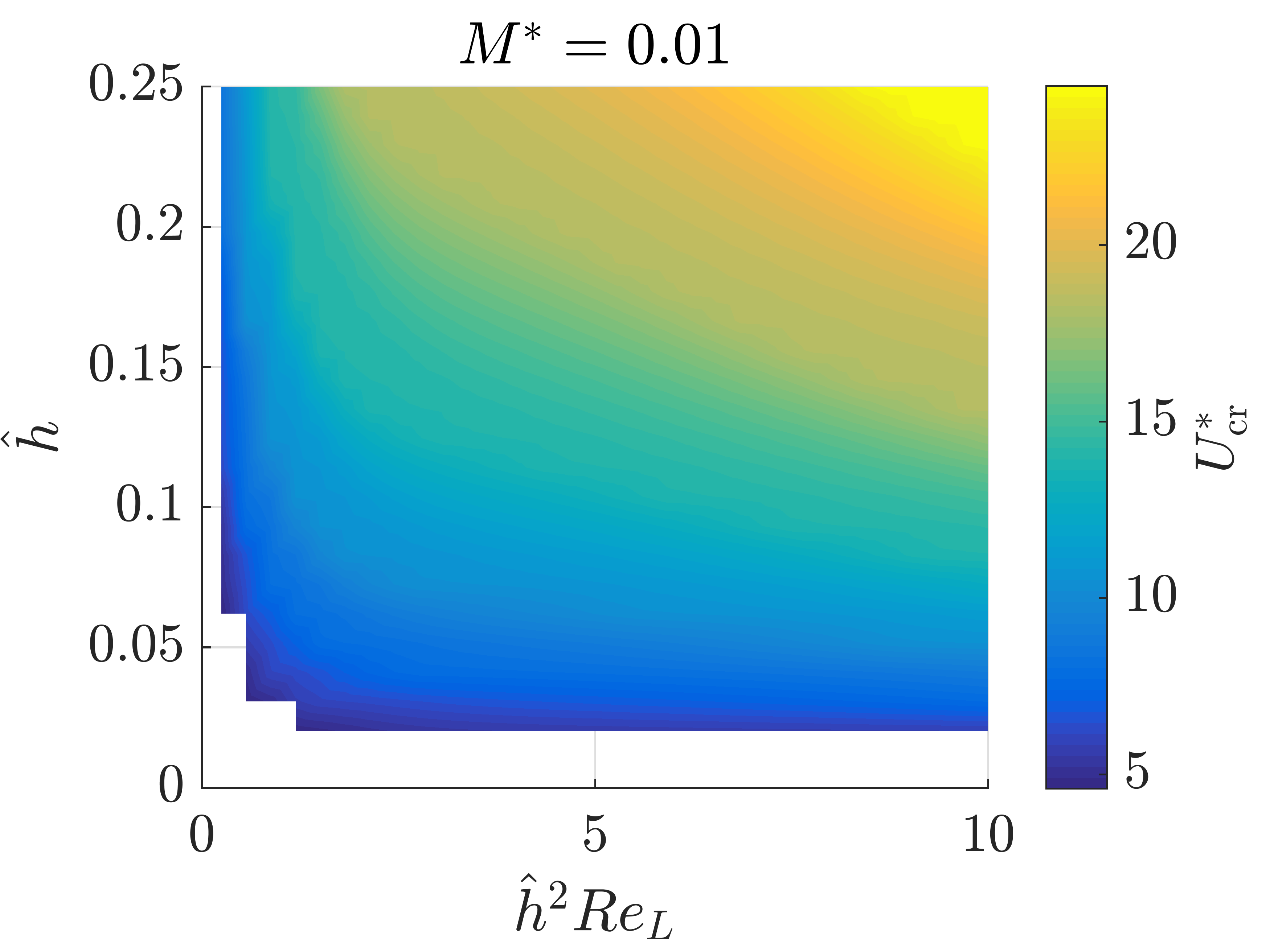}
        \caption{$U^*_{\mathrm{cr}}$ contours vs. $\hat{h}^2 Re_L$ and $\hat{h}$.}  
    \end{subfigure}%
    ~ 
    \begin{subfigure}[t]{0.5\textwidth}
        \centering
   	\includegraphics[width=1\textwidth]{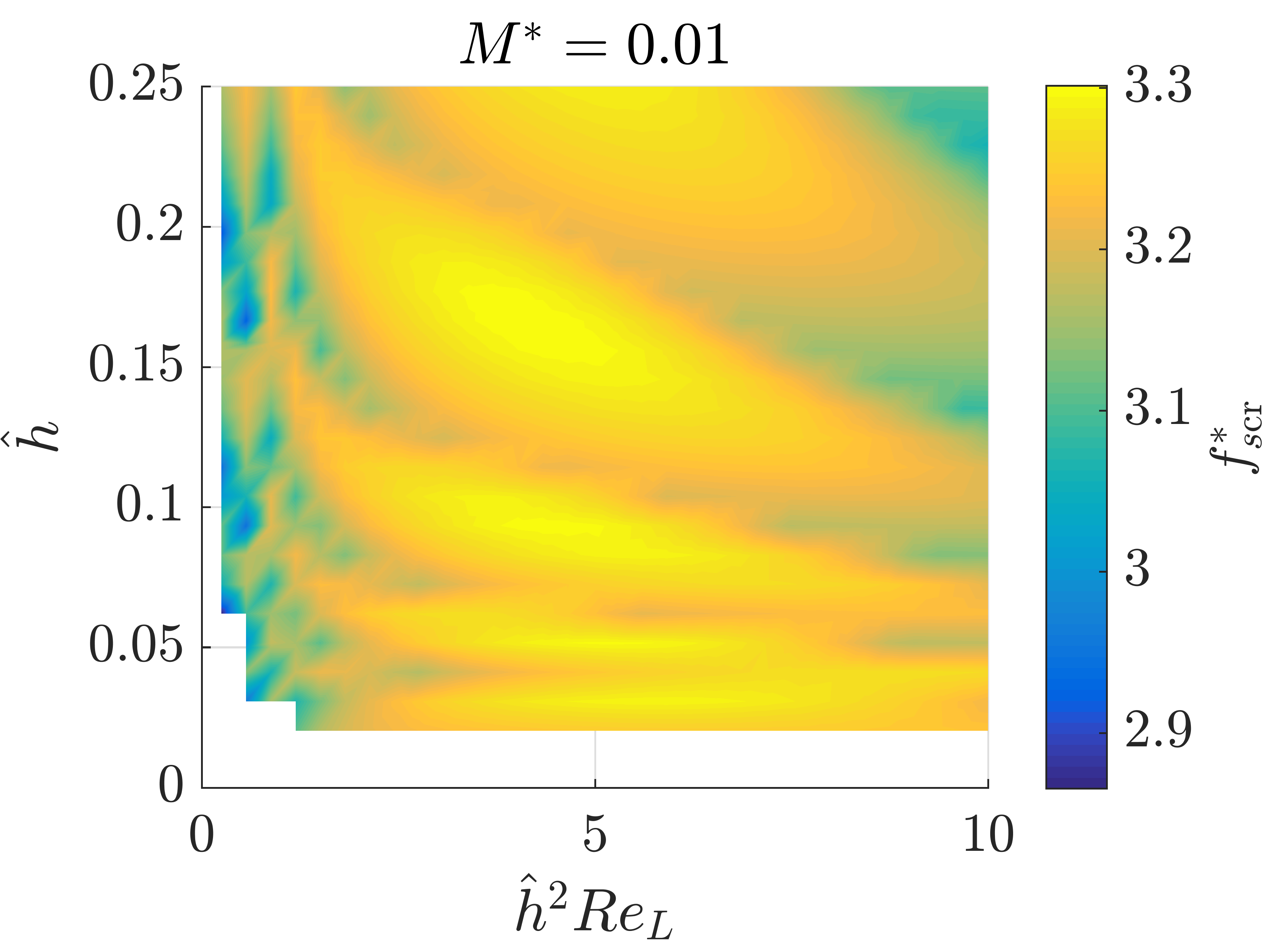}
        \caption{$f^*_{s\mathrm{cr}}$ contours vs. $\hat{h}^2 Re_L$ and $\hat{h}$.}  
    \end{subfigure}
    \caption{Quasi-1D predicted critical flutter values as a function of $\hat{h}^2 Re_L$ and $\hat{h}$ at $M^* =\Mstvar$.}  \label{fig:Mst_p01_hhatRehhat2var_CritVals}
\end{figure}


\renewcommand{\casenumvar}{2 } 
\renewcommand{\Mstvar}{0.1 } 

\begin{figure}[H]
    \centering
    \begin{subfigure}[t]{0.5\textwidth}
        \centering
   	\includegraphics[width=1\textwidth]{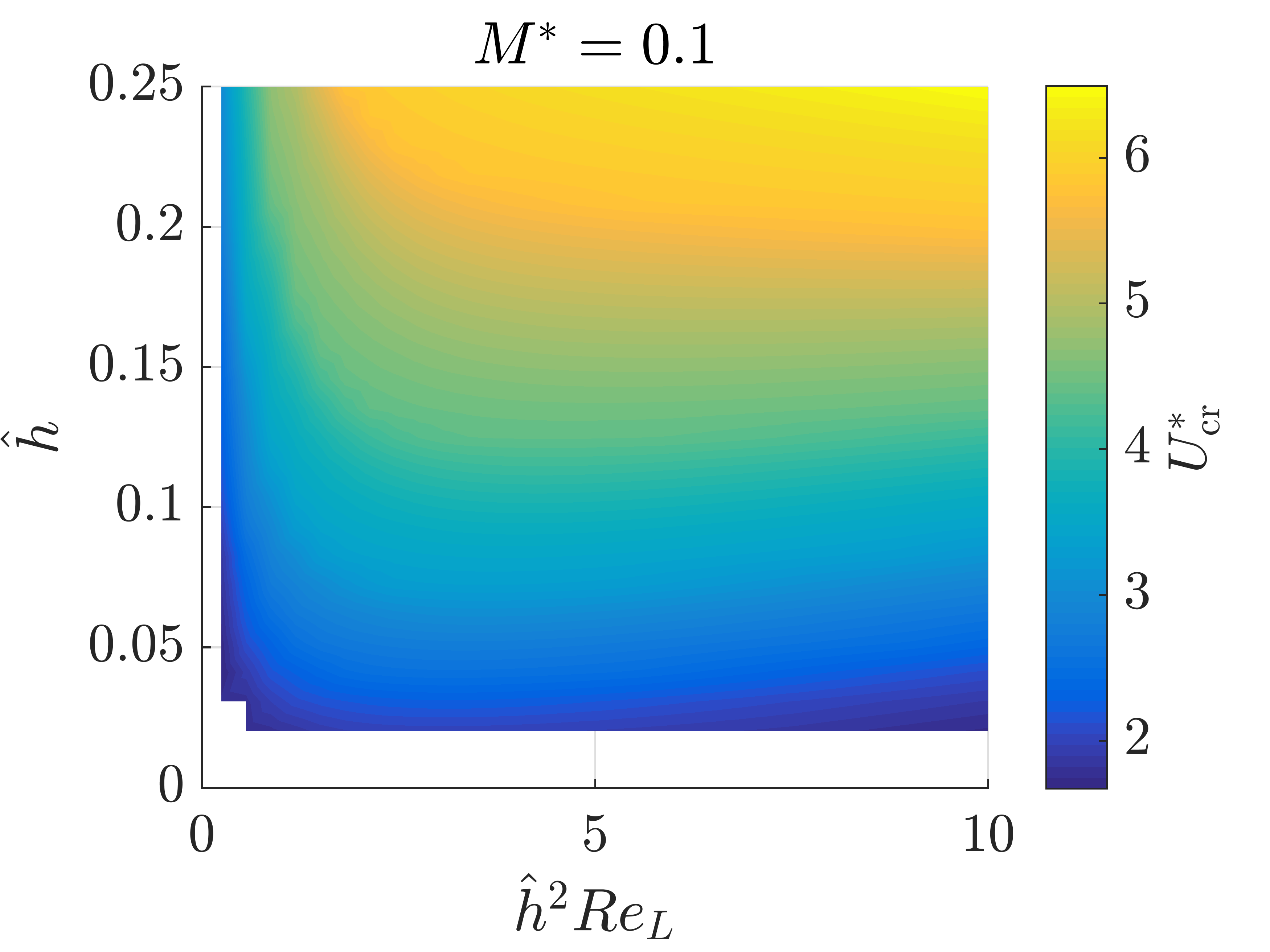}
        \caption{$U^*_{\mathrm{cr}}$ contours as a function of $\hat{h}^2 Re_L$ and $\hat{h}$.}  
    \end{subfigure}%
    ~ 
    \begin{subfigure}[t]{0.5\textwidth}
        \centering
   	\includegraphics[width=1\textwidth]{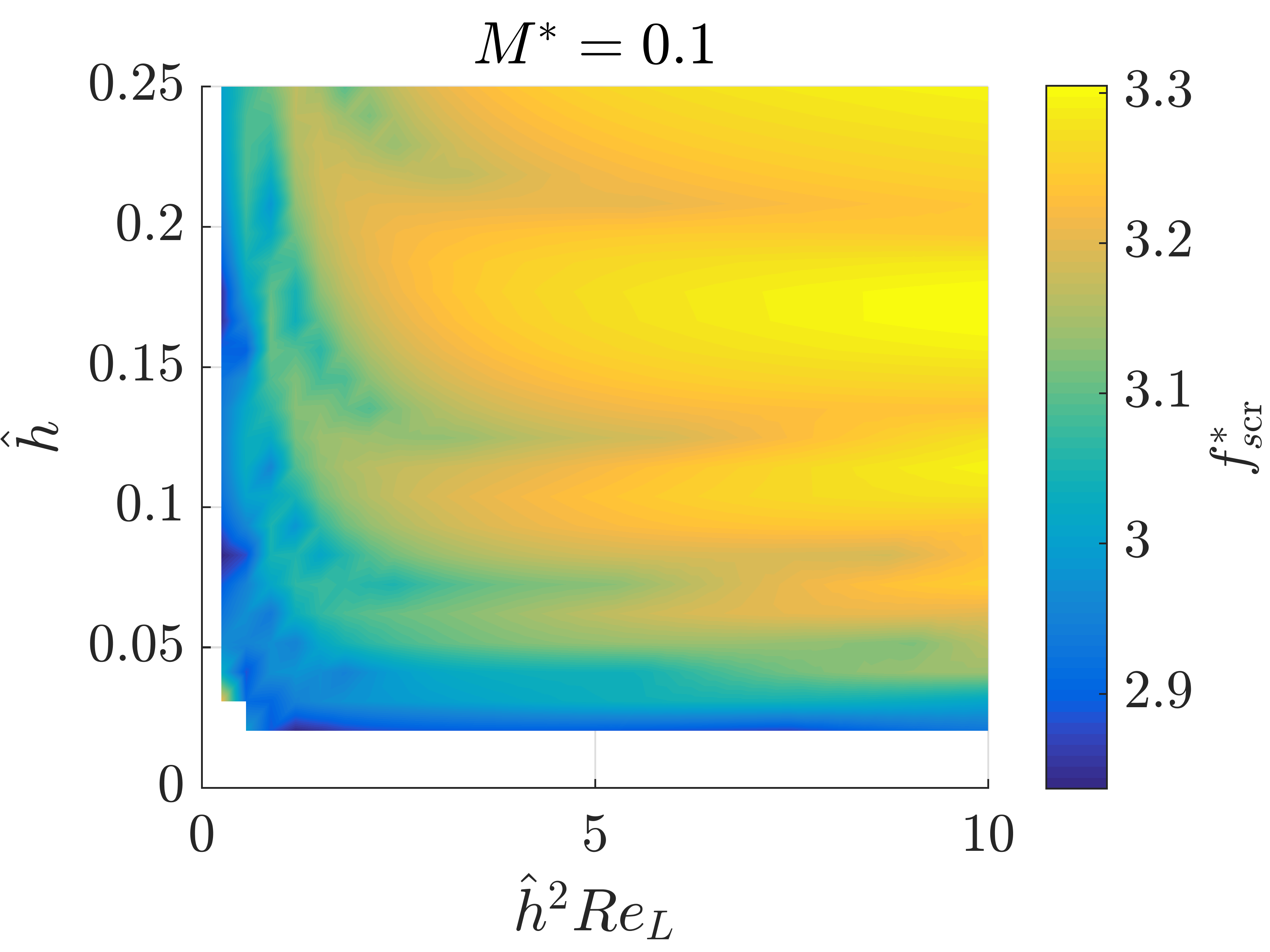}
        \caption{$f^*_{s\mathrm{cr}}$ contours as a function of $\hat{h}^2 Re_L$ and $\hat{h}$.}  
    \end{subfigure}
    \caption{Quasi-1D predicted critical flutter values as a function of $\hat{h}^2 Re_L$ and $\hat{h}$ at $M^* =\Mstvar$.}  \label{fig:Mst_p1_hhatRehhat2var_CritVals}
\end{figure}


\section{Conclusion}
In this study, we derived a model for the fully coupled fluid-structure dynamics of a cantilever in channel flow, and predicted the flutter stability boundary in a constant, symmetric channel.  
The model takes a quasi-one-dimensional approach based on a slowly varying, thin-gap approximation ($h_0'^2 \ll 1$, $\delta'^2 \ll 1$, and $\frac{h}{L } = \hat{h} \ll 1$), and is closed by assuming that the velocity profile across the gap is unchanging, as would strictly be true in the lubrication limit where ${\hat h}^2 Re_L \ll 1$.  We examined the range of validity of the model by performing full numerical simulations of the two-dimensional incompressible Navier-Stokes equations coupled to the large-deflection Euler-Bernoulli beam.

Our results show that the model validity bounds are flexible: by enforcing $\hat{h} \ll 1$ strictly, the model is able to predict flutter properties at both large $\hat{h}^2 Re_L$ and $M^*$ values.  The agreement with FSI DNS values at high $\hat{h}^2 Re_L$ is surprising; yet it is corroborated by the comparison between the inviscid model flutter boundary by \citeauthor{Shoele2015} \citep{Shoele2015} as $\hat{h}^2 Re_L \gg 1$. This suggests that as the viscous term diminishes, the modeled inertial term gives rise to similar balances between the quasi-1D model presented and the inviscid models at small $\hat{h}$ values. 
Notable also is that increasing $\hat{h}^2 Re_L$ at small $\hat{h}$ values presents first a stabilizing effect for both heavy and light cantilevers, followed by a destabilizing effect for light cantilevers.  Heavy cantilevers appear to always become more stable as $\hat{h}^2 Re_L$ increases.  

Furthermore, the model is shown as a reliable alternative to expensive fluid-structure numerical simulations, with the potential of handling various geometries of the flow passage, including asymmetric ones.  It may serve as the first instance to explore the parametric space for device designs or understanding fluid-structure resonance phenomena in internal flows.    

\section*{Acknowledgement}
The authors would like to thank Andres Goza for his invaluable insights on fluid-structure physics and numerical methods that largely comprise the code base for the DNS simulations in this work. We would also like to acknowledge Bosch Energy Research Network (BERN) grant 13.01.CC17 and the NASA Jet Propulsion Laboratory for their support of this research.

\appendix
\section{Two Dimensional Model Coefficients} \label{chp:appendixA}

The linear fluid-structure coupled operator in equation \ref{eq:Eps1LinearEq} is, 
\begin{equation} \label{eq:Eps1FSISystem}
\resizebox{.9 \textwidth}{!}{$
\mathbf{A}
= \mathbf{
\begin{bmatrix}
    0 & \mathbf{1} & 0 & 0 \\
    M^{-1} K & M^{-1} C & M^{-1} T^{\mathrm{bot}} & M^{-1} T^{\mathrm{top}}  
    \\
    -[E_q^{\mathrm{bot}} + B_q^{\mathrm{bot}} (M^{-1} K)] & -[D_q^{\mathrm{bot}} + B_q^{\mathrm{bot}} (M^{-1} C)] & -B_q^{\mathrm{bot}} (M^{-1} T^{\mathrm{bot}} ) &[G_q^{\mathrm{bot}} - B_q^{\mathrm{bot}} (M^{-1} T^{\mathrm{top}} )] 
    \\
    E_q^{\mathrm{top}} + B_q^{\mathrm{top}} (M^{-1} K) & D_q^{\mathrm{top}} + B_q^{\mathrm{top}} (M^{-1} C) & G_q^{\mathrm{top}} + B_q^{\mathrm{top}} (M^{-1} T^{\mathrm{bot}} ) & B_q^{\mathrm{top}} (M^{-1} T^{\mathrm{top}} )  \\
\end{bmatrix} }
$},
\end{equation}
where, 
\begin{equation} \label{eq:Eps1MCKCoeffs}
\resizebox{.9 \textwidth}{!}{$
\begin{aligned}
M_{ij} &= \frac{1}{N_{ij}} \int_0^L\left(  M_{\mathrm{s} i}(x) +   M_{\mathrm{f} i}^{\mathrm{bot}}(x) + M_{\mathrm{f} i}^{\mathrm{top}}(x)   \right) g_j(x) dx, \ 
C_{ij} =  -\frac{1}{N_{ij}}\int_0^L\left( C_{\mathrm{f} i}^{\mathrm{bot}}(x) + C_{\mathrm{f} i}^{\mathrm{top}}(x) \right) g_j(x) dx, \\ 
K_{ij} &=  -\frac{1}{N_{ij}}\int_0^L\left( K_{\mathrm{s} i}(x) + K_{\mathrm{f} i}^{\mathrm{bot}}(x) + K_{\mathrm{f} i}^{\mathrm{top}}(x) \right) g_j(x) dx,  \
N_{ij} = \int_0^L g_i(x) g_j(x) dx. 
\end{aligned}
$}
\end{equation}
exist in $\mathbb{R}^{n+1 \ \times \ n+1}$, and 
\begin{equation} \label{eq:Eps1TCoeffs}
T_j^{\mathrm{bot}} = \frac{1}{N_{ij}}\int_0^L T_{\mathrm{f}}^{\mathrm{bot}}(x)  g_j(x) dx, \  
T_j^{\mathrm{top}} = - \frac{1}{N_{ij}} \int_0^L\ 
T_{\mathrm{f}}^{\mathrm{top}}(x) g_j(x) dx,
\end{equation}
exist in $\mathbb{R}^{n+1 \ \times \ 1}$.
The coefficients are defined in terms of the basis expansion as follows.  Let 
\begin{equation} \label{eq:IntDef}
F_{\alpha,ijk}^{lmn}(x) = \int_{0}^{x} \frac{\mathcal{L}_i \left[ g_{\alpha}^l \left(\tilde{x} \right) \right]}
{\mathcal{L}_j \left[ h_e^m\left(\tilde{x} \right) \right]} \mathcal{L}_k \left[ h_e^n\left(\tilde{x} \right) \right] d\tilde{x},
\end{equation}
where $g_{\alpha}$ is the basis function of index $\alpha$ defined in equation \ref{eq:ClampedFreeBC_BasisExpansion}, $h_e$ is the equilibrium channel gap, $l$, $m$, and $n$ are their exponents, and $\mathcal{L}_i \left[ \cdot \right]$ is the linear operator of order $i \in \mathbb{Z} : > -2$, defined as

\begin{equation} \label{eq:LinOpDef}
\mathcal{L}_i \left[ \cdot \right] =
\begin{cases}
\frac{d^i}{d \tilde{x}^i} \left( \cdot \right) & \text{if } i \geq 0\\
\int_{0}^{\tilde{x}} \left( \cdot \right) dx' & \text{if } i = -1
\end{cases}.
\end{equation}
We also define the ratio,
\begin{equation} \label{eq:Frat}
R(x) = \frac{F_{\alpha,000}^{010}(x)}{F_{\alpha,000}^{010}(L)}.
\end{equation}

Given equations \ref{eq:IntDef}, \ref{eq:LinOpDef}, and \ref{eq:Frat}, the following are the coefficients from equation \ref{eq:Eps1MCKCoeffs}: added fluid mass,
\begin{equation} \label{eq:Mfi_2D}
M_{\mathrm{f}i} = \rho_f \left[ F_{i,-100}^{110}(L) R(x) - F_{i,-100}^{110}(x) \right],
\end{equation}
fluid damping,
\begin{equation}  \label{eq:Cfi_2D}
\begin{aligned}
C_{\mathrm{f}i} = \rho_{f} q_{x0} \left\{ \frac{\zeta_{\mathrm{out}}}{h_e^2(L)} F_{i,-100}^{100}(L) R(x) + 
\left( \frac{f_0}{2} + \frac{q_{x0} \eta}{4} \right) \left( F_{i,-100}^{130}(L) R(x) - F_{i,-100}^{130}(x) \right) + 
\right. \\ \left.
2 \xi_x \left[ F_{i,-101}^{131}(x) - F_{i,000}^{120}(x) + \left( F_{i,000}^{120}(L) - F_{i,101}^{131}(L) \right) R(x) \right]
\right\},
\end{aligned}
\end{equation}
fluid stiffness,
\begin{equation}  \label{eq:Kfi_2D}
\begin{aligned}
K_{\mathrm{f} i} = \rho_{f} q_{x0}^2  \left\{ \xi_x  \right[ 3 F_{i,001}^{141}(x) - F_{i,100}^{130}(x) + \left( F_{i,100}^{130}(L) - 3F_{i,001}^{141}(L) \right) R(x) \left] + 
\right. \\ \left.
\frac{3 f_0}{4} \left( F_{i,000}^{140}(L) R(x) - F_{i,000}^{140}(x) \right) + 
\left( \frac{\zeta_{\mathrm{in}} g_i(0)}{ h_e^3(0)} + \frac{\zeta_{\mathrm{out}} g_i(L)}{ h_e^3(L)} \right) R(x) - \frac{\zeta_{\mathrm{in}} g_i(0)}{h_e^3(0)}
\right\},
\end{aligned}
\end{equation}
and boundary flow rate forcing,  
\begin{equation}  \label{eq:Tfi_2D}
\begin{aligned}
T_{\mathrm{f} i} = \rho_{f}  q_{x0} \left\{ \left( \frac{\zeta_{\mathrm{in}}}{h_e^2(0)} + \frac{\zeta_{\mathrm{out}}}{h_e^2(L)} \right) R(x) - 
\frac{\zeta_{\mathrm{in}}}{ h_e^2(0)} + 
2 \xi_x \left(  F_{i,001}^{031}(x) - F_{i,001}^{031}(L) R(x)  \right) + 
\right. \\ \left.
\left( \frac{f_0}{2} + \frac{q_{x0} \eta}{4} \right) \left( F_{i,000}^{030}(L) R(x) - F_{i,000}^{030}(x) \right)
\right\}.
\end{aligned}
\end{equation}

The structure coefficients are 
\begin{equation} \label{eq:Eps1StructCoeffs}
\begin{aligned}
M_{\mathrm{s} i} = \rho_s h_b g_i(x), \
K_{\mathrm{s} i} = E I_{\mathrm{2D}} \frac{\mathrm{d}^4}{\mathrm{d} x^4} g_i(x).
\end{aligned}
\end{equation}

The boundary flow-rate forcing components of matrices in \ref{eq:Eps1FSISystem} are, 

\begin{equation} \label{eq:Bqi_2D}
B_{qi} = -\frac{F_{i,-100}^{110}(L)}{F_{i,000}^{010}(L)},
\end{equation}

\begin{equation} \label{eq:Dqi_2D}
D_{qi} = \frac{q_{x0}}{F_{i,000}^{010}(L)} \left\{ 2 \xi_x \left( F_{i,-101}^{131}(L) - F_{i,000}^{120}(L) \right) -
\left( \frac{f_0}{2} + \frac{q_{x0} \eta}{4} \right) F_{i,-100}^{130}(L) - \frac{\zeta_{\mathrm{out}}}{h_e^2(L)} F_{i,000}^{100}(L)
\right\},
\end{equation}

\begin{equation} \label{eq:Eqi_2D}
E_{qi} = \frac{q_{x0}^2}{F_{i,000}^{010}(L)} \left\{ \xi_x \left( 3 F_{i,001}^{141}(L) - F_{i,100}^{130}(L) \right) - 
\frac{3 f_0}{4} F_{i,000}^{140}(L) - \frac{\zeta_{\mathrm{in}} g_i(0)}{h_e^3(0)} - \frac{\zeta_{\mathrm{out}} g_i(L)}{h_e^3(L)}
\right\},
\end{equation}

\begin{equation} \label{eq:Gq_2D}
G_q = \frac{q_{x0}}{F_{i,000}^{010}(L) } \left[ 2 \xi_x F_{i,001}^{031}(L) - \left( \frac{f_0}{2} + \frac{q_{x0} \eta}{4} \right) F_{i,000}^{030}(L) - 
\frac{\zeta_{\mathrm{in}}}{h_e^2(0)} - \frac{\zeta_{\mathrm{out}}}{h_e^2(L)}
\right].
\end{equation}

For laminar flow, substituting $f_0$ from equation \ref{eq:FrictionTerm_Shimoyama}, we can show that
\begin{equation}
\eta = \left. \frac{\mathrm{d} f}{\mathrm{d} Q_{x}} \right|_{Q_{x} = q_{x0}} = 
-\frac{48}{Re_{\bar{h}}} \frac{1}{q_{x0}} = -\frac{f_0}{q_{x0}},
\end{equation}
and that viscous coefficients scales as 
\begin{equation}
f_0 \frac{L}{\bar{h}} = \frac{48}{Re_{\bar{h}}} \frac{L}{\bar{h}} 
\sim \left( \hat{h}^2 Re_L \right)^{-1}.
\end{equation}

\newpage

\bibliographystyle{model2-names}
\bibliography{Tosi_ModelingConstChannelFSI.bib}







\end{document}